\documentclass[a4paper,fleqn,usenatbib]{mnras}

\usepackage{mathptmx}
\usepackage[T1]{fontenc}
\usepackage{ae,aecompl}
\usepackage{graphicx}        
\usepackage[dvipsnames]{xcolor}
\usepackage{amsmath}
\usepackage{tabularx}
\usepackage{arydshln}
\usepackage{float}

\usepackage{stackengine} 
\usepackage{csquotes}

\usepackage{amssymb} 
\usepackage{amsmath} 
\usepackage{tikz}
\usepackage{times}
\usepackage{color}

\usepackage{textcomp}
\usepackage{graphicx}
\graphicspath{{Figures/}}
\usepackage{subfigure} 
 
\newcommand{\gccm}{\textrm{g\,cm}^{-3}}

\newcommand{\km}{\textrm{km}}
\newcommand{\ms}{\textrm{ms}}

\def\part_n{\partial_\perp}

\def\Gclas{{\mathcal G}_{\rm (C)}}
\def\Grel{{\mathcal G}_{\rm (R)}}
\def\csrel{c_{{\rm s}_{\rm (R)}}}
\def\cscla{c_{{\rm s}_{\rm (C)}}}

\begin{document}

\title{Neutron star collapse and gravitational waves with a non-convex equation of state}
\author[M.~A.~Aloy,  J.~M.~Ib\'a\~nez, N.~Sanchis-Gual, M.~Obergaulinger, J.~A.~Font, S.~Serna and A.~Marquina]
{Miguel A.~Aloy$^{1}$, Jos\'e M.~Ib\'a\~nez$^{1,2}$, Nicolas Sanchis-Gual$^{1}$, Martin Obergaulinger$^{1}$, 
\newauthor 
 Jos\'e A.~Font$^{1,2}$, Susana Serna$^{3}$, and Antonio Marquina$^{4}$
\\
$^1$Departamento de
  Astronom\'{\i}a y Astrof\'{\i}sica, Universitat de Val\`encia,
  Dr. Moliner 50, 46100, Burjassot (Val\`encia), Spain
  \\
$^2$Observatori Astron\`omic, Universitat de Val\`encia, C/ Catedr\'atico 
  Jos\'e Beltr\'an 2, 46980, Paterna (Val\`encia), Spain
  \\
$^3$Departament de Matem\`atiques, Universitat Aut\`onoma de Barcelona, Edifici C, 08193 Bellaterra (Barcelona), Spain
   \\
$^4$ Departamento de
Matem\'aticas, Universitat de Val\`encia,
Dr. Moliner 50, 46100, Burjassot (Val\`encia), Spain 
\\
}

\date{\today}

\maketitle

\begin{abstract}
  The thermodynamical properties of the equation of state (EoS) of
  high-density matter (above nuclear saturation density) and the
  possible existence of exotic states such as phase transitions from
  nuclear/hadronic matter into quark-gluon plasma, or the appearance
  of hyperons, may critically influence the stability and dynamics of
  compact relativistic stars. From a theoretical point of view,
  establishing the existence of those states requires the analysis of
  the ``convexity'' of the EoS. We show indications of the existence
  of regions in the dense-matter EoS where the thermodynamics may be
  non-convex as a result of a non-monotonic dependence of the sound
  speed with the rest-mass density. When this happens,
  non-conventional dynamics may develop. In this paper we investigate
  the effects of a phenomenological, non-convex EoS on the equilibrium
  structure of stable compact stars and on the dynamics of unstable
  neutron stars that collapse gravitationally to black holes, both for
  spherically symmetric and uniformly-rotating configurations. We show
  how the dynamics of the collapse with a non-convex EoS departs from
  the convex case, leaving distinctive imprints on the gravitational
  waveforms. The astrophysical significance of these results for
  microphysical EoSs is discussed.
\end{abstract}

\begin{NoHyper}
\section{Introduction}
\label{section:intro}

A long-standing, fundamental, and still open issue in relativistic
astrophysics is the knowledge of the equation of state (EoS)
describing the thermodynamical properties of high-density matter,
i.e.~matter at densities above nuclear-matter. Such extreme conditions
are achieved in the cores of neutron stars. Theoretical progress
towards the understanding of this issue relies on electromagnetic
observations and heavy-ion experiments
(see~\citet{Glendenning2000,Heiselberg2000,Glendenning2001,Weber2005,Camenzind2007,Haensel2007}
and references therein). With the recent observations of gravitational
waves from mergers of binary black holes and binary neutron
stars~\citep{Abbott1,Abbott2,Abbott:2017vtc,GW170814-prl,Abbott:2017a,GW170608,Abbott_2017ApJ...848L..12}
a new channel to collect complementary information and improve our
understanding of the dense-matter EoS has already opened \citep[see,
e.g. the recent constraints obtained
in][]{Abbot_2018PhRvL.121p1101,Margalit_2017ApJ...850L..19,
  Annala_2018PhRvL.120q2703, De_2018PhRvL.121i1102,
  Malik_2018PhRvC..98c5804, Most_2018PhRvL.120z1103M,
  Radice_2018ApJ...852L..29, Raithel_2018ApJ...857L..23,
  Zhou_2018PhRvD..97h3015}. However, despite the ongoing efforts, the
issue has not been fully addressed thus far.

There are many reasons why this matter must be worked out. First,
because the properties of the dense-matter EoS and the possible
existence of exotic states such as phase transitions (PTs) to
quark-gluon plasma or associated with the presence of hyperons in the
core of neutron stars, may critically influence the stability and
dynamics of these objects. Furthermore, a third family of compact
stars, more compact and denser than neutron stars, and originated by
the appearance of quark phases in the core of neutron stars, has been
long suggested~\citep{Schertler2000,Glendenning2000a}. More recently,
the observations of two high-mass pulsars, PSR
J1614-2230~\citep{Demorest2010} and PSR
J0348-0432~\citep{Antoniadis2013}, has also placed severe constraints
on the dense-matter EoS. In particular, the softening of the EoS due
to the presence of hyperons or PTs to quark matter or boson
condensates is prone to affect the stability of neutron
stars~\citep{Bednarek2012,Zdunik2013}. The possibility of different
types of phases, i.e.~neutrons and quarks, coexisting in dense matter
is currently under intense scrutiny (see~\citet{Buballa2014} and
references therein). Moreover, the possible existence of hybrid stars
has recently been considered by~\citet{Bejger2017}.

The dense-matter EoS also plays a fundamental role in the evolution
(on a hydrodynamical timescale) of archetypal scenarios of
relativistic astrophysics such as core-collapse supernovae, short- and
long-duration progenitors of gamma-ray bursts, the cooling of
proto-neutron stars, the formation of stellar-mass black holes (BHs),
or the merger of compact-binary systems.  In particular, and in the
context of PTs, the dynamics of neutron star cores
collapsing to BHs has been analyzed numerically in spherical symmetry
by~\citet{Abdikamalov2009} and by~\citet{Peres2013}. In the former
work the collapse is induced by a PT from hadronic
matter to deconfined quark matter, while in the latter the collapse is
induced by a PT to hyperonic matter. The corresponding
extensions of these works including rotation can be found
in~\citet{Dimmelmeier2009,Bejger2012} and~\citet{Peres2013b}.



From a theoretical point of view, the existence of such exotic states
of matter in the dense-matter EoS also requires the analysis of the
``convexity'' of the EoS. Relevant contributions towards the knowledge
of the properties of non-convex thermodynamics induced by some EoS
were made in the pioneering works of~\citet{Bethe1942},
\citet{Zeldovich1946}, and~\citet{Thompson1971}. In
particular~\citet{Thompson1971} introduced the concept of {\it
  fundamental derivative} in gas dynamics. Nowadays, fluids which
display a region of negative values of the fundamental derivative are
called Bethe-Zel'dovich-Thompson fluids, or BZT fluids
(see~\citet{Voss2005} and references therein). A classical example is
provided by a Van der Waals EoS. In this EoS, besides the mixing
regime where different phases coexist, there is a region of
non-convexity, where the fundamental derivative is
negative~\citep{Menikoff1989}. BZT fluids have drawn some attention in
the last fifteen years due to their potential applications in industry
\citep[see, e.g.][]{Cinnella2008,Guardone2010}.  Unlike a regular
fluid, a BZT fluid might condense on isentropic
compression\footnote{As a side remark, we point out that BZT flows may
  show a non-convex dynamics in which compound waves as, for example,
  rarefaction shocks, can develop during their evolution
  \cite[see][]{Argrow:1996,Guardone2002,Voss2005,Cinnella2006,Serna_Marquina:2014}.}.

The extension to relativistic fluid dynamics of previous studies on
BZT fluids in the framework of {\it classical} fluid dynamics was
accomplished by~\citet{Ibanez2013}. This work presented the conditions
under which the hyperbolic system of relativistic Euler equations is
convex. The authors considered a perfect fluid obeying a causal EoS
and the results were obtained by analyzing the properties of the
characteristic fields of the relativistic hydrodynamics
equations. Following~\citet{Thompson1971} the conditions were given in
terms of the so-called (classical) fundamental derivative,
$\Gclas$.

A classical, and somewhat academic, example of a thermodynamical
system in which the adiabatic index displays a non-monotonous
behaviour with the density is the region around the neutron-drip point
in cold catalyzed dense matter~\citep{Shapiro1983}. Similar regions
appear also in EoSs derived from a field-theoretical model for nuclear
and neutron matter, both at zero temperature~\citep{Diaz-Alonso1985}
and at finite temperature~\citep{Marti1988}. Moreover, the most
popular EoSs used in recent hydrodynamical simulations of compact
stars display again, in some regions of the space of thermodynamical
parameters, a non-monotonous behaviour of the adiabatic exponent with
the density \citep[as can be seen in the fittings reported
by][]{Haensel2002,Haensel2004,Haensel2007,Bauswein2010}. Those regions
are good candidates to develop non-convex thermodynamics.

Other examples can be found at densities much higher than nuclear
saturation density ($n_0 \approx 0.16$ fm$^{-3}$) at which
nuclear/hadronic matter undergoes a transition into a quark-gluon
plasma (QGP). The nature of the finite-temperature QCD transition
remains ambiguous~\citep{Aoki2006} and may even evolve from a
crossover transition at low baryon number density to a first-order PT
at high baryon number density with the existence of a critical point.
 Using QCD lattice techniques, the HotQCD
Collaboration~\citep{Bazavov2014} and the Wuppertal-Budapest
Collaboration~\citep{Borsanyi2014}, have reported results
about the EoS characterizing the transition from the hadronic phase
into the QGP phase. Their results, which favor the crossover nature of
the transition, in the continuum extrapolated EoS and in the
phenomenologically relevant range of temperature, $130-400$ MeV, show
similarities regarding the trace anomaly, pressure, energy density and
entropy density. The energy density in the crossover region,
$145\le T$ (MeV) $\le 163$, is a factor of about 1.2 to 3.1 times the
energy density at nuclear saturation density, and the sound speed has
a minimum $\simeq 0.38$ \citep{Bazavov2014} within the former
interval.  Following a different strategy, which combines the
knowledge of the EoS of hadronic matter at low densities with the
observational constraints on the masses of neutron
stars,~\citet{Bedaque2015} conclude that the speed of sound of dense
matter is not a monotonous function of the energy density, with local
maximum and minimum above and below $1/\sqrt{3}$, respectively.  As we
show in Sec.\,\ref{section:GGL-EoS}, a non-convex region in the space
of thermodynamical parameters appears where the adiabatic index is an
strong enough decreasing function of the density. This is equivalent
to demanding that the classical local sound speed be a sufficiently
steep decreasing function of the density. Roughly, regions where the
sound speed is not a monotonic function of the density are suitable to
develop a non-convex thermodynamics. Indeed, as we shall see in this
paper, these regions may also develop a distinctive hydrodynamic
behaviour in the course of the collapse of unstable neutron star-like
configurations.

Motivated by the above indications of the existence of possible
regions in the dense-matter EoS where the thermodynamics can be
non-convex, we present in this paper a numerical study of the
structure and dynamics of compact stellar configurations described by
a BZT fluid. We choose a particularly simple form of the EoS, namely
an ideal gas EoS with an adiabatic index which depends on the density
\citep{Ibanez:2017TUBOS}. While this phenomenological EoS can only be
regarded as a {\em toy-model}, it serves nonetheless to exemplify the
particularities that appear when the EoS is non-convex. For our study
we consider two different situations, firstly, the equilibrium
structure of stable compact stars and, secondly, the dynamics of
unstable  neutron stars that collapse gravitationally to BHs, both for spherically 
symmetric and uniformly-rotating initial 
configurations. A future study using actual microphysical EoS
from nuclear physics will be presented elsewhere.

This paper is organized as follows: Section\,\ref{section:Micro-EoS}
shows that non-convex thermodynamics may exist in various
microphysical EoS of common use in astrophysical scenarios such as
massive stellar core collapse. In most cases, this convexity loss is
associated to the existence of first-order
PTs. Section~\ref{section:GGL-EoS} describes our toy-model, non-convex
EoS. This EoS is employed to obtain the results presented in the
following three sections.  Section~\ref{section:TOV-GGL-EoS} discusses
the structure of spherically-symmetric relativistic stellar
equilibrium configurations, while Sections~\ref{section:Collapse-1d}
and \ref{section:Collapse-GGL-EoS} analyse the dynamics of unstable
configurations which promptly collapse producing a central BH in
spherical and axial symmetry, respectively. Since, as we show below,
the effects of convexity loss are bound to very compact collapsing
cores, the observational signature of this \emph{anomalous}
thermodynamics may potentially be best noticed on the
gravitational-wave signature. Thus, from the collapsing, axisymmetric,
rotating cores we present in Section~\ref{section:Collapse-GGL-EoS},
we calculate their gravitational-wave emission aiming at identifying
features that differentiate convex dynamics from non-convex ones.
Finally, the conclusions of our work are presented in
Section~\ref{section:summary}.

\section{Non-convexity in microphysical EoS employed in stellar core collapse}
\label{section:Micro-EoS}

In classical fluid dynamics, the convexity of a thermodynamical system is determined by the EoS~\citep{Menikoff1989} and, more specifically, by the so-called {\it fundamental derivative}, $\Gclas$ 
\begin{equation}
\Gclas
:= - \frac{1}{2} \,V \,\displaystyle{\frac{\displaystyle{\left.\frac{\partial^2 p}{\partial V^2}\right|_s}}{\displaystyle{\left.\frac{\partial p}{\partial V}\right|_s}}}\,,
\label{G1}
\end{equation}
where $V:= 1/\rho$ the specific volume, $\rho$ the rest mass density, $p$ the pressure and $s$ is the specific entropy. The fundamental derivative measures the convexity of the isentropes in the $p-V$ plane. If $\Gclas > 0$ then the isentropes in the $p-V$ plane are convex and the rarefaction waves are expansive.

The relationship between the classical and the relativistic fundamental derivatives was found in~\citet{Ibanez2013} and is given by
\begin{equation}                                                                
\Grel= \Gclas - \frac{3}{2} \,\csrel^2 \,,
\label{G6}
\end{equation} 
where $\csrel$ is the relativistic sound speed, related to the classical definition of the sound speed,
\begin{equation}
\cscla^2 \, = \,
\displaystyle{\left.\frac{\partial p}{\partial \rho} \right|_s \, , }
\label{eq:sound_class}
\end{equation}
through the relation $\cscla^2 =h \csrel^2$, where $h = 1 + \epsilon +
p/\rho$ is the specific enthalpy and $\epsilon$ the specific energy.

Equation\,(\ref{G1}) can be cast in two different forms \citep[][see
also \citealt{Ibanez:2017TUBOS}]{Menikoff1989} which are useful to
understand the physical origin of the sign of the fundamental
derivative (i.e. the root of the convexity loss). In terms of the
adiabatic index, $\Gamma_1$,
\begin{equation}
\Gamma_1 := \displaystyle{\left.\frac{\partial \ln p}{\partial \ln \rho} \right|_s }
=
\displaystyle{ \frac{\rho}{p}\, \cscla^2}
\label{Gamma1deff}
\end{equation}
and its density derivatives (at constant entropy) one finds
\begin{equation}
{\mathcal G}_{\rm (C)}
= \displaystyle{
\frac{1}{2} \left( 1 + \Gamma_1 + \left.\frac{\partial  \ln \Gamma_1}{\partial \ln \rho} \right|_s \right),
}
\label{eq:G2}
\end{equation}
\noindent
and in terms of the derivatives of the sound speed
\begin{equation}
{\mathcal G}_{\rm (C)}
= \displaystyle{ 1 + \left.\frac{\partial \ln{c_{\rm s_{(C)}}}}{\partial \ln{\rho}} \right |_s}.
\label{eq:G3}
\end{equation}
Thus, it is clear that a necessary condition for the fundamental
derivative to be negative is that
$\displaystyle{\left. \partial \Gamma_1 / \partial \rho \right|_s \, <
  \, 0}$. Alternatively, it is sufficient for $\Gclas <0$ that
$\displaystyle{\left. \partial \ln{\cscla} / \partial \ln{\rho}
  \right|_s \, < \, -1}$.

We remind the reader that thermodynamics places no constraint on the
sign of the fundamental derivative \citep{Menikoff1989}. A system with
a negative fundamental derivative may be thermodynamically stable as
long as $\Gamma_1 \ge 0$, which implies that the energy per unit mass
remains strictly convex (as a function of $V$) along an isentrope.

\subsection{Sample of microphysical EoS}
\label{section:Micro-EoS-Sample}

We have performed a survey of a few nuclear-matter EoS which can be
found in the Compstar Online Supernovae Equations of State
(CompOSE)\footnote{http://compose.obspm.fr} looking for regions of the
parameter space in which either the relativistic or the classical
fundamental derivative become negative. We do not aim to exhaustively
check all the possible dense-matter EoSs. Instead, we shall see that
some of the EoSs we consider here (all of which have been used in the
context of stellar core collapse) display regions where the
thermodynamics is 
potentially
non-convex. The EoS from the CompOSE database have been included to
sample cases in which baryons are treated as non-relativistic
particles or, alternatively, they are included within a suitable
relativistic theory. Also, we have considered variants of the latter
cases where different parameter sets of a relativistic mean-field
(RMF) theory are available and employed in astrophysical
simulations. Finally, different variants of the EoS account for the
possibility of transitions to quark matter or include more exotic
particles such as hyperons. We note that the tables employed to
compute the fundamental derivatives are evaluated at exactly the same
values of the baryon number density, $n$, and charge fraction
$Y_q=n_q/n$ ($n_q$ is the charge density of strongly interacting
particles) as in the CompOSE database. This is an important point,
since high-order derivatives of the thermodynamic variables (like
those needed for the calculation of the fundamental derivatives) may
display small amplitude, high-frequency oscillations associated to the
discretization of the EoS table. While this problem is minor in
regions where the relativistic fundamental derivative is positive and
significantly different from zero, it may affect the determination of
a ``physically sound'' non-convex region when the fundamental
derivatives are close to zero. We warn the reader on the ``numerical''
loss of convexity associated to insufficiently fine thermodynamic
discretization of some tabulated EoS when the adiabatic index is
non-constant \citep{Vaidya:2015A&A...580A.110V}.%
\footnote{As a technical note, we point out that to reduce the
  numerical noise in the evaluation of the fundamental derivatives we
  tabulate the EoS as a function of $n$, of $Y_q$ and of the entropy
  per baryon, $s$, using the CompOSE public software. This is
  specially useful since we make use of derivatives at constant
  entropy in the expressions of $\Gclas$ (Eq.\,\eqref{eq:G3}; below)
  and $\Grel$ (Eq.\,\eqref{G6}).}

The EoS of \cite{Lattimer:1991NuPhA.535..331} with compression modulus
$K = 220$\,MeV (LS220) is considered here in two variants:
\texttt{LS220} and \texttt{LS220($\Lambda$)}. The difference between
both cases is the introduction of $\Lambda-$hyperons in the second one
(following
\citealt{Oertel:2012PhRvC..85e5806,Gulminelli:2013PhRvC..87e5809}). The
hyperon-nucleon interaction is taken from the model by
\cite{Balberg:1997NuPhA.625..435}.  The
\cite{Lattimer:1991NuPhA.535..331} EoS assumes that the nuclear
interaction is an effective non-relativistic Skyrme type model without
momentum dependence.
Nucleons are treated as non-relativistic particles; $\alpha-$particles as hard spheres 
forming an ideal Boltzmann gas. As the density increases, nuclei
dissolve into homogeneous nuclear matter above saturation density. The
(assumed first-order) PT to bulk nuclear matter is
treated by a Maxwell construction. Photons and electrons/positrons are
included as a free gas. The low density extension, below the validity
range of the original Lattimer and Swesty EoS is based on a nuclear
statistical equilibrium model by \cite{Oertel:2012PhRvC..85e5806}. A
first application of this EoS in the supernova context is described in
\cite{Peres:2013PhRvD..87d3006}. As for the \texttt{LS220($\Lambda$)}
EoS, it has been broadly employed in the literature
(e.g. \citealt{Obergaulinger:2014MNRAS.445.3169,Obergaulinger_Aloy:2017}).

The EoS of \cite{Shen:1998PThPh.100.1013,Shen:1998NuPhA.637..435}
(dubbed STOS, hereafter) uses the Thomas-Fermi and variational
approximations with a RMF model. It has been considered here in two
variants. The first one \texttt{STOS}, with only baryonic
contributions (no leptons or photons included). The second variant of
the Shen et al. EoS, \texttt{STOS(B165)}, includes a transition to
quark matter
\citep{Sagert:2009PhRvL.102h1101,Sagert:2010JPhG...37i4064,Fisher:2011ApJS..194...39}.
The transition from the hadronic to the quark phase is done via a
Gibbs construction \citep[as
in][]{Drago_1999JPhG...25..971,Nakazato_2008PhRvD..77j3006} and
employing a bag model for the quark phase with a bag constant of
$B^{1/4} = 165$\,MeV and a strong interaction constant
\citep{Alford_2007Natur.445E...7} $\alpha_s = 0.3$. In both cases the
EoS employs a non-linear RMF model with the TM1 parametrization
\citep{Sugara:1994NuPhA.579..557} of the effective interaction.  Only
neutrons, protons, alpha particles and a single heavy nucleus as well
as electrons/positrons and photons are considered.
The \texttt{STOS(B165)} EoS yields a maximum gravitational mass of
$1.67 M_\odot$ \citep{Sagert:2010JPhG...37i4064} and, therefore, it is
not compatible with the maximum masses observed for neutron
stars. However, this EoS received some attention in the past decade
since it my leave an effect on the supernova explosion mechanism due
to the formation of a secondary shock wave induced by the QGP PT
\citep{Sagert:2009PhRvL.102h1101}.\footnote{Recently,
  \cite{Fischer_2018NatAs...2..980} has shown that the transition to
  the QGP phase may be the engine of supernova explosions in blue
  supergiants.} The high degree of isospin asymmetry and the presence
of temperatures of a few MeV in the early post-bounce phase of core
collapse may induce the transition to the quark phase already around
the saturation number density.

We have also included cases in which the hadronic EoS is based on the
statistical model of \cite{Hempel:2010NuPhA.837..210} (HS) and with
RMF interactions of different types.  These are the EoS tagged with
\texttt{BHB($\Lambda\phi$)}, \texttt{HS(DD2Y)}, \texttt{HS(TMA)},
\texttt{HS(TM1)}, \texttt{HS(NL3)}, \texttt{SFHO}, \texttt{SFHX}. The
first and second ones, \texttt{BHB($\Lambda\phi$)}
\citep{Banik:2014ApJS..214...22} and \texttt{HS(DD2Y)}
\citep{Marques_2017PhRvC..96d5806}, asume RMF interactions DD2
\citep{Typel:2010PhRvC..81a5803}, and include $\Lambda-$hyperons
interacting via $\phi$ mesons, neutrons, anti-neutrons, protons,
anti-protons, lambdas, anti-lambdas, and nuclei. The EoS tagged with
\texttt{HS(TMA)}, \texttt{HS(TM1)}, \texttt{HS(NL3)}, \texttt{SFHO},
and \texttt{SFHX} include RMF interactions with parameterizations TMA
\citep{Toki:1995NuPhA.588..357}, TM1
\citep{Sugara:1994NuPhA.579..557}, NL3
\citep{Lalazissis:1997PhRvC..55..540}, SHFo
\citep{Hempel:2010NuPhA.837..210} and SFHx
\citep{Hempel:2010NuPhA.837..210}, respectively, and contributions
from neutrons, anti-neutrons, protons, anti-protons, electrons,
positrons, photons, and nuclei. Applications of HS EoS for various
different RMF interactions in supernova simulations can be found in
\cite{Hempel:2012ApJ...748...70} and
\cite{Steiner:2012nuco.confE..38}.

The EoS of \cite{Shen:2011PhRvC..83c5802}, to which we will refer as
GSHen in the following, is based on a RMF model to self-consistently
calculate non-uniform matter at intermediate density and uniform
matter at high density. At low densities, a virial expansion for a
non-ideal gas of nucleons and nuclei is used to obtain the EoS. Three
variants of the GSHen EoS are included in our sample:
\texttt{GSHen(FSU1)}, \texttt{GSHen(FSU2)} and
\texttt{GSHen(NL3)}. The differences between them are due to the
distinct approximations employed within the RMF model. They employ
either the FSUGold (FSU1; \citealt{Todd-Rutel:2005PhRvL..95l2501}) or
FSU2 parameters sets of \cite{Shen:2011PhRvC..83f5808}, in the first
two cases, or the NL3 parameter set of
\cite{Lalazissis:1997PhRvC..55..540}.


The SU(3) Chiral Mean Field
EoS~\citep[CMF;][]{Dexheimer08,Schurhoff10,Dexheimer15} is a
non-linear realization of the sigma model which includes pseudo-scalar
mesons as the angular parameters for the chiral transformation. In the
particular variant of this EoS we have chosen,
\texttt{CMF($\Lambda$)}, it includes nucleons and hyperons as degrees
of freedom (and in the case we consider here also free
leptons). Within the model, baryons are mediated by vector-isoscalar,
vector-isovector, scalar-isoscalar, and scalar-isovector mesons
(including strange quark-antiquark states). We have also considered a
variant of the previous EoS that includes also quarks, tagged with
\texttt{CMF($\Lambda$B)}, which has been recently applied in numerical
simulations of mergers of binary neutron stars
\citep{Most_2018arXiv180703684M}.

\subsection{Study of the non-convexity}
\label{section:Micro-EoS-non-convexity}

\subsubsection{Non-convexity at phase transitions}
\label{subsec:phasetransitions}
\begin{figure*}
\begin{center}
\includegraphics[width=\textwidth]{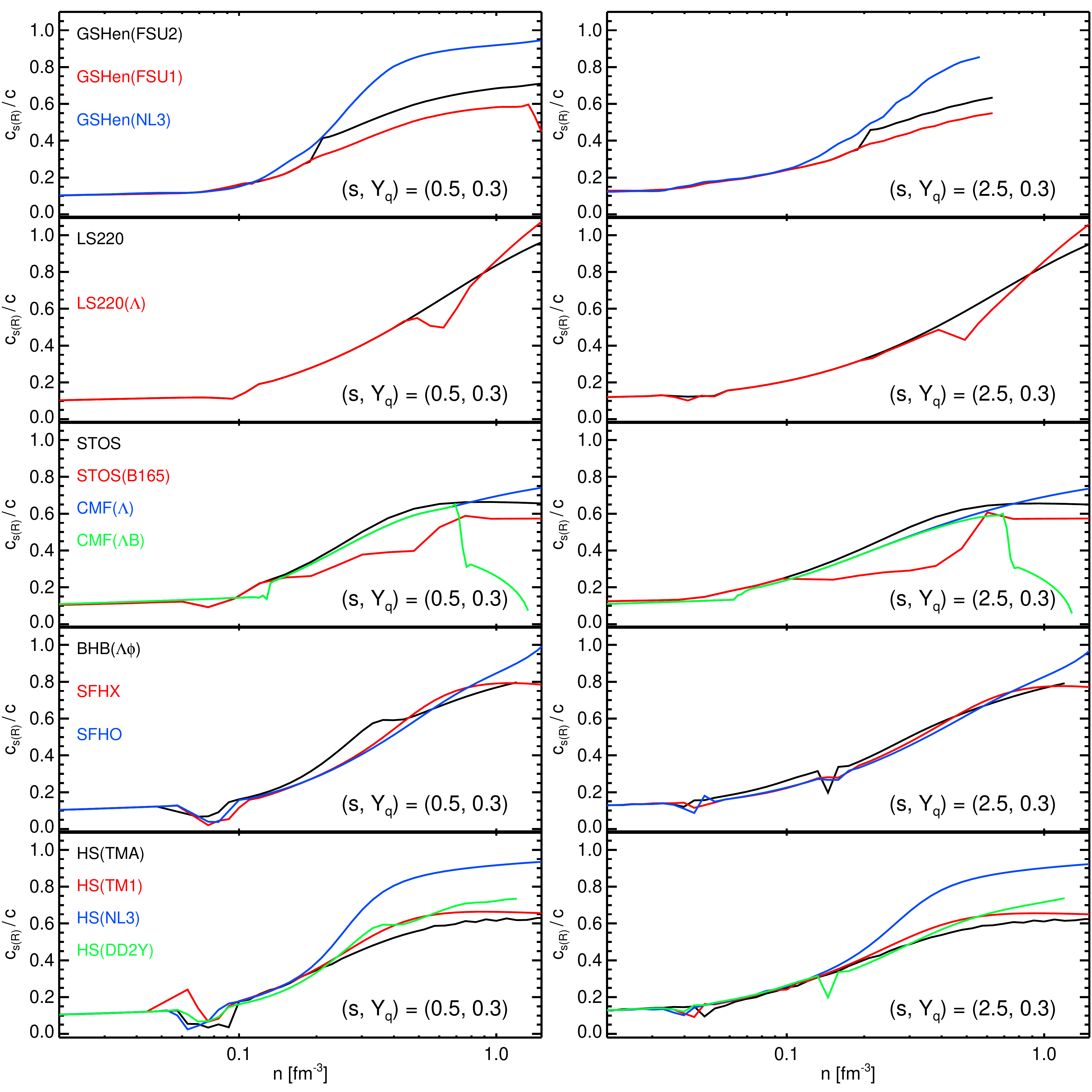}
\caption{Relativistic sound speed for various nuclear matter EoSs as a
  function of the baryon number density computed for a fixed value
  $Y_q=0.3$ along the isentropes $s=0.5$ (left panels) and $s=2.5$
  (right panels). The legends \texttt{GSHen(FSU1)},
  \texttt{GSHen(FSU2)}, \texttt{GSHen(NL3)} refer to the GSHen EoS
  \citep{Shen:2011PhRvC..83c5802} including different
  parameterizations of the RMF. Models dubbed with
  \texttt{LS220($\Lambda$)} and \texttt{LS220} correspond to the LS220
  EoS \citep{Lattimer:1991NuPhA.535..331} including hyperons or not
  including them, respectively. The tags \texttt{STOS} and
  \texttt{STOS(B165)}, refer to the STOS EoS
  \citep{Shen:1998PThPh.100.1013,Shen:1998NuPhA.637..435}, the latter
  one including a transition to a quark matter.
  \texttt{BHB($\Lambda\phi$)}, \texttt{HS(TMA)}, \texttt{HS(TM1)},
  \texttt{HS(NL3)}, \texttt{HS(DDY2)}, \texttt{SFHO} and \texttt{SFHX}
  correspond to the hadronic EoS based on the HS statistical model and
  implementing RMF interactions of different types. Finally,
  \texttt{CMF($\Lambda$)}, corresponds to the hadronic CMF model,
  while \texttt{CMF($\Lambda$B)} corresponds to the same EoS, but
  including also quarks.}
\label{fig:Compose-sound}
\end{center}
\end{figure*}

\begin{figure*} 
\begin{center} \includegraphics[width=\textwidth]{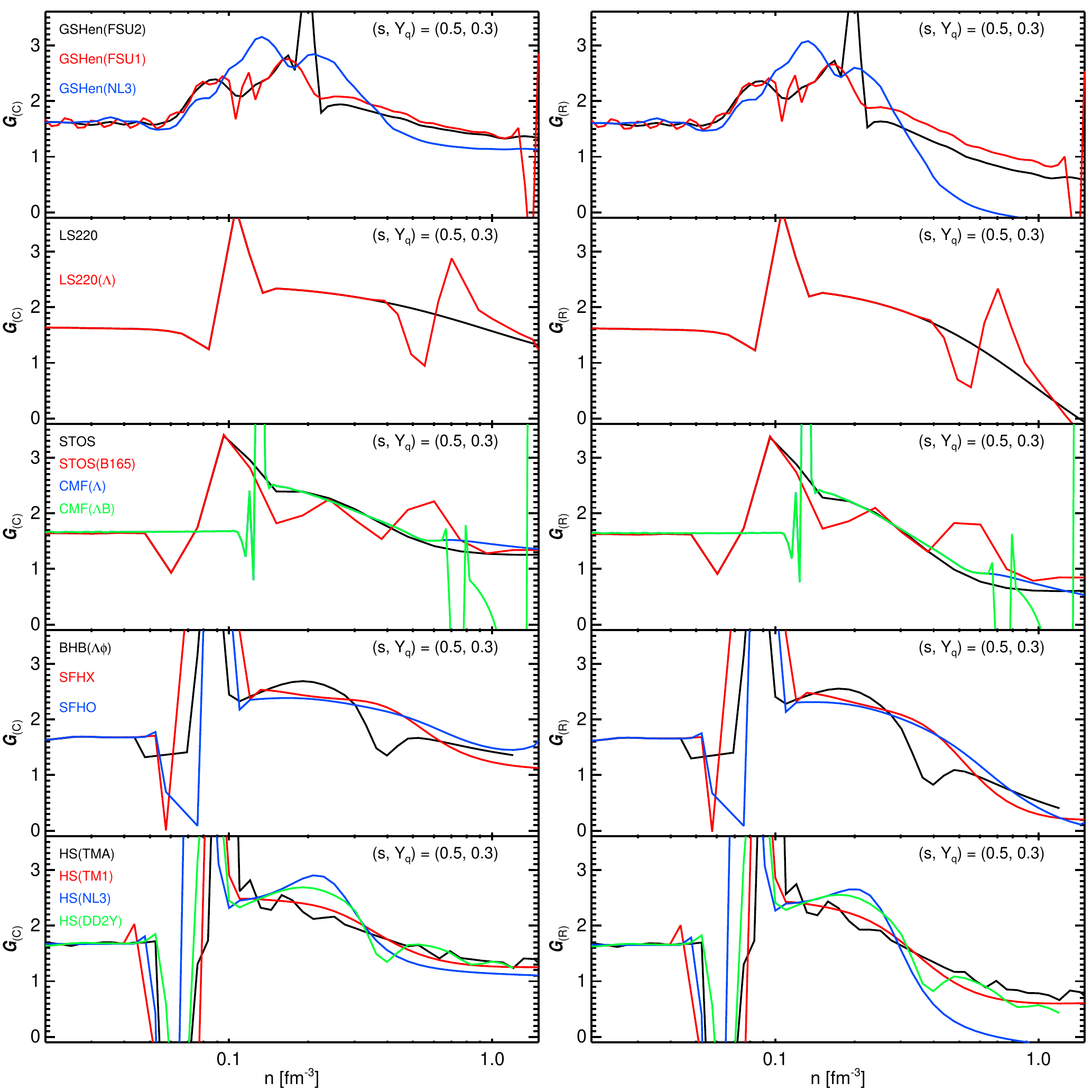} 
  \caption{Classical (left) and relativistic (right) fundamental
    derivative of a few selected dense-matter EoS from CompOSE as a
    function of the baryon number density $n$. For all EoS we fix the
    value of the entropy per baryon to $s=0.5$ and of the charge
    fraction $Y_q=0.3$. }
\label{fig:Compose-funder1}
\end{center}
\end{figure*}

%
\begin{figure*}
\begin{center}
\includegraphics[width=\textwidth]{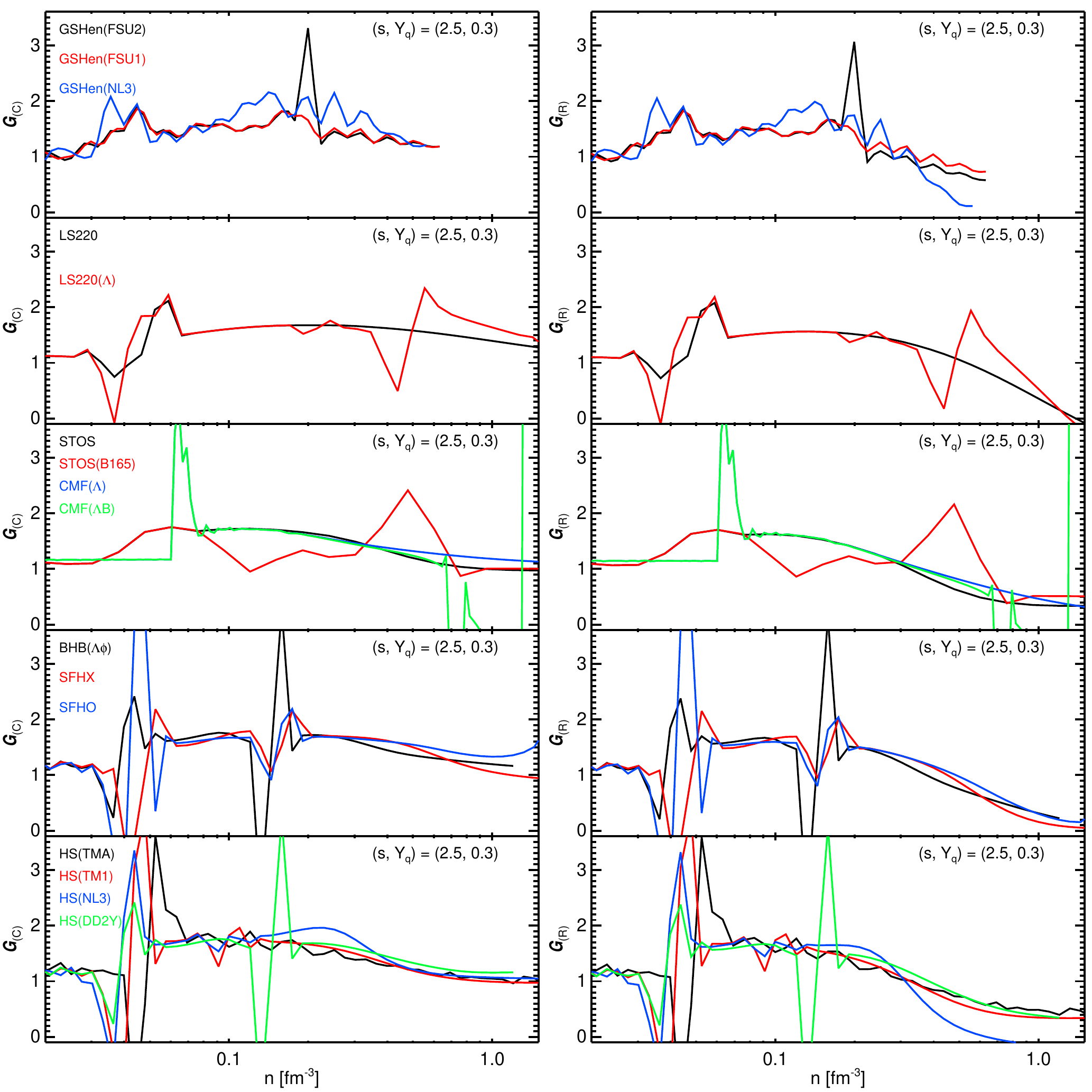}
\caption{Same as Fig.\,\ref{fig:Compose-funder1} but computed at an
  entropy per baryon $s=2.5$.}
\label{fig:Compose-funder2}
\end{center}
\end{figure*}

The nature of PTs taking place under the conditions met in the
collapse of massive stellar cores has elicited a long debate in the
scientific community \citep[see, e.g.][and references
therein]{Hempel_2013PhRvC..88a4906}. Following the nomenclature of
\cite{Iosilevskiy_2010arXiv1005.4186}, first-order PTs in nuclear
matter are ``non congruential'' (NCPT), since they involve the
coexistence of two or more macroscopic phases with different chemical
composition. For instance, in the hadron-quark transition, there can
be different types of quarks \citep[see,
e.g.][]{Nakazato_2008PhRvD..77j3006}. The previous property contrasts
with the ``congruential'' nature of first-order PTs in pure substances
(e.g. the vapour-liquid PT in water) and introduces additional degrees
of freedom, which modify their thermodynamic properties. A very
remarkable feature of NCPTs is that they are not isobaric for a fixed
temperature, i.e. they happen for a range of pressures corresponding
to the range of local concentrations of species involved in the system
\citep[c.f.][]{Hempel_2013PhRvC..88a4906}. It is also known that an
NCPT may be ``forced-congruential'' by assuming, e.g. local charge
neutrality in each phase of the coexistence regime independently
\citep{Iosilevskiy_2010arXiv1005.4186}, in which case the
thermodynamics is akin to congruential first-order PTs.

Let us examine the impact on the fundamental derivative of undergoing
a first-order PT.  Very generally, away from PTs (regardless of
whether they are first-order or continuous), $\Gamma_1>1$ and displays
a slow variation along isentropes
($|\partial \ln\Gamma_1 / \partial \ln\rho |_s|\ll 1$), so $\Gclas>1$
according to Eq.\,\eqref{eq:G2}. This situation may qualitatively
change near congruential or forced-congruential first-order PTs, where
$\Gclas$ can be negative \citep[e.g.][]{Menikoff1989}, since the sound
speed in a mixed-phase is smaller than in a pure phase under the
assumption of mechanical equilibrium.\footnote{However, these
  assumptions may break down if the transition between two phases is
  not instantaneous, so that the system may be out of equilibrium
  during the PT \citep[c.f.][]{Menikoff1989}.} By definition
\citep[see, e.g.][ch.\,9]{Callen_1985thermodynamics}, in a first-order
PT the entropy and the volume are discontinuous, while both the Gibbs
and the Helmholtz free energies as well as the pressure are
continuous, but not their first derivatives with respect,
e.g. temperature or density (more precisely, along coexistence curves
the free energies suffer jump discontinuities in their derivatives).
Correspondingly, isentropes exhibit kinks across the coexistence
curves that translate into jumps of both the sound speed and the
adiabatic index \citep[see, e.g. Fig.\,7 in][]{Menikoff1989}. The jump
in any of these two quantities entails a $\delta$-function singularity
in $\Gclas$ (see Eqs.\eqref{eq:G2} or \eqref{eq:G3}). The coefficient
of the $\delta$-function is negative (and thus, $\Gclas<0$) if the
sound speed decreases with density when crossing from a pure to a
mixed phase, and positive (i.e. $\Gclas >0$) otherwise. The former
case corresponds to a retrograde saturation boundary, while the latter
one is said to be a normal boundary in the terminology of
\cite{Thompson_1986JFM...166...57}.

We now may extend the arguments of \cite{Menikoff1989} to NCPTs.  The
extra complexity induced by the existence of various phases with
several globally conserved, net quantum numbers \citep[e.g. baryon
number, electric charge, strangeness, etc. see,
e.g.][]{Hempel_2013PhRvC..88a4906} does not change the fact that, if
the PT is of first-order type, by definition, the entropy is
discontinuous, i.e. there is a finite latent heat released/absorbed
during the transition. Since the discontinuity in the entropy is the
source of negative values of the fundamental derivative, we shall
conclude that the arguments of \cite{Menikoff1989} also apply to
first-order NCPTs.
\paragraph{Transition from inhomogeneous to homogeneous nuclear
  matter.}
Applied to the transition from inhomogeneous to homogeneous nuclear
matter (happening for $n\lesssim 0.1$\,fm$^{-3}$), starting from low
number-densities, the sound speed decreases when the homogeneous phase
begins to appear in the matter (as can be seen in the bottom panels of
Fig.\,\ref{fig:Compose-sound}), hence marking the location of
saturation conditions where $\Gclas<0$ (i.e. the coexistence boundary
is retrograde). Likewise, as the transition to homogeneous matter ends
at higher densities, the sound speed increases with density across the
coexistence curves and the $\delta$-function singularity in the
fundamental derivative yields $\Gclas>0$ (that is, the coexistence
boundary is normal). Obviously, the discrete thermodynamic conditions
at which nuclear EoSs are tabulated do not necessarily coincide with
the locus of coexistence curves in the phase space. Thus, instead of
$\delta$-function discontinuities in $\Gclas$ one produces a smeared
transition where the numerically discretized fundamental derivative
becomes positive, even if in the continuous case $\Gclas<0$ along the
retrograde coexistence curves. As a consequence, if the PT from
inhomogeneous to homogeneous nuclear matter is of first-order kind, we
expect that there should be a single point along an isentrope crossing
the PT were $\Gclas<0$ in the continuous limit. Also, due to the
tabular nature of the EoSs here considered, it may happen that the
spreading of the negative $\delta$-function discontinuity in the
fundamental derivative along two or more consecutive tabular values on
the same isentrope results in a finite (i.e. not pointwise) number
density interval where $\Gclas<0$. It is also evident that the
numerical discretization of derivatives exhibiting jumps across
coexistence curves is potentially (very) noisy and, this is the root
of the large oscillations observable in $\Gclas$ and $\Grel$ in
Figs.\,\ref{fig:Compose-funder1} and \ref{fig:Compose-funder2} (see,
especially, the bottom panels of these figures). This considerations
lead us to suggest that the tabulation of nuclear matter EoSs should
try to adapt to properly capture large gradients in the fundamental
derivative, specially when $\Gclas$ becomes negative or approaches
zero. In other words, EoS tables should be more densely populated with
nodal points near regions where $\Gclas<0$.
  
  The character of the transition from inhomogeneous to homogeneous
  nuclear matter (first-order or continuous) is still a matter of
  debate.  Extended (and somewhat controversial) discussions on the
  treatment of this PT can be found in the literature
  \citep[e.g.][]{Ducoin_2007,Raduta_2010PhysRevC.82.065801,Pais_2014PhRvC..90f5802,Nandi_2017PhRvC..95f5801},
  however, we believe that a consensus on the physically soundest
  assumptions has not been reached in the Nuclear Physics community
  yet. Furthermore, an insufficiently fine tabulation of an EoS table
  implementing any Maxwell or Gibbs construction may also result in
  numerical loses of convexity (see
  App.\,\ref{sec:numerical_artifacts}). Hence, we examine the topic of
  convexity loss bearing this limitation in mind and back up our
  results by examining the monotony properties of the speed of
  sound. The previous quantity corresponding to a lower order
  thermodynamic derivative of the Helmholtz free energy potential than
  the fundamental derivative and being, hence, less prone to develop
  spurious numerical oscillations.

  Besides the previous considerations, we emphasize that the negative
  values attained by the classical and relativistic derivatives may
  result from the treatment of PT in the EoS. Alternative treatments
  of a PT (corresponding to physically different types of PT) may
  yield a convex thermodynamic behaviour. For instance,
  \cite{Pons1999} suggested that employing a Gibbs construction
  instead of a Maxwell construction in an EoS including hyperons may
  prevent the formation of discontinuities, keeping finite the
  compressibility.

  We outline that the sampled hadronic EoSs based on the HS statistic
  model which do not contain hyperons (\texttt{HS(TMA)},
  \texttt{HS(TM1)}, \texttt{HS(NL3)}, \texttt{SFHO} and \texttt{SFHX})
  seem to loose convexity in a narrow range of baryonic number
  densities with a typical width $\Delta n \sim 0.02\,$fm$^{-3}$.
The large variations of $\Gclas$ and $\Grel$ in the
  inhomogeneous to homogeneous nuclear matter transition found in the
  previous EoSs are associated to remarkably non smooth behavior of
  the sound speed, which sinks significantly
  ($\csrel \lesssim 5\times10^{-3}c$) in the range
  $0.06\,\text{fm}^{-3}\lesssim n \lesssim 0.09\,\text{fm}^{-3}$,
  precisely, in the mixed phase (as we have indicated above). %
  The PT from non-uniform to uniform nuclear matter is
  treated with the same Maxwell construction as in the case of the
  \texttt{LS220} EoS according to \cite{Hempel:2010NuPhA.837..210}. We
  observe, however, that the fundamental derivatives become negative
  in all three variants of the HS EoS independently of the entropy per
  baryon (Figs.\,\ref{fig:Compose-funder1} and
  \ref{fig:Compose-funder2}; bottom panels).

The different variants of the GSHen EoS display neither a negative
fundamental derivative nor very large oscillations in the region of
transition from nuclei to nuclear
matter. \cite{Shen:2011PhRvC..83c5802} claim that their construction
for the PT accounts for the Coulomb
interactions. According to the former authors, Coulomb interactions
are non-negligible in large astrophysical systems (in contrast to
small systems such as the ones found in heavy-ion collisions) and
result in a non-uniform phase (across the PT) where the
average proton density equals the electron density. For that
non-uniform phase \cite{Shen:2011PhRvC..83c5802} obtain a monotonic
increase of the adiabatic index, $\Gamma_1$, with the number density,
contrasting with the decrease in $\Gamma_1$ shown by the
LS220. Without entering into a deeper discussion on whether the
aforementioned PT is of first-order or of any other
kind, for what matters this paper, none of the two EoSs (LS220 and
GSHen) shows a loss of convexity in the transit from non-uniform to
uniform nuclear matter (unless $\Lambda$ hyperons are included
  in the \texttt{LS220($\Lambda$)} EoS; see below).

\paragraph{Hyperon phase transition.}
We note that the appearance of hyperons in the
\texttt{BHB($\Lambda\phi$)} EoS generates a non-convex thermodynamics
if the entropy per baryon is large enough at baryonic number densities
$n\sim 0.15$\,fm$^{-3}$
(Fig.\,\ref{fig:Compose-funder2}). \cite{Banik:2014ApJS..214...22}
claim that they ``did not find any indication for a first-order phase
transition in connection to the appearance of $\Lambda$
hyperons''. This conclusion is extracted on the basis of the
smoothness of the pressure growth with baryon density even after the
appearance of $\Lambda$ hyperons. However, a close look to their
Fig.\,9 reveals that the entropy is non-smooth precisely where
hyperons appear (in their case at baryon densities
$\lesssim 10^{15}\,\text{gr}\,\text{cm}^{-3}$). This behaviour is
reflected in the non-smoothness of the sound speed in two different
number density intervals, of which, the one happening at higher
entropy per baryon ($s=2.5$) and $n\simeq 0.15\,\text{fm}^{-3}$ yields
negative values of the fundamental derivatives. This is a first
evidence of thermodynamic convexity loss connected to the hyperonic
phase.

The incorporation of hyperons in the \texttt{LS220($\Lambda$)} EoS
brings a loss of convexity at entropies per baryon $s=2.5$
(Fig.\,\ref{fig:Compose-funder2}) in the transition from inhomogeneous
to homogeneous nuclear matter (note also the behaviour of the sound
speed at number densities
$0.4\,\text{fm}^{-3}\lesssim n \lesssim 0.6\,\text{fm}^{-3}$ in
Fig.\,\ref{fig:Compose-sound} for $s=2.5$). For values $s=3.5$ this
convexity loss is not found (because this larger entropy is also
associated to larger temperatures at which the transition to
homogeneous nuclear matter disappears), and thus the results at low
values of $s$ are linked to the Maxwell construction across a
first-order PT (see the discussion in
Sect.\,\ref{subsec:phasetransitions}). We note that the variant of the
LS220 EoS that does not incorporate hyperons (\texttt{LS220}) shows
positive fundamental derivatives there.

In contrast to the \texttt{BHB($\Lambda\phi$)} and
\texttt{LS220($\Lambda$)} EoSs, convexity is not lost in the case of
the CMF model implemented in the \texttt{CMF($\Lambda$)} EoS, which
also incorporates hyperons.
In the \texttt{CMF($\Lambda$)}, the nuclear PT takes place as an
smooth crossover due to the requirements of beta equilibrium and
charge neutrality \citep[c.f.][]{Dexheimer08}, at least that the
entropies per baryon we consider here, and consistently, the numerical
fundamental derivatives remain positive and smooth as can be seen in
Figs.\,\ref{fig:Compose-funder1} and \ref{fig:Compose-funder2}.

The \texttt{HS(DDY2)} EoS also incorporates hyperons using the same
recipe for its incorporation as in the \texttt{BHB($\Lambda\phi$)} EoS
and, thus, one may observe the same behaviour for the former EoS as
for the latter in terms of loss of convexity.

%
  %

\paragraph{Quarks phase transition.}
Our knowlege of the QCD phase diagram suggests the existence of a QGP
PT within the range of densities and temperatures on reach of core
collapse events \citep[c.f.][see also
\citealt{Haensel2007}]{Oertel:2012PhRvC..85e5806}.  The incorporation
of a quark phase in the \texttt{STOS(B165)} EoS displays a noticeable
impact on the fundamental derivatives, which, however, never become
negative in spite of the fact that the hadron-quark transition is of
first-order type in this EoS (note the difference in the slopes of the
Gibbs free energy in the hadronic and quark phases observed in
\cite{Nakazato_2008PhRvD..77j3006}; Fig.\,3, the right panels).  We
should bear in mind that both the \texttt{STOS} and
\texttt{STOS(B165)} are more coarsely tabulated than most of the other
EoSs considered here. Hence, we shall conclude that the fact that we
do not observe negative values of $\Gclas$ in the latter EoS is the
result of an insufficiently fine tabulation close to the
thermodynamical boundaries of the PT to quark matter. This conclusion
is reinforced by the comparison of the EoS \texttt{CMF($\Lambda$)} and
\texttt{CMF($\Lambda$B)}, which only differ in the incorporation of a
first-order NCPT to quarks in the latter. As we can see in
Figs.\,\ref{fig:Compose-funder1} and \ref{fig:Compose-funder2}, for
$n\gtrsim 0.6\,\text{fm}^{-3}$, there is a rather broad region where
$\Gclas<0$, corresponding to the pronounced drop of the sound speed at
such number densities (Fig.\,\ref{fig:Compose-sound}).

\subsubsection{Genuinely relativistic convexity loss}
The loss of convexity at $n\gtrsim 0.8\,\text{fm}^{-3}$ and low
entropy per baryon is a very robust finding in the case of
\texttt{GSHen(NL3)} (Fig.\,\ref{fig:Compose-funder1}; blue line in the
upper row). This is, indeed, a genuinely relativistic effect since it
happens due to the large value of the sound speed
($0.8 < \csrel / c <1$; Fig.\,\ref{fig:Compose-sound}) and the
corresponding action of the term ``$-3 \csrel^2/2$'' in
Eq.\,(\ref{G6}). We note that, differently from the case of the LS220
EoS (see Sec.\ref{subsec:otherregions}), this happens in a fully
causal region for the \texttt{GSHen(NL3)}. Remarkably, the same
convexity-loss is observed for the \texttt{HS(NL3)} EoS
(Fig.\,\ref{fig:Compose-funder1}; blue line in the bottom row), which
shares with the \texttt{GSHen(NL3)} EoS the same RMF parameterization
(NL3). Nevertheless, in the case of \texttt{HS(NL3)} the relativistic
fundamental derivative is also negative at larger entropies per baryon
(Fig.\,\ref{fig:Compose-funder2}; blue line in the bottom row). Using
the \texttt{GSHen(NL3)} table available at the CompOSE database, it
was not possible to obtain thermodynamic values along the $s=2.5$
isentrope for $n\gtrsim 0.55\,\text{fm}^{-3}$ in the case of
\texttt{GSHen(NL3)}. In contrast, for the \texttt{HS(NL3)} table, one
may compute values along the same isentrope up to
$n\lesssim 4.79\,\text{fm}^{-3}$. This happens because of the
different tabular limits of both EoSs. While the \texttt{GSHen(NL3)}
table provides nodal points in the ranges
$10^{-8}\,\text{fm}^{-3}\le n\lesssim 1.5\,\text{fm}^{-3}$ and
$0.16\,\text{MeV} \lesssim T \lesssim 75\,\text{MeV}$, the
\texttt{HS(NL3)} table extends further the previous ranges to
$10^{-12}\,\text{fm}^{-3}\le n\lesssim 10\,\text{fm}^{-3}$ and
$0.1\,\text{MeV} \lesssim T \lesssim 158\,\text{MeV}$. Particularly,
the larger values of $T$ tabulated in the \texttt{HS(NL3)} EoS allow
prolonging the $s=2.5$ isentrope to larger number densities per
baryon. In spite of this technical limitation, and regarding that at
high densities and entropies, i.e. for homogeneous matter, both EoS
should provide the same results (since the underlying models -NL3- are
the same), we also conclude that also the \texttt{GSHen(NL3)} EoS is
(relativistically) non-convex at high number-density and entropy per
baryon.

\subsubsection{Non-physical loss of convexity}
\label{subsec:otherregions}
None of the two variants of the \cite{Lattimer:1991NuPhA.535..331} EoS
are causal at high enough densities. This has the implication that the
sound speed predicted by these EoSs is larger than the speed of light
(see the region with $n\gtrsim 1.3\,$fm$^{-3}$ in
Fig.\,\ref{fig:Compose-sound}). Consistently, the relativistic
fundamental derivative becomes negative (see
Figs.\,\ref{fig:Compose-funder1} and \ref{fig:Compose-funder2} at
baryonic number densities $n\gtrsim 1.3\,$fm$^{-3}$). This happens
because of the action of the term ``$-3 \csrel^2/2$'' in
Eq.\,(\ref{G6}), which becomes a dominant (negative) contribution as
the sound speed approches (and eventually overtakes) the speed of
light \citep[see, e.g.][]{Ibanez2013}. The loss of convexity displayed
by both LS220 EoSs in non-causal (high-density) thermodynamic regions,
is non-physical.

The large amplitude oscillations observed for the \texttt{GSHen(FSU1)}
EoS at high number densities are likely numerical artifacts (see
App.\,\ref{sec:numerical_artifacts}).

\section{A phenomenological non-convex EoS }
\label{section:GGL-EoS}

The traditional (simple) way to mimic the complex thermodynamical processes taking place inside a collapsing stellar core in simulations of hydrodynamical supernovae leading to the formation of compact objects, or during the merger of neutron stars in a compact binary system, is to consider EoSs of polytropic-type. Some examples include (i) a polytropic EoS where `gamma' is a discontinuous function of the density~\citep{vanRiper1978}, (ii) the piecewise-polytropic approximation~\citep{Mueller1985}, and (iii) the hybrid polytropic EoS, in which the pressure is composed of a cold component, $p_c$, described by a polytrope of adiabatic index $\Gamma_c$, and an ideal-gas component which incorporates the thermal effects, $p_t$ \citep[see e.g.][]{Maione2016}. We name the latter EoS `PolyTh' and present a detailed analysis of its properties in Appendix~\ref{appB}.

In order to explore the fundamental traits of a relativistic non-convex dynamics induced by a non-convex thermodynamics we use a phenomenological EoS introduced in \cite{Ibanez:2017TUBOS}. Here we recap the essentials of the analysis performed on \cite{Ibanez:2017TUBOS} regarding the non-convex properties that this EoS possess. We begin by the expression of the pressure $p$, which obeys an ideal-gas-like EoS of the form
\begin{eqnarray}
p = (\gamma - 1) \rho \epsilon\,,
\label{GGL-1a} 
\end{eqnarray}
where
$\gamma$ depends on the density according to the following law:
\begin{eqnarray}
\gamma
 := 
\gamma_0
 + 
{\cal K} \, \exp\left(\displaystyle{-\,\frac{x^2}{\sigma^2} }\right)
\,,\,\,\,
{\cal K} :=  \gamma_1 - \gamma_0 \,,\,\,\, 
x  :=  \rho - \rho_1\,, 
\label{GGL-1b} 
\end{eqnarray}
and where $\epsilon$ and $\rho$ are, respectively, the specific
internal energy and the rest-mass density. 
We note that in \cite{Ibanez:2017TUBOS} the pressure contains an
additive term of the form $B\rho$, which depends on another free
parameter of the EoS. Hereafter, we restrict to the case $B=0$.
\cite{Ibanez:2017TUBOS}
proposal for $\gamma$ in Eq.~(\ref{GGL-1b}) can be considered as a
generalization of the classical prescription used in early studies of
core-collapse supernovae \citep[see e.g.][]{vanRiper1978}. The
function $\gamma(\rho)$ in Eq.\,(\ref{GGL-1b}) reaches a maximum at
$\rho = \rho_1$, a value we designate as $\gamma_1 =
\gamma(\rho_1)$. Let us notice that $\rho_1$ plays the role of a
simple scale factor for the density, if we express the \emph{width} of
the Gaussian law ($\sigma$) in units of $\rho_1$ too, convention we
adopt in the following.
The EoS defined by Eqs.\,(\ref{GGL-1a}) and (\ref{GGL-1b}) will be
named hereafter `GGL-EoS' (for Gaussian Gamma Law). The reference
parameters we chose to analyse its properties are $\gamma_0 = 4/3,
\gamma_1=1.9, \sigma = 1.1$, and $\rho_1 = 10^{15}$\,g\,cm$^{-3}$. The
values of $\gamma_0$ and $\gamma_1$ mimic the behaviour of collapsing
dense matter during the standard prompt mechanism of hydrodynamical
supernovae, before and after core bounce \citep[see,
e.g.][]{Janka_etal:2012}.
\begin{figure}
\centering
\includegraphics[width=\columnwidth]{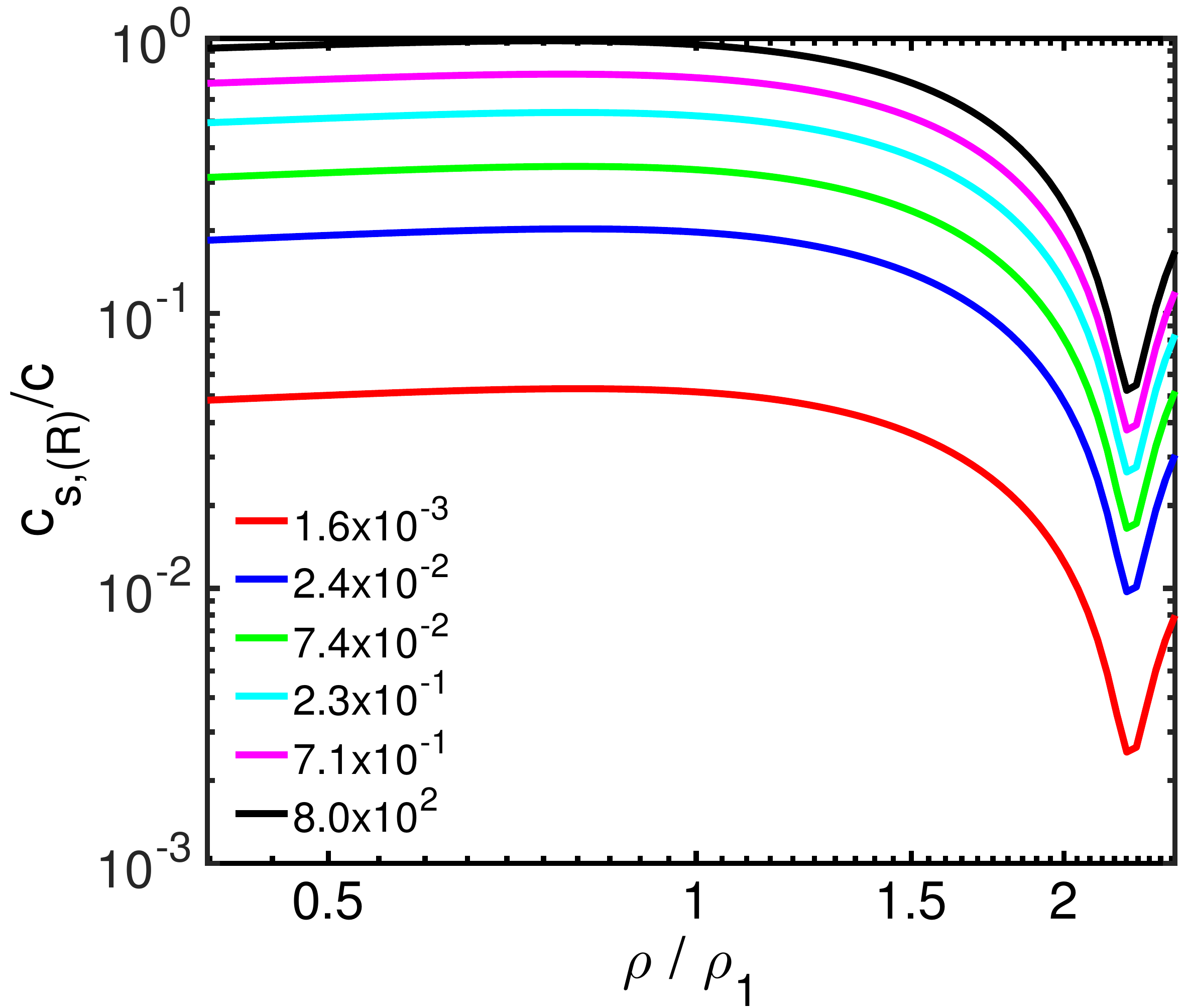}
  \caption{Relativistic speed of sound  versus density for the GLL-EoS ($\gamma_0 = 4/3,  \gamma_1=1.9,  \sigma = 1.1, \rho_1 = 10^{15}$\,g\,cm$^{-3}$). The curves are parameterized by the specific internal energy, using the particular values indicated in the legend (in units of $c^2$).}
\label{fig:cs_rel}
\end{figure}
\begin{figure}
\centering
\includegraphics[width=\columnwidth]{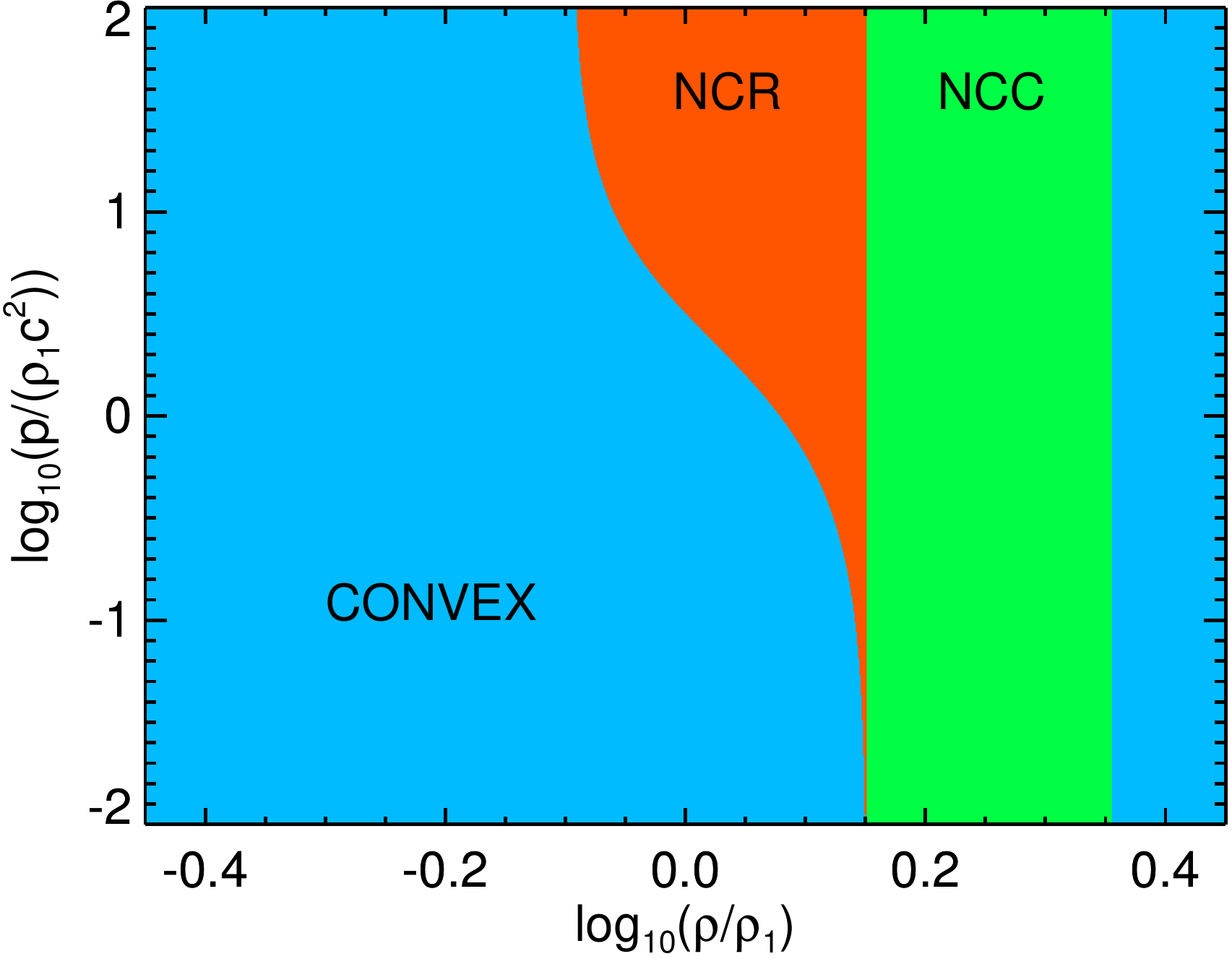}
  \caption{Regions of the $P-\rho$ plane of the GLL-EoS with parameters $\gamma_0 = 4/3,  \gamma_1=1.9,  \sigma = 1.1,$ and $\rho_1 = 10^{15}$\,g\,cm$^{-3}$, displayed in different colors according to their convexity (lack thereof) properties. The blue colored area corresponds the region of the parameter space where both $\Gclas>0$ and $\Grel>0$, i.e., where the EoS exhibits a convex-thermodynamics. The red colored area corresponds to the region of the parameter space where $\Gclas>0$ and $\Grel \le 0$, i.e., the EoS is non-convex from the relativistic point of view (NCR). Finally, the green colored area corresponds to the region of classical non-convexity (NCC), in which $\Gclas\le 0$ and $\Grel\le0$.}
\label{fig:P-rho_GLL-EoS}
\end{figure}

Applying the definition of the classical speed of sound (Eq.\,\ref{eq:sound_class}) 
to our GGL-EoS, we obtain
\begin{equation}  
\cscla^2  = 
\gamma  \left(\frac{p}{\rho} +\epsilon \frac{d\ln\gamma}{d\ln\rho} \right) 
 = 
\gamma \,\, \epsilon \,\, \left(\gamma - 1 + \frac{d\ln\gamma}{d\ln\rho} \right)\,.
\label{cs2cla}
\end{equation}  
For later reference, we also write the specific enthalpy for the GGL-EoS:
\begin{equation}
h   =   1   +   \gamma \epsilon .
\label{eq:hGGL}
\end{equation}
Figure\,\ref{fig:cs_rel} shows, in logarithmic scale, the relativistic speed of sound (in units of the speed of light $c$) as a function of the density, parameterized by the specific internal energy. We note that the parameterization used in the GGL-EoS avoids non-causality (i.e., yields $\csrel < c$) and leads to very low values of $\csrel$ for densities much higher than $\rho_1$. As the legend of Fig.~\ref{fig:cs_rel} indicates, $\csrel$ is an increasing function of $\epsilon$.

The explicit expressions for the adiabatic index, $\Gamma_1$
\citep[see e.g.][]{Chandra1939}, and the fundamental derivatives for
the GGL-EoS have been obtained in \cite{Ibanez:2017TUBOS}. The
adiabatic index (Eq.\,\ref{Gamma1deff}), which is in general $\Gamma_1 \ne  \gamma$, reads
\begin{equation}
  \Gamma_1
  =
  \displaystyle{
    \gamma \, 
    \left\{ 1 +\left(\frac{\rho \epsilon}{p}\right) \frac{d\ln\gamma}{d\ln\rho} 
    \right\} \,.
  }
\label{Gamma1}
\end{equation}

The classical fundamental derivative for our GGL-EoS is:
\begin{equation}
\Gclas = 
{\mathcal G}^+  +
\displaystyle{
\frac{\gamma \, \epsilon}{2 \cscla^2 }
\left( \gamma \frac{d\ln \gamma}{d\ln \rho} + 
\frac{d^2\ln \gamma}{d(\ln \rho)^2} \right) 
}\,,
\label{G_cla}
\end{equation}
where 
\begin{equation}
{\mathcal G}^+ := 
\displaystyle{
\frac{1}{2} \left\{ 1 + \gamma + \,\, \left(\frac{d\ln \gamma}{d\ln \rho} \right) \right\}\,.
}
\end{equation}
From the above equation and the expression for $\cscla^2$ given by
Eq.\,(\ref{cs2cla}) it is easy to conclude that $\Gclas$ is
independent of $\epsilon$.


 
Figure\,\ref{fig:P-rho_GLL-EoS} shows the regions of the $P-\rho$ plane in which the GLL-EoS is divided in terms of the character of the thermodynamics. We observe that with the exception of a small region around and above $\rho_1$, the EoS is convex, i.e., the classical and relativistic fundamental derivatives, satisfy $\Grel > 0$ and $\Gclas > 0$ (blue region in Fig.\,\ref{fig:P-rho_GLL-EoS}). In the green region $\Gclas \le 0$, and, as a result, so is $\Grel \le 0$. This is a non-convex classical (NCC) region of the EoS. There is also a non-convex relativistic (NCR) region in which only  $\Grel <0$, while $\Gclas > 0$  (red region inf Fig.\,\ref{fig:P-rho_GLL-EoS}).

\begin{figure}
\centering
\includegraphics[width=\columnwidth]{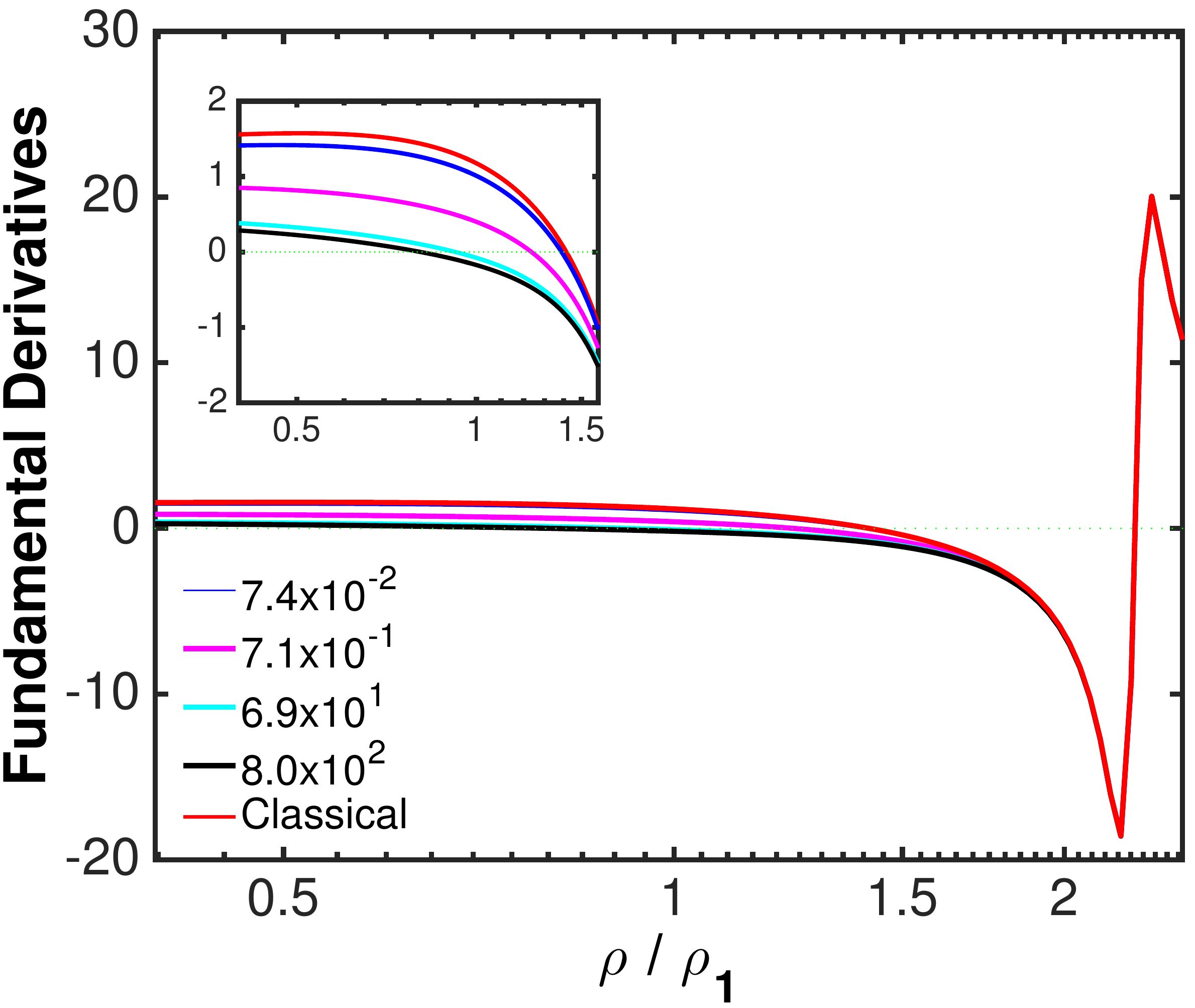}
\caption{Classical and relativistic fundamental derivatives as a function of density for the GGL-EoS. The classical fundamental derivative is indicated by a red thick line. The relativistic fundamental derivative is parameterized by the specific internal energy, with the particular values indicated in the legend. The inset shows a detail of the region around $\rho = \rho_1$. }
\label{fig:FDs-rho}
\end{figure}
	
Figure\,\ref{fig:FDs-rho} shows the two fundamental derivatives, classical (red thick line) and relativistic, as a function of density, being $\Grel$ parameterized by the specific internal energy.  The inset shows a detail of the region around $\rho = \rho_1$ in order to highlight that, according to Eq.\,(\ref{G6}), $\Gclas$ is an upper bound of $\Grel$. Furthermore, the inset clearly displays the existence of regions for which $\Grel \le 0$ and $\Gclas \ge 0$ simultaneously. We point out the qualitative similarity between the behaviour of the fundamental derivatives of the GGL-EoS around $\rho=\rho_1$ compared with that of a number of microphysical EoSs at high enough number density (see Figs.\,\ref{fig:Compose-funder1} and \ref{fig:Compose-funder2}). Note that, in the case of the GLL-EoS, the convexity-loss around $\rho=\rho_1$ is not related to any PT, which does not exist in the phenomenological EoS. This point is relevant inasmuch as the convexity loss of several microphysical EoSs at baryon densities around the PT to uniform nuclear matter may result from the explicit construction employed to deal with the mixed phase in a thermodynamically consistent way.

%
\section{Equilibrium configurations}
\label{section:TOV-GGL-EoS}
%

We turn next to analyze spherically-symmetric configurations of relativistic stars in equilibrium that satisfy the GGL-EoS  introduced in the previous section. The relationship between the specific internal energy and the rest-mass 
density follows from the first law of thermodynamics for adiabatic processes. The corresponding ordinary differential equation, for our GGL-EoS, can be written as
\begin{equation}
\displaystyle{\frac{d\, \ln\,\epsilon}{d\,\ln\,\rho} \,=\, \gamma(\rho) \, - \, 1 }\,.
\end{equation}
The integration constant in the above equation can be defined from
the polytropic form $(p=\kappa_{_{{\rm dnr}}} \, \rho^{\Gamma_{_{{\rm dnr}}}})$ 
of the EoS for a degenerate ideal Fermi gas of electrons at very low densities, i.e.~the degenerate non-relativistic regime (dnr), 
where $\kappa_{_{{\rm dnr}}}\,=\,1.0036\times 10^{13}\,Y_e^{5/3}$ (in CGS units, and $Y_e=1/2$). In practice, and in order to obtain values of the maximum gravitational mass (see below) compatible with current observational data, we have verified that an optimal value is $\kappa_{_{{\rm dnr}}}\,=\,2.0072\times10^{12}\,Y_e^{5/3}$ (i.e.~a reduction factor of $1/5$).

The resulting tabulated relationship between the specific internal
energy and the rest-mass density is fitted with a potential law
\begin{equation}
\epsilon = \kappa_{\rm ad} \, \rho^b \,,
\label{GGL-2} 
\end{equation}
where $\epsilon$ is given  in units of $c^2$. The fitting parameters are $\kappa_{\rm ad} = 4.2266 \times 10^{-10}$ and $b = 0.58584$ for a fitting interval $\rho\in [\rho_1^{-1}, 10\rho_1]$. Equations\,(\ref{GGL-1a}), (\ref{GGL-1b}) and (\ref{GGL-2}) define completely our GGL-EoS. Figure\,\ref{fig:pc-rhoc} shows the GGL-EoS used to construct the static equilibrium models in this section and employed in the dynamical evolutions of rotational collapse of neutron stars to BHs in the next one. 

\begin{figure}
\centering
\includegraphics[width=\columnwidth]{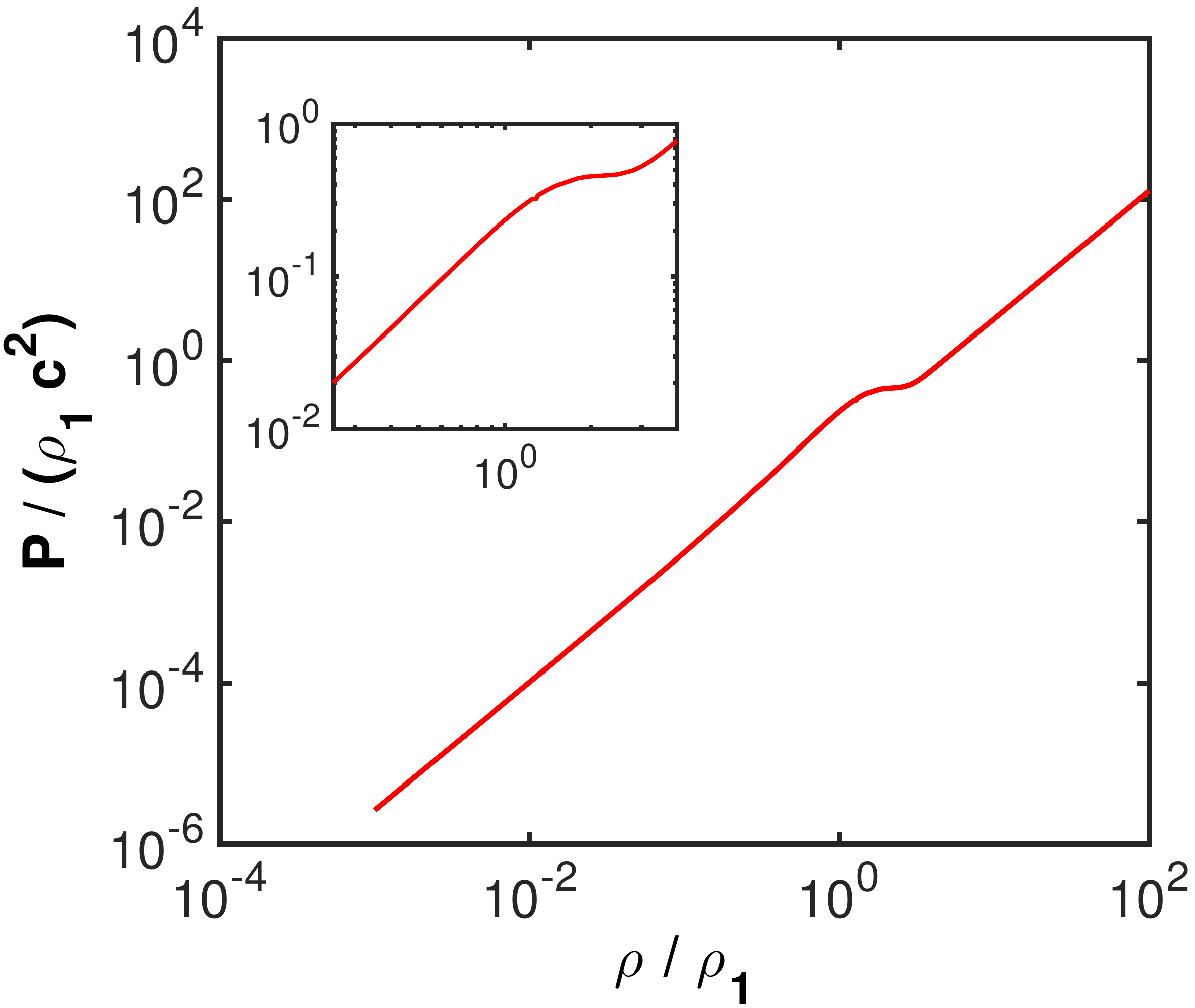}
\caption{GGL-EoS: pressure versus density,
parameterized by the specific internal energy given by the fit shown in Eq.~(\ref{GGL-2}). The inset shows a detail
around the value $\rho_1$.}
\label{fig:pc-rhoc}
\end{figure}
%

In order to obtain spherically-symmetric relativistic equilibrium
configurations that obey the GGL-EoS we solve the
Tolman-Oppenheimer-Volkoff (TOV) equation. The gravitational
mass $M_{\rm G}$ of the equilibrium configurations, parameterized by
the central density $\rho_{c}$, is shown in Fig.~\ref{fig:MG-rhoc-1a}
as a function of the radius. It reaches a maximum $M_{\rm G}^{\rm max}
= 2.536\,M_{\odot}$, at a central density $\rho_{\rm c}^{\rm crit}
\approx 1$ (units of $\rho_1$), being the corresponding radius $R
\approx 16.6$ km. The inset displays $M_{\rm G}$ versus the central
density. Models with central densities in the interval $\rho_{\rm
  c}^{\rm crit} \le \rho_{\rm c}/\rho_1 \lesssim 1.4 $, define a small
plateau in the $M_{\rm G}(\rho_{\rm c})$ curve where this function is
strictly decreasing, i.e.~there is no local maximum. By construction,
the models are isentropic and therefore they satisfy the well-known
static stability criterion against radial
oscillations~\citep{Bardeen1966}: the stability region is the one at
central densities below the critical one, $\rho_{\rm c}^{\rm crit}$,
at which the gravitational mass has an absolute maximum. Also shown in
Fig.\,\ref{fig:MG-rhoc-1a} is the region bounded by the Schwarzschild
radius (black dots in the upper-left corner of the figure).

 \begin{figure}
  \centering
    \includegraphics[width=\columnwidth]{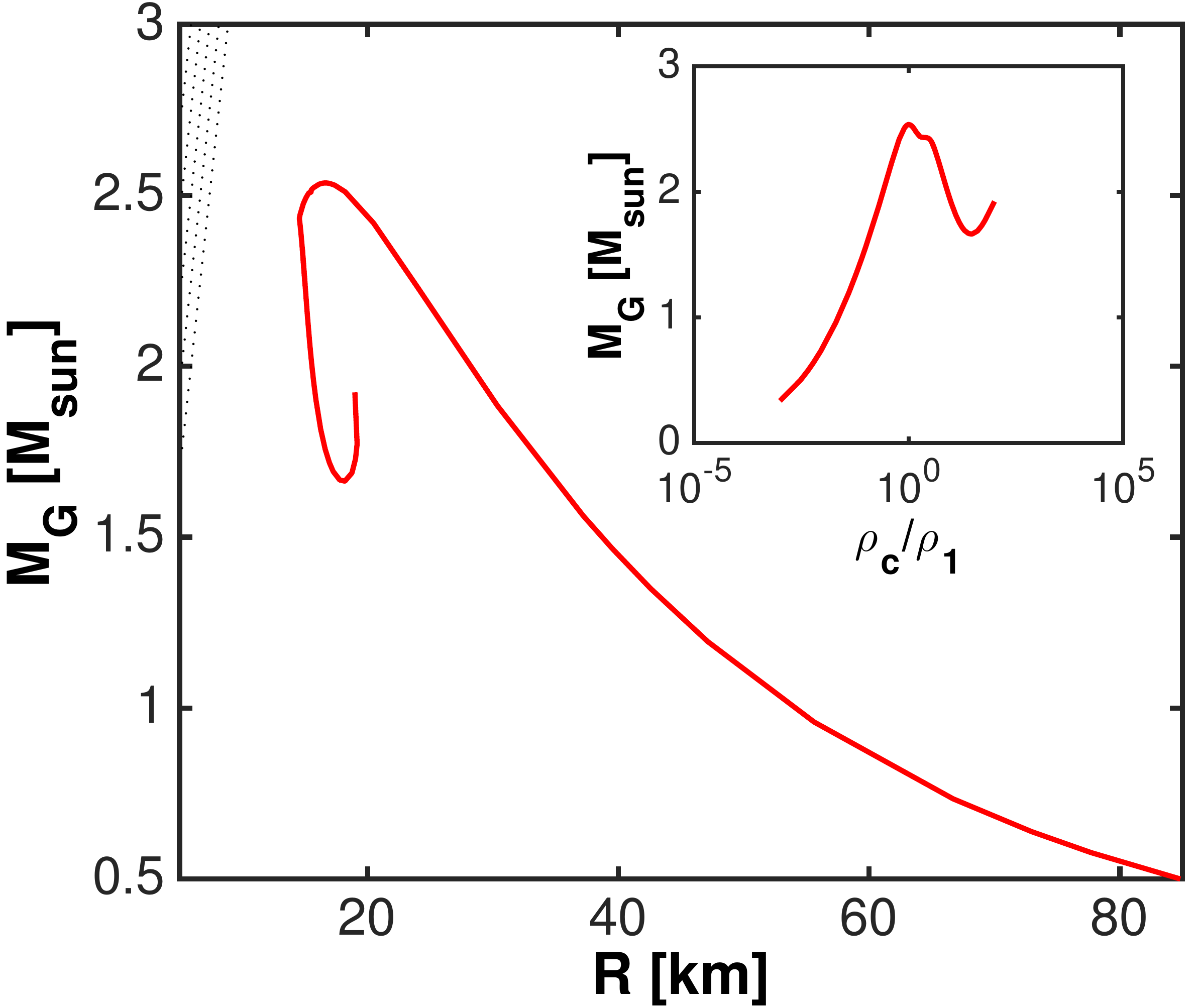}
  \caption{
Gravitational mass  versus radius  of the equilibrium configurations for our GGL-EoS, parameterized by the central density. The black dotted region in the upper-left corner corresponds to $R \le 2 \, M_G$. The inset shows the gravitational mass  versus the central density.}
  \label{fig:MG-rhoc-1a}
\end{figure}

Moreover, the specific internal energy and the specific enthalpy are, by definition, increasing functions of the density.
Therefore, their maxima (in radius) are reached at the centre of the configuration. For the critical model, the central values  of $\epsilon$ and $h$ are, respectively, $\epsilon=0.26$ and $h=1.49$. Thus, the critical model is, from a thermodynamical point of view, only moderately relativistic. Consistent with our GGL-EoS, the relativistic speed of sound at the centre of the equilibrium configurations is not a monotonic function of the central density. Its central value  for the critical model is $\csrel=0.546$. This value is an upper bound for all the equilibrium models. Let ${\cal C} := {\rm {max}}\,\,(2Gm/(rc^2))$ be the maximum value of the compactness parameter, in radius, for each model. For our 
GGL-EoS the models reach an absolute maximum of ${\cal C} = 0.59$ at $\rho_{\rm c} = 2\,\rho_1$, being its
value at the critical central density ${\cal C} = 0.53$.

Figure\,\ref{fig:FD-radius-1} shows the two fundamental derivatives,
${\mathcal G}_{\rm (R)} $ and $ {\mathcal G}_{\rm (C)}$, as a function
of the radius, for an equilibrium model obeying the GGL-EoS and with
central density $\rho_{\rm c}/\rho_1 = 1.4$. Due to the particular
form of our GGL-EoS, equilibrium models with central densities larger
than the critical one will develop non-convex thermodynamics.  For the
sequence of equilibrium models we compute, there exists a small
interval of central densities, namely $1.3 \lesssim \rho_{\rm
  c}/\rho_1 \lesssim 1.4$, in which ${\mathcal G}_{\rm (R)} < 0$ and $
{\mathcal G}_{\rm (C)} > 0$. This is shown in
Fig.\,\ref{fig:FD-radius-1} for the particular case $\rho_{\rm
  c}/\rho_1 = 1.4$. As a result, in such a narrow region of
central-density values, the innermost cores of our models can develop
non-convex thermodynamics induced by purely relativistic effects. The
dynamical collapse of these objects, if perturbed, would rapidly
trigger the presence of compound waves induced by such non-convex
thermodynamics. Alternatively, the presence of non-convex relativistic
regions may also induce a non-standard dynamics as a result of the
non-monotonic dependence of the sound speed with the rest-mass
density. We investigate the aforementioned possibilities in the next
sections.

\begin{figure}
\centering
    \includegraphics[width=\columnwidth]{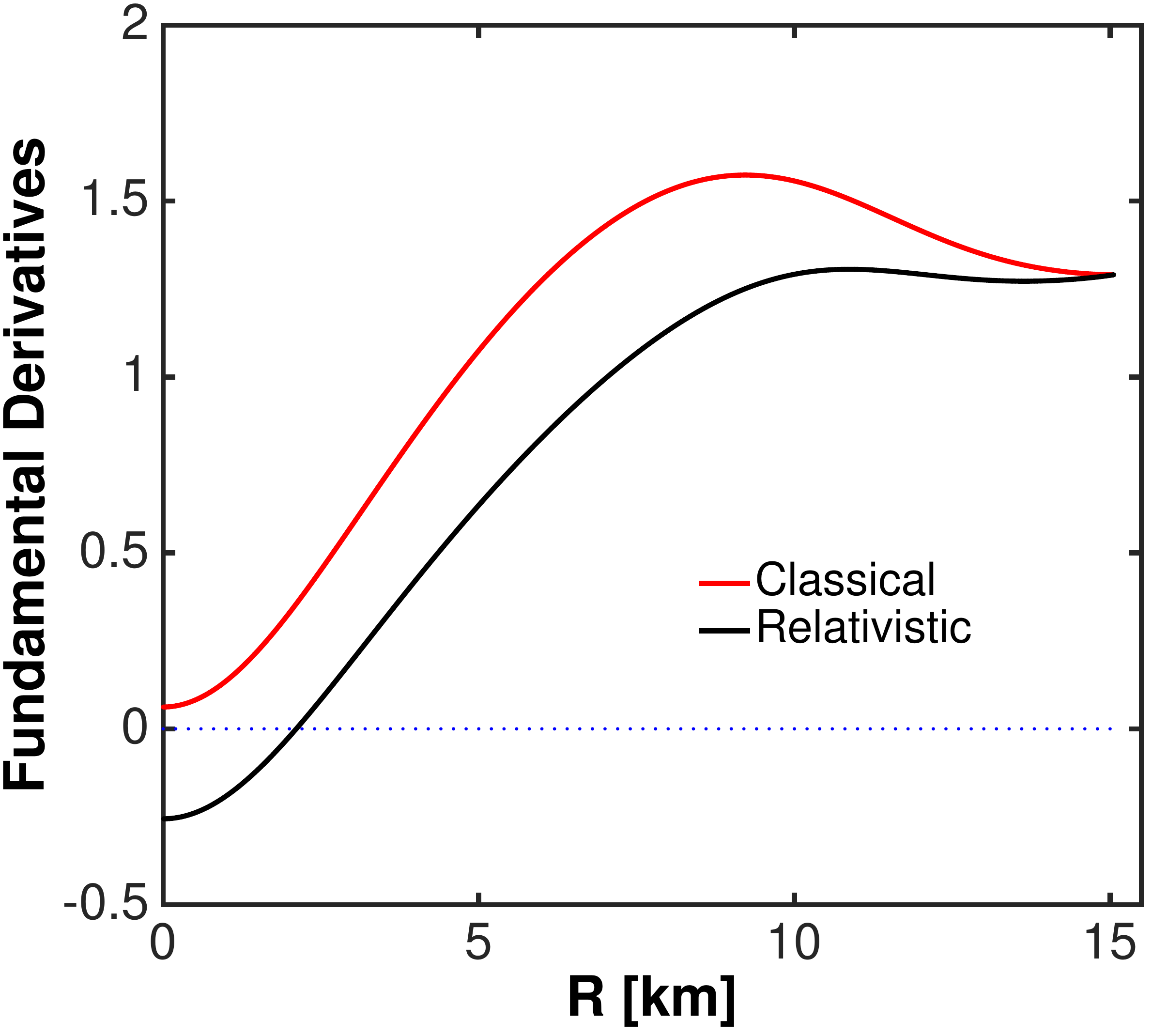}
\caption{
Radial profiles of the classical (red curve) and relativistic (black curve) fundamental derivatives for an equilibrium model with $\rho_{\rm c}/\rho_1 = 1.4$ and obeying the GGL-EoS.
}
  \label{fig:FD-radius-1}
\end{figure}


\section{Gravitational collapse in spherical symmetry}
\label{section:Collapse-1d}

\begin{table*}
  \centering
  \begin{tabular}{l|lcccccccc}
    \hline\hline
    model & IC & $\gamma_1$ &$\rho_1$& EoS & $\rho_{\rm c}$ & $M_0$ & $R_0$ &  $\xi$ &$t_{\rm BH}$\\
    name   &       &                       &$10^{15}\,$gr\,cm$^{-3}$&          &$10^{15}\,$gr\,cm$^{-3}$& $M_\odot$ &  [km] &          &[ms]
    \\ \hline
    P-1.9G & P & 1.9 &  $1.0$ & GGL &  $2.18$ & 1.39& 10.0&1.39 &0.066
    \\ 
    P-1.9S & P & 1.9 &  $1.0$ & SGGL &  $2.18$ & 1.39& 10.0&1.39&0.186
    \\ 
    P-1.6G & P & 1.6 &  $1.0$ & GGL &  $2.18$ & 1.39& 10.0&1.39&0.063
    \\ 
    P-1.45G & P & 1.45 &  $1.0$ & GGL &  $2.18$ & 1.39& 10.0&1.39&0.063
    \\ 
    G-1.9G & G & 1.9 &  $1.0$ & GGL &  $2.18$ & 1.28& 10.6&1.21&0.117
    \\ 
    S-1.9S & S & 1.9 &  $1.00$ & SGGL &  $2.18$ & 1.33& 11.3&1.18&0.166
    \\ \hline
    P-1.9G1& P & 1.9 &  $0.9356$ & GGL &  $2.046$ & 1.98& 11.8&1.68 &0.068
    \\ 
    P-1.9G2& P & 1.9 &  $1.00$ & GGL &  $2.046$ & 1.98& 11.8&1.68 &0.073
    \\ 
    P-1.9G3& P & 1.9 &  $1.31$ & GGL &  $2.046$ & 1.98& 11.8&1.68 &0.093
    \\ 
    P-1.9G4& P & 1.9&  $1.559$ & GGL &  $2.046$ & 1.98& 11.8&1.68 &0.111
    \\ 
    \hline\hline
  \end{tabular}
  \caption{
    List of spherically symmetric, non-rotating models.  For each model (name in the first
    column), the second column states the initial conditions (P, G,
    and S standing for the polytropic model and the ones computed with the
    GGL-EoS and SGGL-EoS, respectively).  The further columns characterize the EoS
    used in the simulation: the parameters $\gamma_1$ and $\rho_1$ are given in the
    third column and fourth columns, while in the fifth column we list
    that the variant of the GGL-EoS employed in the
    run. Finally, in the last four columns, we provide the
    mass, the radius and the compactness 
    of the initial configuration, as well as the time of formation of the BH, respectively.
  }
  \label{Tab:1d-models}
\end{table*}

Within the framework of the GGL-EoS, the most promising scenario for
encountering non-convex effects is the collapse of a star with a
central density, $\rho_{\mathrm{c}}$, similar to or above of $\rho_1$.
We explore this possibility first in spherically symmetric simulations
of toy models for neutron stars, comparing two equations of state.
Four models were simulated with the GGL-EoS with different parameters
and two with a modified version thereof, which we call Semi-GGL-EoS
(SGGL-EoS).  It consists of the GGL-EoS, but with a flat rather than
decaying adiabatic index above $\rho_1$:
\begin{equation}
\begin{split}
  \gamma^{\mathrm{SGGL}} &= \gamma^{\mathrm{GGL}} ( \min (\rho, \rho_1))\\
  &= \gamma_0 + {\cal K} \, \exp\left(-
    \frac{\min(\rho,\rho_1)-\rho_1}{\sigma^2} \right).
\end{split}
\label{eq:SGGL}
\end{equation}
This EoS maintains the stiffening of the GGL-EoS at $\rho = \rho_1$,
but its avoidance of  the non-convexity at high densities allows us to
gauge the importance of non-convex dynamics.

We consider three different initial models, all of which have been
constructed by solving the TOV equation, albeit using different EoSs:
\begin{enumerate}
\item The first one is a star with a polytropic EoS, $p=\kappa
  \rho^\gamma$, with a single adiabatic index $\gamma=2$ for all
  densities and $\kappa = 8.422 \times 10^{4}$ in CGS units.
\item The second model was computed with the GGL-EoS,
    following the prescription developed in
    Sec.\,\ref{section:TOV-GGL-EoS}, but with the
  following parameters: $\gamma_0 = 4/3, \gamma_1 = 1.9, \sigma = 1.1,
  \rho_1 = 10^{15} \, \gccm$.
\item In the third type of model, we use the SGGL-EoS (again
    following the prescription developed in
    Sec.\,\ref{section:TOV-GGL-EoS}) with the same parameters as
  in the point (ii).
\end{enumerate}
For numerical reasons, we endow our initial configurations
  with a power-law decaying atmosphere for values of the rest-mass density
  $\rho \lesssim 10^{-10}\rho_{\rm c}$, where $\rho_c$ is the central
  rest-mass density. This atmosphere possesses a dynamically
  negligible mass. Irrespective of the type of initial model, we
simulate the models with the GGL-EoS or SGGL-EoS.

All initial models have the central rest-mass density in the range
$\rho_{\mathrm{c}} \approx 2.05\times 10^{15} - 2.18 \times 10^{15} \,
\gccm$, which is about twice the parameter $\rho_1$. The mass
  of the initial configurations are either $M_0\simeq 1.39\,M_\odot$ or
  $M_0\simeq 1.98\,M_\odot$ (see Table~\ref{Tab:1d-models}).  These two
  masses roughly bracket the mass of the iron cores of massive stars
  (from which collapse a neutron star remnant may result) with main
  sequence masses in the range $10 M_\odot - 120 M_\odot$ and solar
  metallicity \citep{Woosley2007}. They are initially in equilibrium,
but an \emph{ad hoc} reduction of the pressure will trigger
their collapse.  Indeed, the reason to choose three different EoS to
  construct the initial model is that we aim to assess the dynamical
  effects of the \emph{ad hoc} initialization of the collapse on the
  subsequent dynamics. In the polytropic initial models,
the reduction of the pressure is the result of
the switch to the (S)GGL-EoS, while in the initial models with
GGL-EoS it is brought about by a uniform reduction of the internal
energy density by 15\%.  Following \cite{OConnor2011},
  we define a compactness parameter as
%
\begin{equation}
\xi := \frac{ M_0 / M_\odot }{ R_0 / 10 \, \text{km} },
\label{eq:compactness}
\end{equation}
where $M_0$ and $R_0$ are the initial mass and radius, respectively,
of the equilibrium configuration.\footnote{Note the difference in the
  definition of \emph{compactness}, ${\cal C}$, used in
  Sec.\,\ref{section:TOV-GGL-EoS}.} According to this definition, the
initial models built with a polytropic EoS are more compact than
models constructed with the (S)GGL-EoS. The models having larger mass
($M_0=1.98M_\odot$; Table~\mbox{\ref{Tab:1d-models})} feature the
largest compactness. Consistent with the large compactness of our
models, we do not expect them to develop supernova explosions, even if
a detailled neutrino physics and energy transport or magnetic fields
were included in our simulations. Certainly, both of these effects may
slightly change the dynamics, but for the purpose of assessing
exclusively the impact on the dynamics of the convexity loss, we may
neglect them.

  The simulations were performed with a version of the code {\sc
    Aenus} employed in \cite{Obergaulinger_Aloy:2017} and
  \cite{Obergaulinger_Just_Aloy:2018}, but restricted to special
  relativistic hydrodynamics, using fifth-order monotonicity
  preserving reconstruction schemes and an HLL Riemann solver.
  Gravity was incorporated using the pseudo-relativistic TOV potential
  of \cite{Marek_etal__2006__AA__TOV-potential}, which provides a very
  good approximation to full GR in spherical symmetry. Once the center
  collapses to a BH, we excise the innermost region. The excision is
  undertaken by following the evolution of the lapse function,
  $\alpha$ related to the pseudo-relativistic TOV potential, $\Phi$,
  by $\alpha\simeq \exp{(\Phi/c^2)}$. Numerical cells that develop
  $\alpha \le \alpha_{\rm th}:=0.018$ in the course of the evolution
  are \emph{frozen}, except for a gauge transformation, which shifts
  their position from their location when they hit the condition
  $\alpha=\alpha_{\rm th}$, $r=r_{\rm AH}$ (where $r_{\rm AH}$ is the
  radius of the apparent horizon) to $r=0$ by means of a suitable
  radial shift, $\beta_r$, on a time scale $r_{\rm AH}/\beta_r$. The
  latter shift greatly diminishes spurious reflections at the apparent
  horizon location.  The simulation grid consists of 3200 zones
  logarithmically spaced up to an outer radius of 180\,km.  The large
  extent of the grid, much larger than the radius of the initial
  equilibrium configuration, reduces any potential contamination by
  boundary effects.  The minimum grid resolution is $(\Delta
  r)_{\mathrm{min}} = 100 \, \mathrm{m}$.

\begin{figure*}
  \centering
  \includegraphics[width=0.41\linewidth]{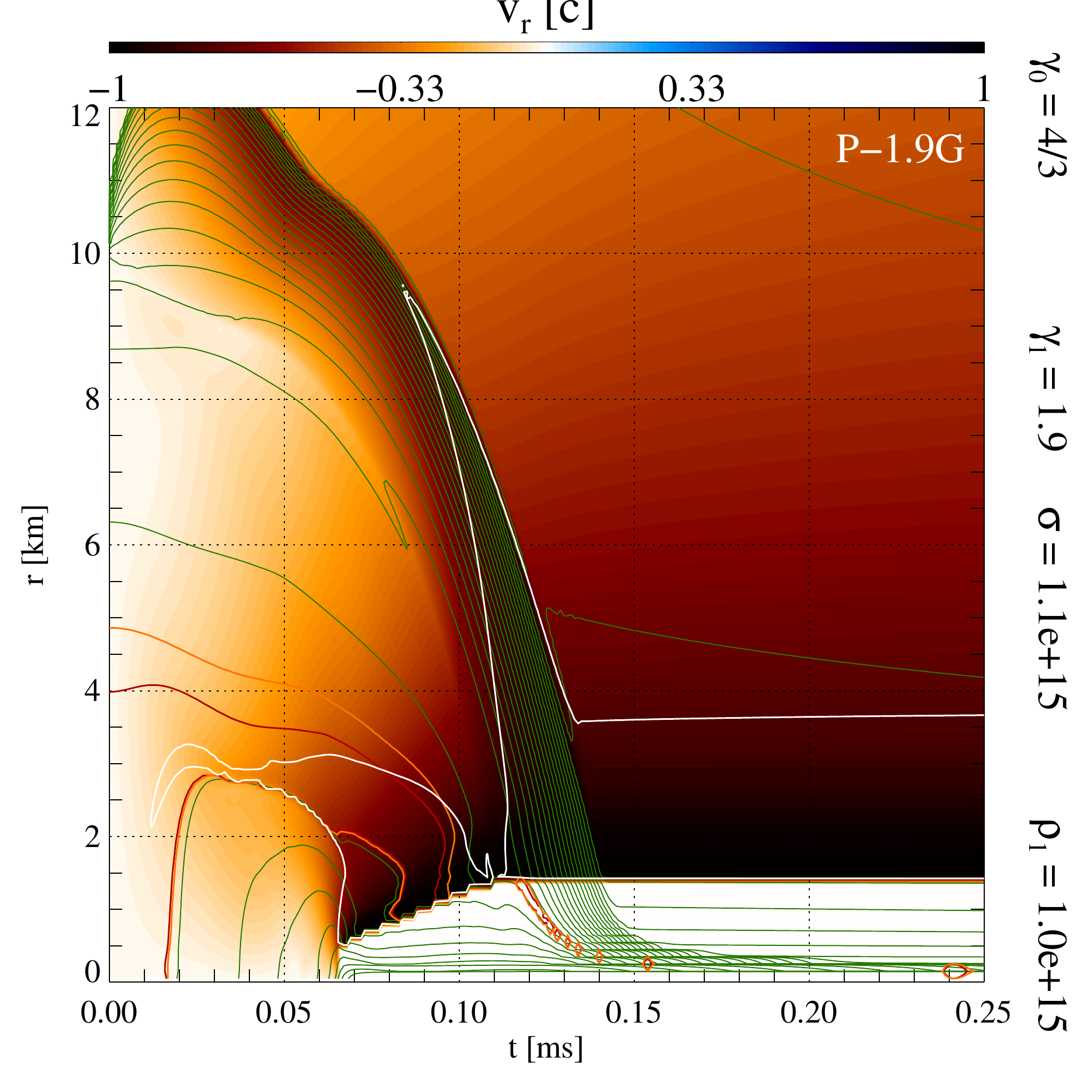}
  \includegraphics[width=0.41\linewidth]{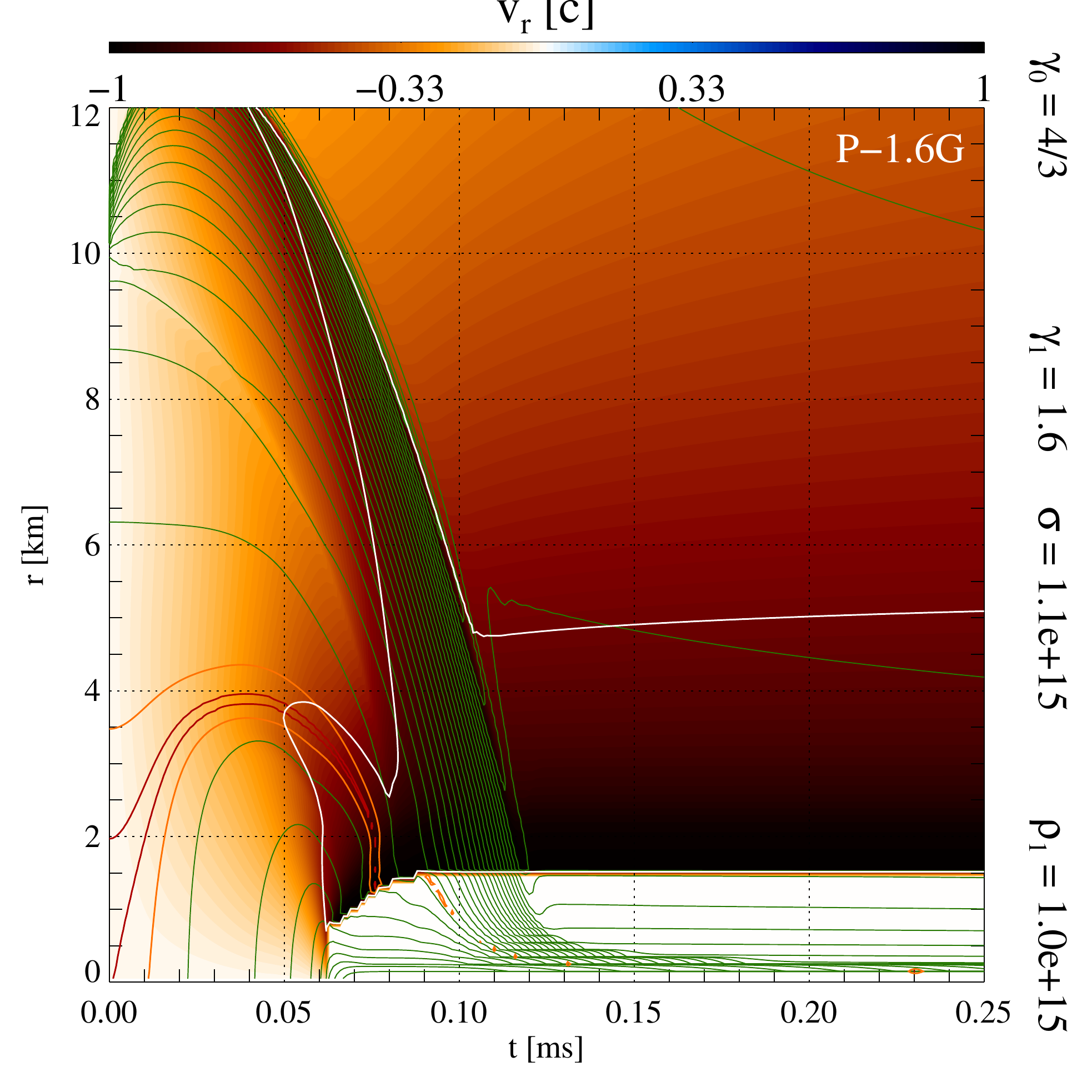}
  \includegraphics[width=0.41\linewidth]{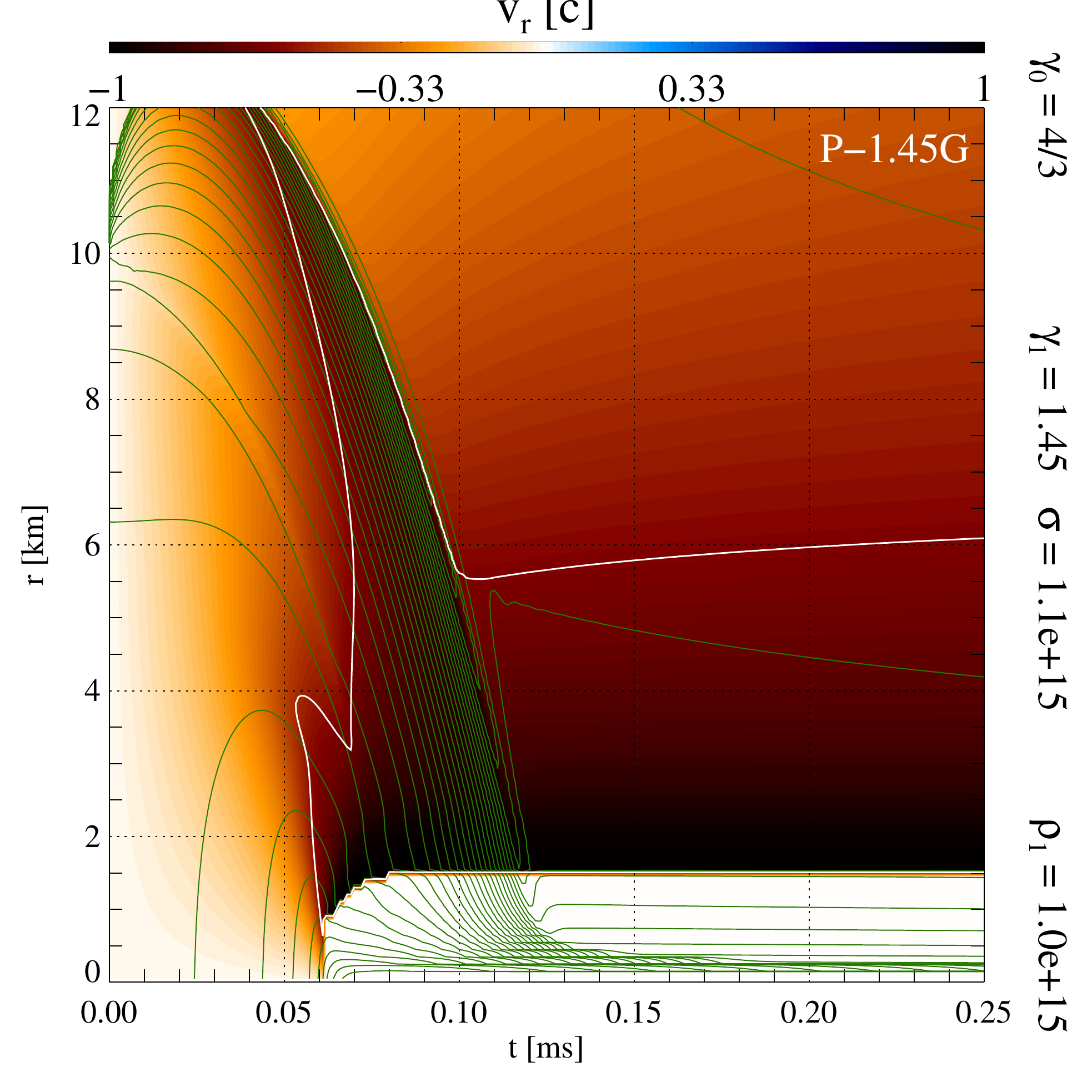}
  \includegraphics[width=0.41\linewidth]{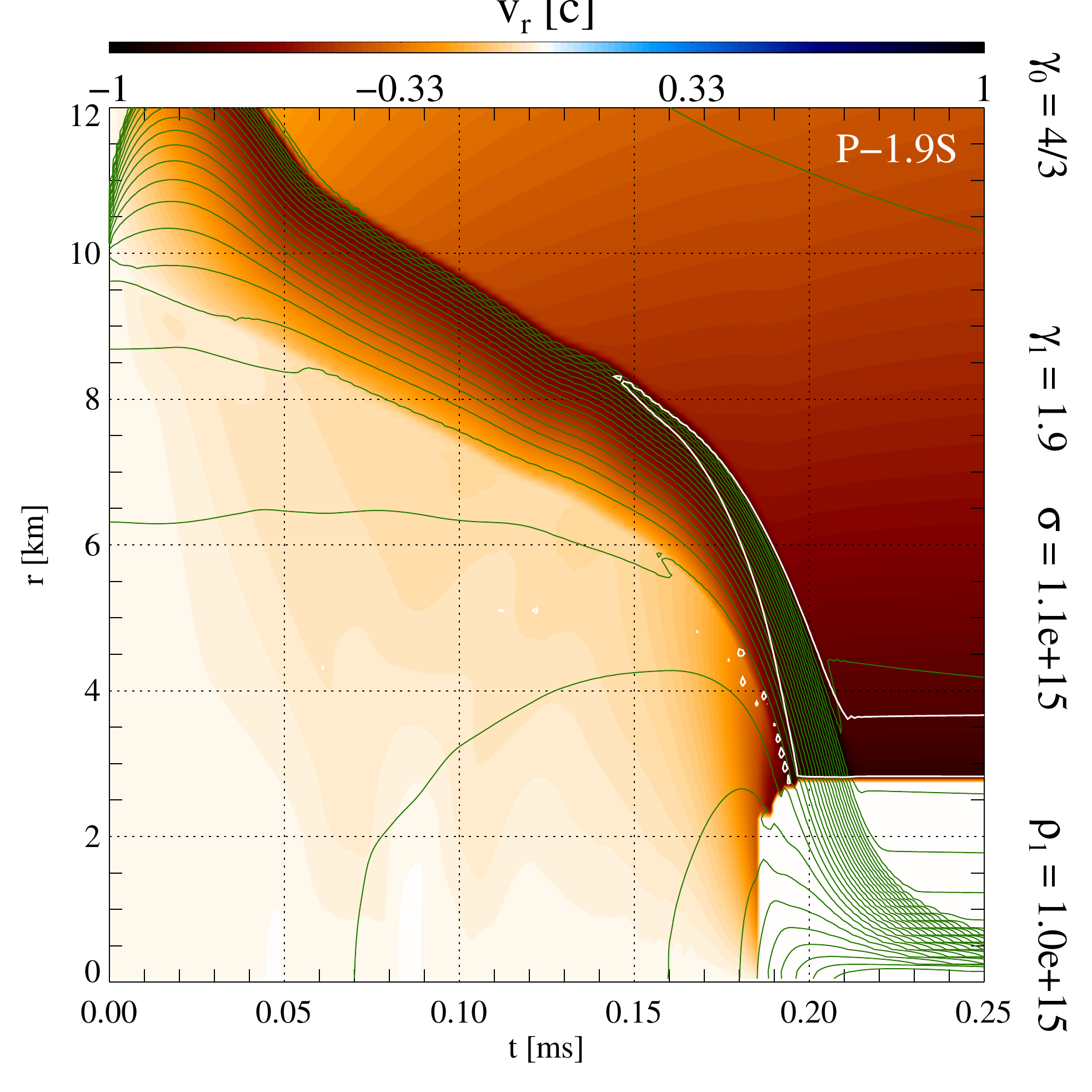}
  \includegraphics[width=0.41\linewidth]{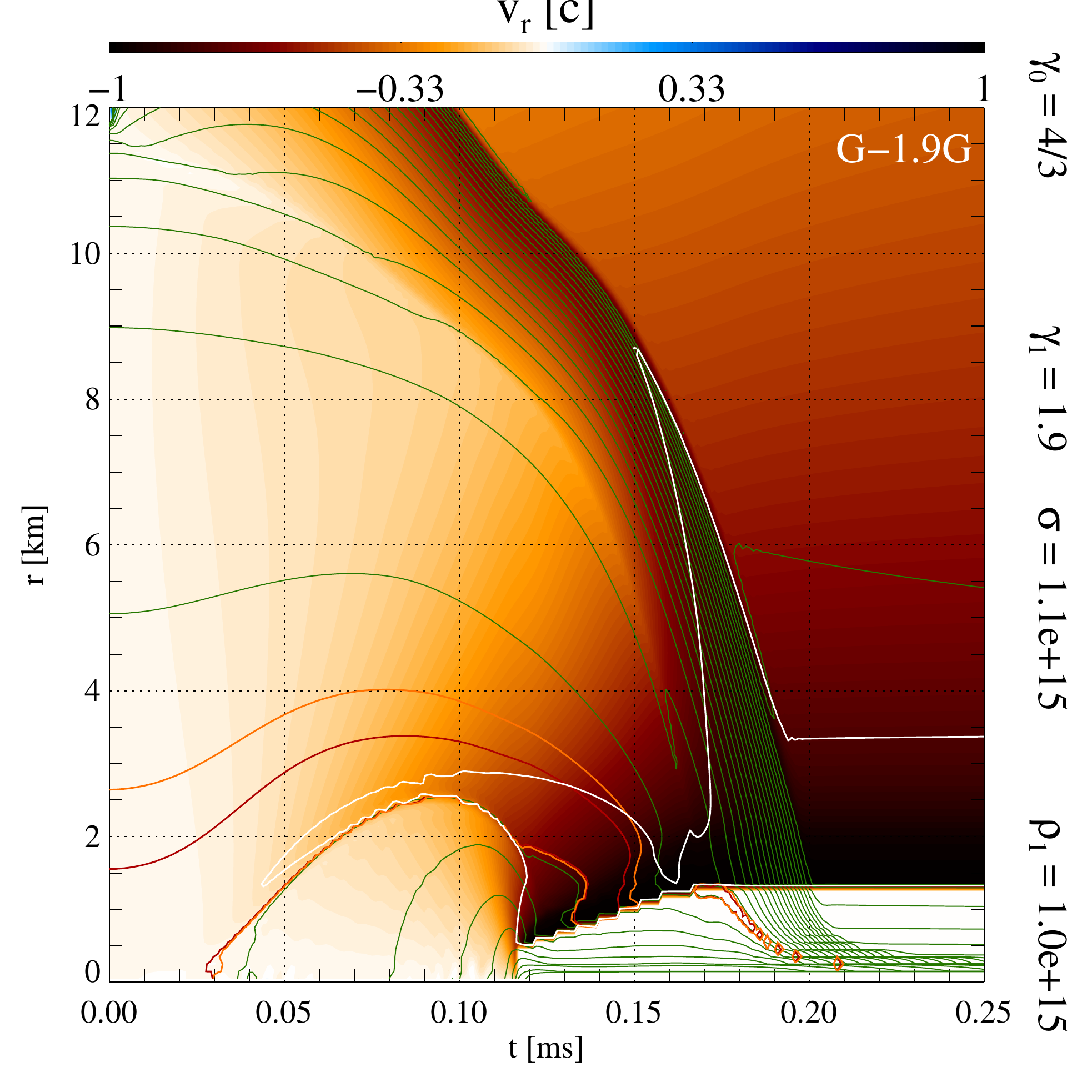}
  \includegraphics[width=0.41\linewidth]{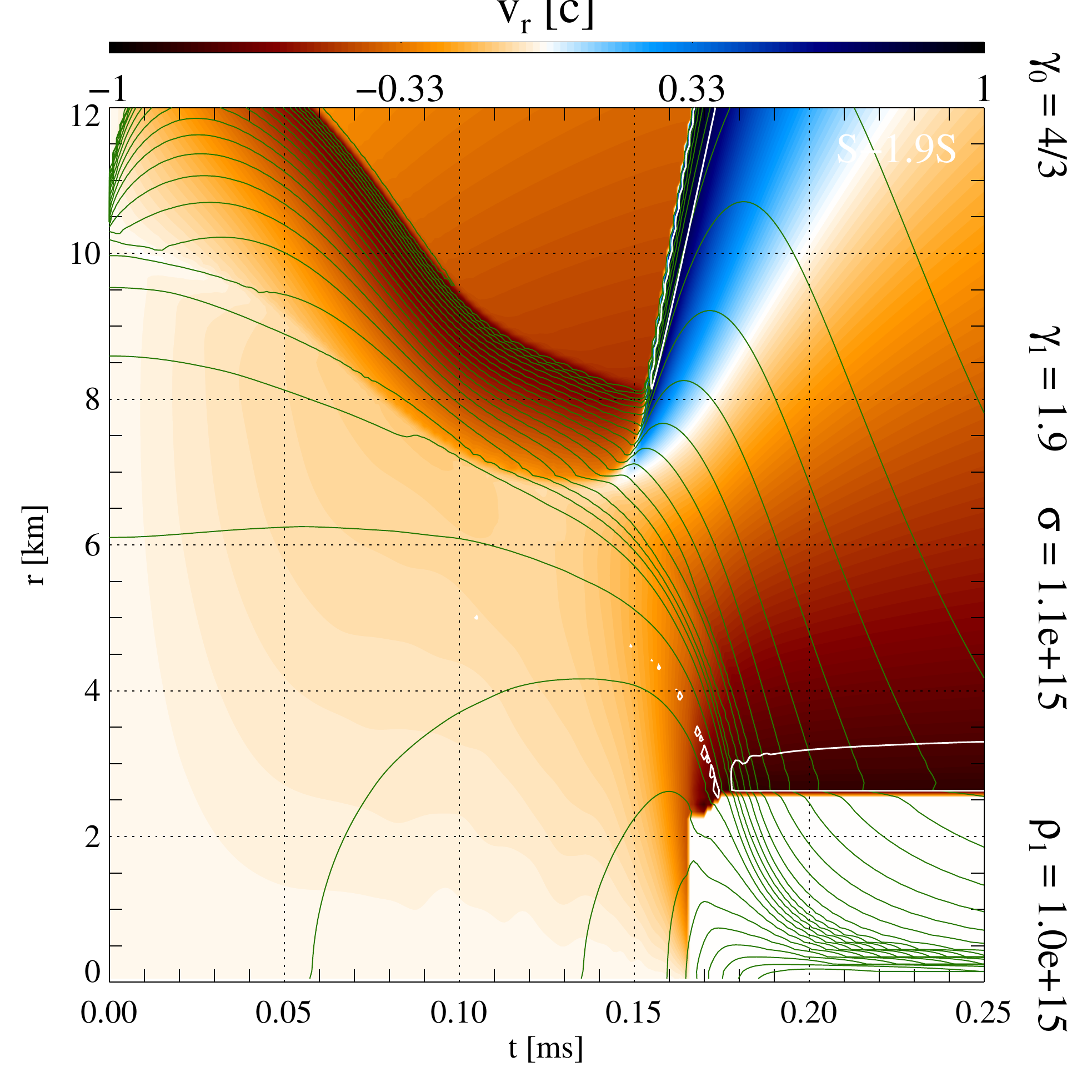}
  \caption{
    Evolution of six different spherically symmetric models, from top
    left to bottom right P-1.9G, P-1.6G, P-1.45G, P-1.9S, G-1.9G, and
    S-1.9S. The diagrams show the velocity in units of the speed of
    light as a function of time and radius.  In addition, contours of
    density (dark green lines) and the boundaries of the
    regions of classical and relativistic non-convexity are displayed
    (classical: dark red, relativistic: orange lines; models with the
    SGGL-EoS do not exhibit such regions) and the locations of the
    sonic point are marked by white lines.  The white region at the
    bottom of each panel is the excised BH. The
      black-blue-white, triangular region displayed in the lower right
      panel corresponds to parts of the self-gravitating configuration
      that bounce and acquire positive radial speeds.
  }
  \label{Fig:1d-collapse-1}
\end{figure*}

We present an overview of the time evolution of the six models holding
the smaller total mass in the spacetime diagrams shown in
Fig.~\ref{Fig:1d-collapse-1}.  All models collapse quickly, as we see
in the high negative velocities (brown shades in
Fig.~\ref{Fig:1d-collapse-1}) and the contracting iso-density
contours.  Black holes are formed promptly, between 0.06 and 0.19 ms
as can be seen from the growth of the white regions for $r<3\,$km in
the aforementioned figure and from the last column of
Table~\ref{Tab:1d-models}. This time scale can be compared with the
light-crossing time of the initial configurations, which range between
0.033 and 0.038 ms.  We note that the surface of the neutron star,
visible as a large concentration of iso-density (dark-green) contours
initially at about 10\,km Fig.~\ref{Fig:1d-collapse-1}, falls towards
the center.  In models with the standard GGL-EoS, the whole neutron
star is accreted, whereas it avoids this fate for the model initially
built and later evolved with the SGGL-EoS, where a shock wave is
launched at the surface and ejects parts of the matter (blue shades in
the lower right panel of Fig.~\ref{Fig:1d-collapse-1}).  This effect
is, however, only circumstantial to our analysis since it is not
connected to the appearance of non-convex regions in the star. It is
and artefact due to the artificial atmosphere that surrounds the
initial configuration, which is necessary for numerical reasons.
Instead, we turn our attention to the central regions before the
formation of the BH.

Model P-1.9G (top left panel of Fig.~\ref{Fig:1d-collapse-1})
possesses regions where the EoS is non-convex right from the
beginning: all gas inside radii of
$r_{\mathrm{ncr}} \approx 4.8 \, \km$ and
$r_{\mathrm{ncc}} \approx 4.0 \, \km$ is relativistically and
classically non-convex (cf.~the orange and red lines).  As the
collapse accelerates, velocities become supersonic and sonic points
form at $t \approx 0.01 \, \ms$ and
$r_{\mathrm{sp}} \approx 2.1 \, \km$ (white lines). Note that
differently from a standard collapse developed with a fully convex EoS
(of which model P-1.9S is an example), two separated sonic points form
relatively close to the stellar center (i.e. detached from the
-artificial- dynamics of the nearly free-falling stellar
surface). This is because of the non-monotonicity of the sound speed
dependence with density (e.g. Fig.\,\ref{fig:cs_rel}). Shortly
afterwards, the density increases sufficiently for the central regions
to become convex.  At $t \approx 0.03 \, \ms$, the inner sonic point
and the boundary between convex and non-convex regions merge.  At this
point, a shock wave forms at this transition (in the spacetime
diagram, it appears as a transition from darker to ligher brown in the
radial velocity maps starting at $r\sim 2.8\,$km, and following the
innermost white line). We highlight the fact that the formed shock is
compressive and not expansive, as one would guess from the fact that
it is produced in a non-convex thermodynamic region. Differentiating
between a compressive and an expansive shock is not straightforward
with an Eulerian numerical method, since the Rankine-Hugoniot jump
conditions do not exactly hold in the discrete numerical solution. The
shock is moving inwards for an observer at rest with respect to the
center of the star (see upper panels of
Fig.\,\ref{Fig:1d-collapse-lagrange}). Thus one could draw the
(erroneous) conclusion that the state \emph{upstream} of the shock is
that located to its left, while the state \emph{downstream} of the
shock is located to its right. If this were the case, subsonic matter
in the region upstream of the shock would cross it and end up in a
supersonic region. That would be interpreted as a sign of an expansion
shock. This conclusion is erroneous because matter is collapsing
(almost free-falling). Observing the lower panels of
Fig.\,\ref{Fig:1d-collapse-lagrange}, the shock clearly progresses
increasing the mass enclosed (to the left of the shock). Thus, the
lower panels unambiguously show that the state upstream of the shock
is located towards larger mass-coordinate (i.e. to the right of the
shock). This state is supersonic and matter crosses the
\emph{compression} shock and accumulates in the downstream subsonic
part of the flow. Since the compactness of our cores is so large, the
collapse is too violent for the shock wave to propagate outwards or
explode the star.  Instead, it remains an accretion shock through
which gas falls towards the center.  Furthermore, it is rather
short-lived and disappears at $t \approx 0.065 \, \ms$ inside the
nascent BH. After BH formation at $t\simeq 0.066\,$ms the sonic point
initially located closer to the center falls on the growing apparent
horizon, which becomes a transonic point thereafter. The second sonic
point, initially located further off center, soon follows the same
fate and touches the apparent horizon at $t\simeq
0.013\,$ms. Meanwhile, the collapsing outer stellar shells speed up
and become supersonic, first close to the infalling surface and a bit
later closer to the apparent horizon. The formation of another sonic
point right at the location where the initial atmosphere is set up (a
point that is also free falling with the rest of the star) is an
artefact of the atmospheric initialization. After most of the mass
falls onto the BH ($t\simeq 0.14\,$ms) this artificial sonic point
remains steady at a distance of $r\simeq 3.6\,$km. From there on, the
dynamics ceases and a steady accretion of the artificial atmosphere
goes on. We stress again that the mass in the atmosphere is totally
negligible with respect to the initial mass of the model.

\begin{figure*}
  \centering
  \includegraphics[width=0.41\linewidth]{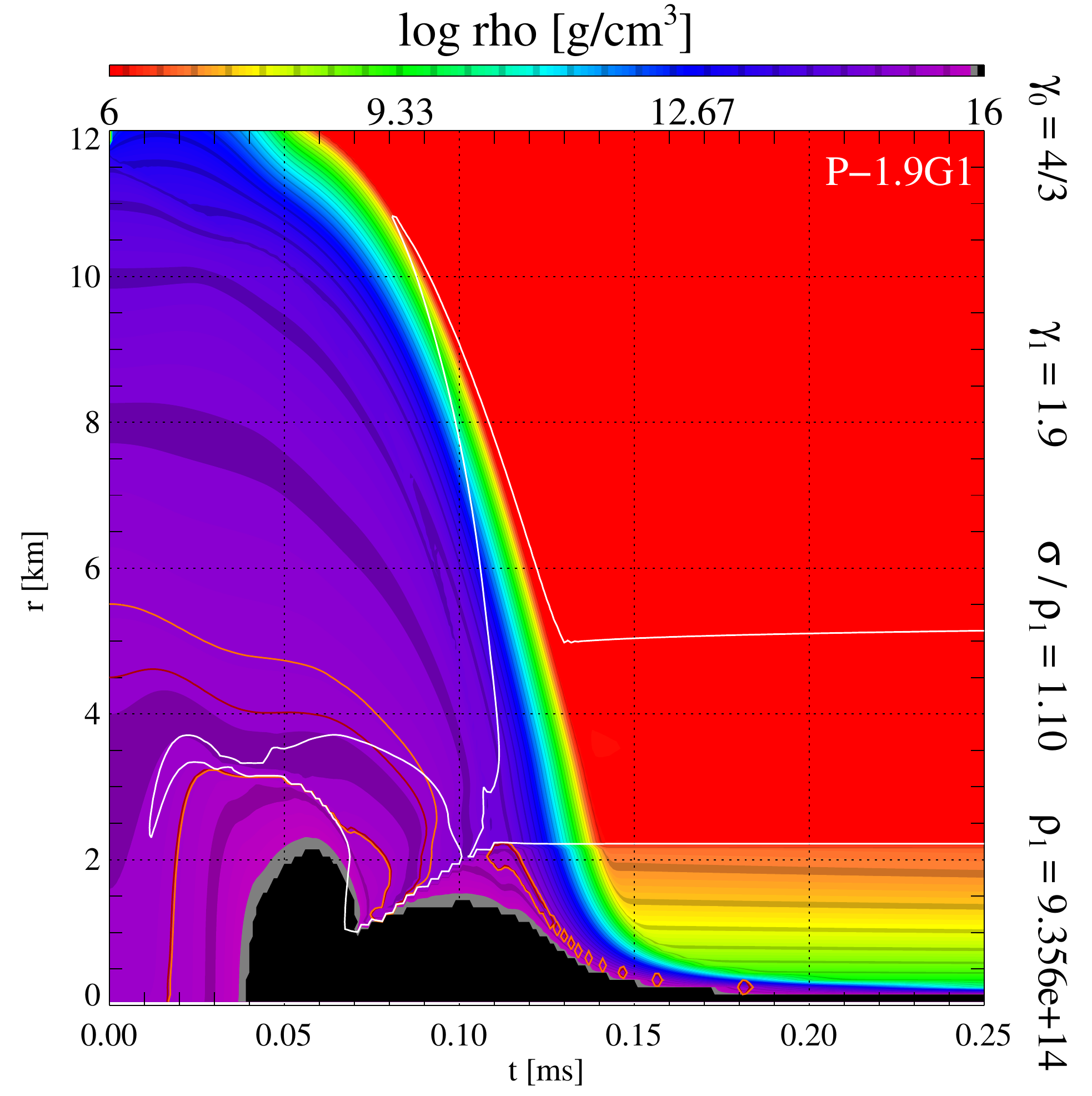}
  \includegraphics[width=0.41\linewidth]{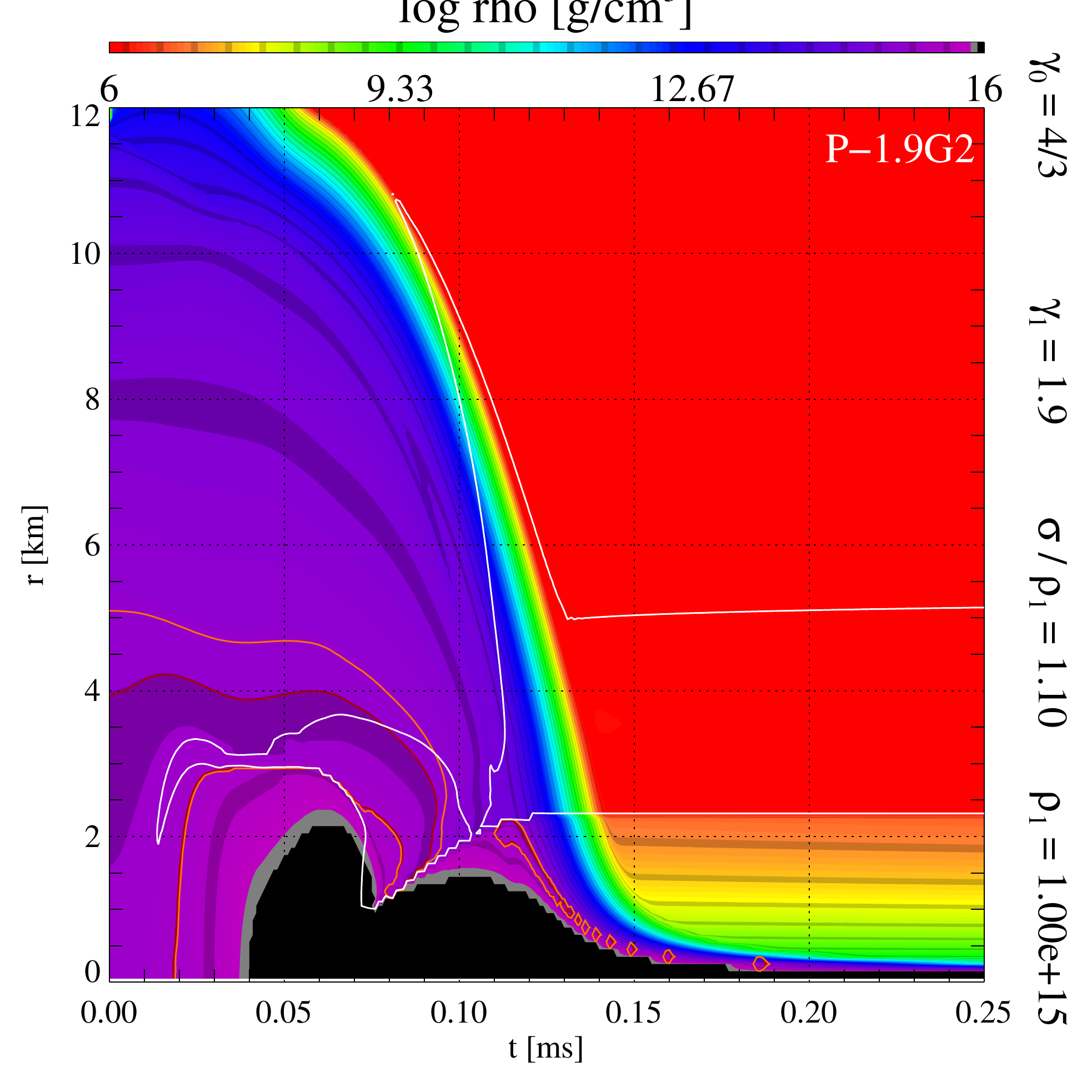}
  \includegraphics[width=0.41\linewidth]{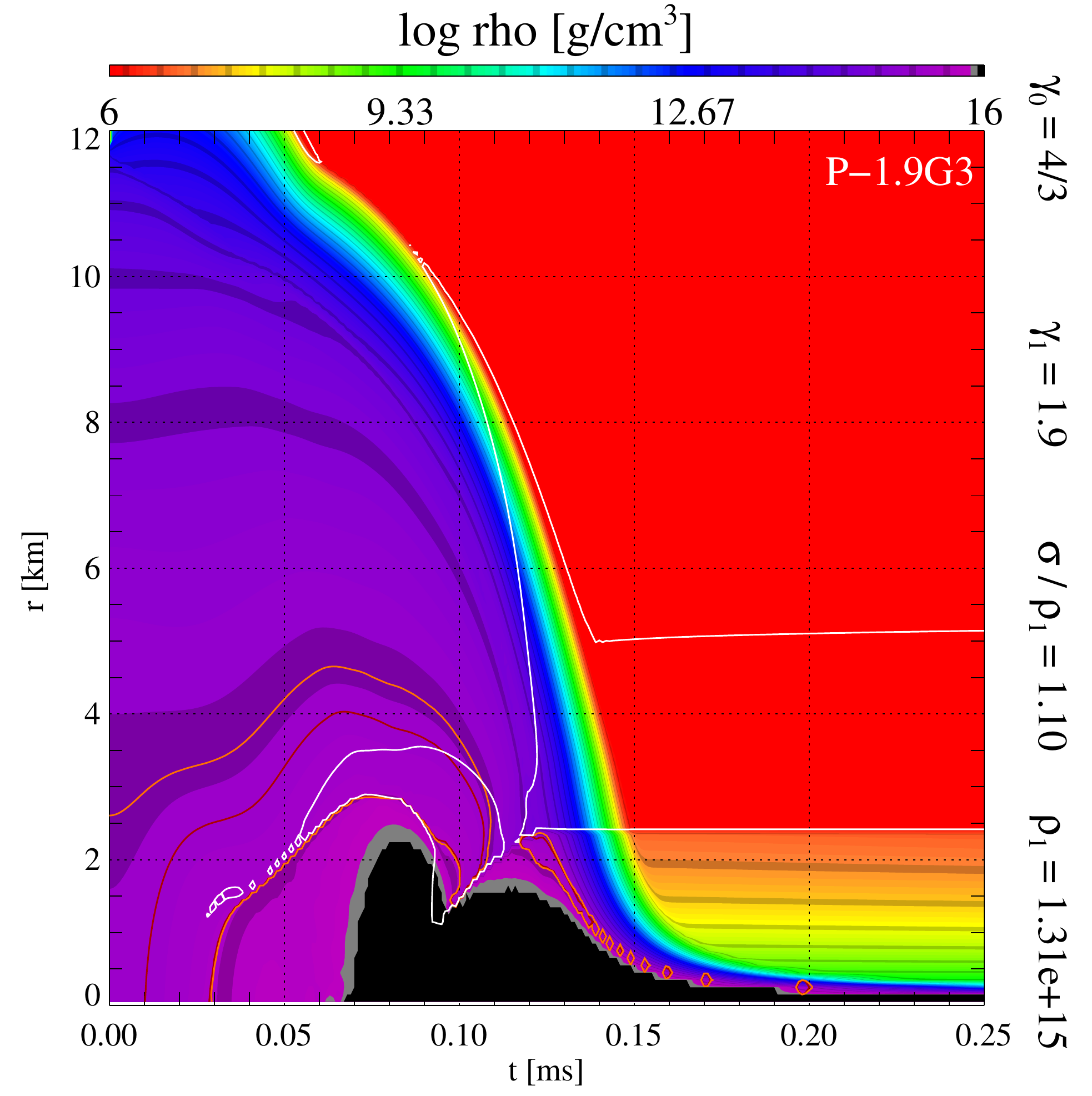}
  \includegraphics[width=0.41\linewidth]{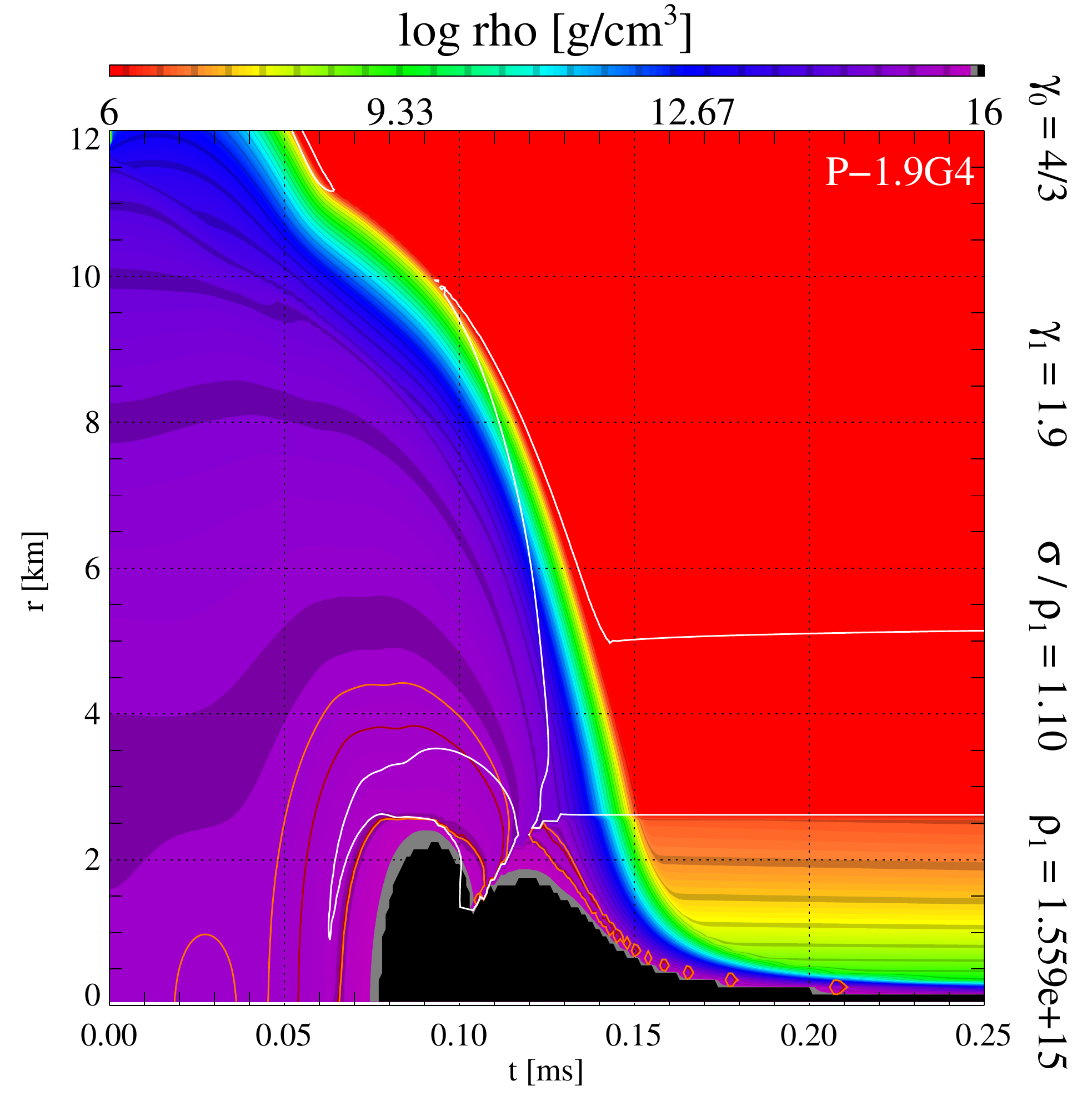}
  \caption{
    Evolution of four different spherically symmetric models, from top left
    to bottom right P-1.9G1, P-1.9G2, P-1.9G3 and P-1.9G4. The diagrams
    show the logarithm of the rest-mass density as a function of time and radius.  In addition, the boundaries of the
    regions of classical and relativistic non-convexity are displayed
    (classical: dark red, relativistic: orange lines) and the locations of the
    sonic point are marked by white lines.  The black region below the
    sonic point located at $r\simeq 2.55\,$km at the
    bottom of each panel are excised from the computational domain
    (it corresponds to the BH). 
  }
  \label{Fig:1d-collapse-1-rho}
\end{figure*}

\begin{figure*}
  \centering
  \includegraphics[width=0.99\linewidth]{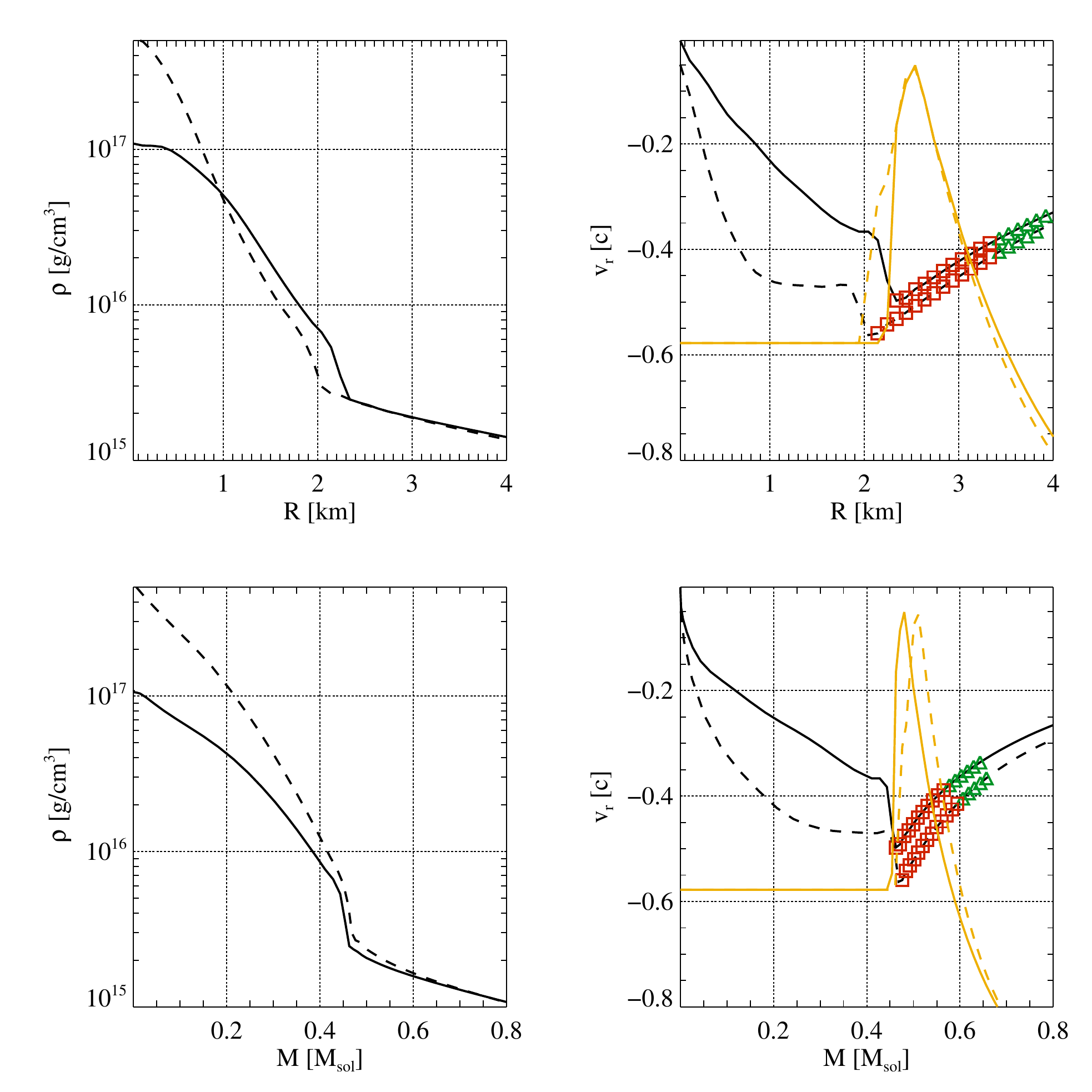}
  \caption{
    Zoom of selected hydrodynamic properties of model
      P-1.9G. (Top panels): As a function of the distance to the
      center of the star we show (left panel) the distribution of the
      rest-mass density and (right panel) the distribution of the
      velocity (black lines) and of the sound speed (orange
      lines). The solid (dashed) lines correspond to a time
      $t=0.060\,$ms ($t=0.064\,$ms). On the right panel, the non-convex
      region is marked with symbols: where $\Gclas<0$ and $\Grel<0$ we
      use red squares, while the region where $\Gclas>0$ and $\Grel<0$
      is displayed with green triangles.  The accretion shock moves
      \emph{inwards} from its position $R\simeq 2.3\,$km at
      $t=0.060\,$ms to $R\simeq 2\,$km at $t=0.064$\,ms. Note that the
      shock forms at the boundary between the regions where the
      classical fundamental derivative changes sign from $\Gclas>0$
      (left to the shock) to $\Gclas<0$ (right to the shock). Left to
      the points where
      $\Gclas=\Grel=0$ there is a sonic point in the fluid (another
      sonic point is located farther away from the center).  (Bottom
      panels): Same as the top panels but as a function of the
      enclosed mass. The accretion shock moves \emph{outwards} from an
      enclosed mass $M\simeq 0.45\,M_\odot$ at $t=0.060\,$ms to
      $M\simeq 0.46\,M_\odot$ at $t=0.064\,$ms.  Comparing the left and
      right panels it is ease to see that the sound speed is not a
      monotonic function of the rest-mass density in the non-convex
      region.
    }
  \label{Fig:1d-collapse-lagrange}
\end{figure*}
Reducing $\gamma_1$ to a value of $\gamma_1 = 1.6$ (model P-1.6G, top
right panel of Fig.~\ref{Fig:1d-collapse-1}) leads to a faster
collapse and reduces the extent of non-convex regions.  Although sonic
points form as in the previous model, they do not align with the
border of the non-convex region.  A compression shock forms, but it is
much weaker than before, hardly noticeable in the space-time diagram.
A further reduction to $\gamma_1 = 1.45$ (model P-1.45G, middle left
panel) entirely removes the non-convex region.  No shock can be
observed, and the collapse proceeds smoothly.  This statement does not
hold for model P-1.9S (middle right panel), where we use the SGGL-EoS
with $\gamma_1 = 1.9$.  In this case, the absence of non-convex
regions is not due to the low value of $\gamma_1$, but to the constant
adiabatic index above $\rho_1$.  This case demonstrates that the
appearance of a (compression) shock wave is not solely connected to
the value of $\gamma_1$, but to the non-convexity. This remark is
relevant in view of the fact that virtually all EoS of nuclear matter
yield values of $\gamma$ significantly larger than $\gamma_1=1.45$ for
rest-mass densities above $\sim 10^{14}\,$gr\,cm$^{-3}$.

The two models with the initial data constructed for the (S)GGL-EoS
(G-1.9G and S-1.9S, bottom panels of Fig.~\ref{Fig:1d-collapse-1})
confirm the findings obtained for the polytropic initial models.  For
the standard GGL-EoS, a compression shock is formed at the inner
border of the convex region, where the inner sonic point is situated.
Similarly to the polytropic initial model, it does not suffice to
explode the star and ultimately ends up in the BH.  The model with the
SGGL-EoS, on the other hand, does not develop a shock wave in the
vicinity of the BH. We stress that the parameters used for both EoS
(GGL and SGGL) are the same. Therefore, this result confirms that when
convexity is not lost, no shocks form in the course of the collapse to
BH. Remarkably, the only difference between, the SGGL EoS and the
GGL-EoS is that the former one avoids the convexity loss preventing
the steep decline of the adiabatic index after its maximum at
$\rho=\rho_1$ (see Eq.\,\eqref{eq:SGGL}). The SGGL-EoS is stiffer than
the GGL-EoS at high densities. In spite of this crucial difference, no
accretion shock forms using the SGGL-EoS (in contrast to the model run
with the GGL-EoS with the same parameters), even if one could argue
that a stiffer EoS is more likely prone to produce \emph{bounces} in
the dynamics with the potential formation of associated shocks.

Models P-1.9G and G-1.9G are evolved with the same GGL-EoS,
  but differ in the initial configuration, which is polytropic (with
  $\gamma=2$) for the former and constructed according to the GGL-EoS
  with $\gamma_{\rm max}=\gamma_1=1.9$ for the latter. This difference in the initial
  configuration yields a temporal shift to the overall dynamics, which
  otherwise is qualitatively the same. We observe a delayed BH formation in model G-1.9G compared
  to model P-1.9G (see Table~\ref{Tab:1d-models}). Furthermore, the
  formation of the shock associated with the existence of two sonic
  points in the collapsing fluid is also present (though delayed) in
  model G-1.9G. Thus, we conclude that building a polytropic initial
  model and then evolving it with the GGL-EoS does not introduce
  major differences either in the qualitative dynamics, nor in the
  final fate of the collapsing core.

  We have also run a series of models having relatively large mases of
  nearly $2 M_\odot$. This series is formed by models P-1.9G1,
  P-1.9G2, P-1.9G3 and P-1.9G4, which have all the same initial
  polytropic model ($\gamma=2$ and $\kappa=100$ in the same units we
  employ later in Sec.\,\ref{section:Collapse-GGL-EoS}, or,
  equivalently $\kappa = 3.46 \times 10^{5}$ in CGS units) but the
  evolution is followed employing the GGL-EoS with different values of
  $\rho_1$ and $\sigma$ (see Table~\ref{Tab:1d-models}). For the
  latter, we fix $\sigma=1.10$ in the former four cases. These models
  are the non-rotating analogs of the models D1 that we will introduce
  in the next section (see Tab.~\ref{table2}). Spacetime diagrams of
  the logarithm of the rest-mass density of all these models are
  displayed in Fig.\,\ref{Fig:1d-collapse-1-rho}. We observe that all
  of them show the same qualitative behaviour as described for the
  reference case P-1.9G. From this series of models we observe that BH
  formation time increases with $\rho_1$ (see
  Table~\ref{Tab:1d-models}). In the spacetime evolution of the
  rest-mass density we observe the much smaller density of the
  surrounding (rarefied) atmosphere (red shades in all panels of
  Fig.\,\ref{Fig:1d-collapse-1-rho}). We also point out that the
  BH-excised region displays a density gradient from the values in the
  atmosphere to the highest densities in the domain (note the regions
  below the inner sonic point displayed with a white contour, which is
  nearly horizontal for $t>0.15\,$ms, in all the panels of the
  figure). This gradient is the result of the radial velocity shift we
  apply inside of the excised region (see above) to concentrate
  effectively all the mass in a volume around $r\simeq 0$. There is,
  however, a small quantitative difference among the P-1.9G1 to
  P-1.9G4 series of models in the time of shock formation, which is
  associated with the loss of convexity of the EoS, as in the previous
  models of lower mass. The region of non-convexity does not appear
  from the very beginning in model P-1.9G4. There is, first a small
  region surrounding the center of the star where the relativistic
  fundamental derivative becomes negative during a brief and
  transitory episode ($0.02\,\text{ms}\,\lesssim t \lesssim
  0.036\,$ms). Due to the adjustment of the central region to the loss
  of convexity, a small oscillation happens and the core slightly
  expands. Since the collapse is ongoing, the oscillation is very
  quickly dumped and, once the density in the vicinity of the stellar
  center grows again above $\simeq \rho_1$, the non-convex region
  begins to grow from the center (at $t\simeq 0.04\,$ms) until it
  reaches a maximum radial extend of $\simeq 4.5\,$km at $t\simeq
  0.07\,$ms. As the shock forms so close to the BH formation time, it
  is even very difficult to detect it as a shock in our numerical
  simulations. We observe that the region where the classical
  fundamental derivative is negative does not appear from the
  beginning in model P-1.9G3. Instead, it appears at $t\simeq
  0.01\,$ms at $r\simeq 2.8\,$km, moves radially outwards a few hundred
  meters (up to $r\lesssim 3\,$km) and then falls back onto the
  BH. Also in the latter model the shock formation is slightly delayed
  with respect to the initiation of the core collapse, though not so
  much as in model P-1.9G4. In model P-1.9G3 the shock forms
  sufficiently early to be clearly captured in our simulations.

  We point out that the numerical code employed in this section is
  different from the one used in the next one for reasons we discuss
  in Section \ref{section:summary}. We have repeated the experiments
  presented in this section with the same fully general relativistic
  hydrodynamics code with which we obtain the results of Section
  \ref{section:Collapse-GGL-EoS} finding that the qualitative results
  as well as the quantitative details are nearly the same. This result
  is reassuring from the methodological point of view since the
  algorithms implemented in both codes are significantly different. We
  also consider the independence of the results with respect to the
  numerical details as a clear hint of their robustness.

\section{Gravitational collapse of rotating neutron stars}
\label{section:Collapse-GGL-EoS}

In order to study the effects of using our non-convex GGL-EoS in a
fully dynamical situation, we consider uniformly rotating neutron star
models that are dynamically unstable to axisymmetric perturbations
and, hence, collapse to BHs on a dynamical timescale. In the previous
section, we have chosen the spherically-symmetric and non-rotating
initial data for this purpose. Here, we rather consider the more
interesting rotating case since it allows to identify the influence of
the non-convex EoS not only on the dynamics of the collapse but {\it
  also} on the gravitational-wave signals produced in the process. In
particular, we use as initial data two uniformly rotating relativistic
star models, dubbed D1 and D4, that have been previously used in a
number of numerical-relativity simulations of neutron star
collapse~\citep{Font2002a,Baiotti2005,Baiotti_etal2007,Giacomazzo_etal_2011}. We construct our initial
rotating stellar models for a polytropic EoS, $p = \kappa \,
\rho^\gamma$, where $\kappa=100$ (in code units, where
$G=c=M_{\odot}=1$) is the polytropic constant and $\gamma=2$ is the
adiabatic index, using the RNS open-access
code~\citep{Stergioulas1995}. The main characteristics of our two
models are reported in Table\,\ref{T3}. Model D1 is slowly rotating
and thus almost spherical, with a ratio of polar-to-equatorial
coordinate radii of $r_{\rm p}/r_{\rm e}=0.95$. Correspondingly, model
D4 is rotating almost at the mass-shedding limit, with $r_{\rm
  p}/r_{\rm e}=0.65$. BU2 is a stable model with $r_{\rm p}/r_{\rm
  e}=0.90$.

The numerical evolution of the initial data entails solving the
coupled system of equations given by Einstein's equations, governing
the dynamics of the gravitational field, and by the hydrodynamics
equations, governing the dynamics of the matter. This is done using
the numerical-relativity code in spherical-polar coordinates described
in~\citet{Baumgarte2013,Montero2014} and that we have used in previous
works \citep[see
e.g.][]{Sanchis-Gual:2015sms,Sanchis-Gual:2017ps}. The Einstein
equations are formulated in the so-called BSSN
formulation~\citep{Baumgarte1998,Shibata1995}. The evolution equations
are integrated using the second-order PIRK
method~\citep{Isabel:2012arx,Casas:2014} which allows to handle
singular terms associated with the choice of curvilinear
coordinates. The derivatives in the spacetime evolution are computed
using fourth-order finite-differences, including fourth-order
Kreiss-Oliger dissipation terms to avoid high-frequency noise.  The
equations of hydrodynamics are formulated in the so-called Valencia
formulation~\citep{Banyuls1997} and solved using the second-order MC
reconstruction scheme and the HLLE approximate Riemann
solver~\citep{Montero:2012yr}. Despite the initial data are built
using a polytropic EoS, they are evolved in our code using the
GGL-EoS, Eqs.\,(\ref{GGL-1a}-\ref{GGL-1b}).  As we have tested
  in the Sec.\,\ref{section:Collapse-1d}, building a polytropic
  initial model and then evolving it with the GGL-EoS does not
  introduce major differences either in the qualitative dynamics, or
  in the final fate of the collapsing core. It simply results in a
  \emph{delayed} dynamics, including the time of BH formation.

\begin{figure}
\begin{center}
\includegraphics[width=0.47\textwidth]{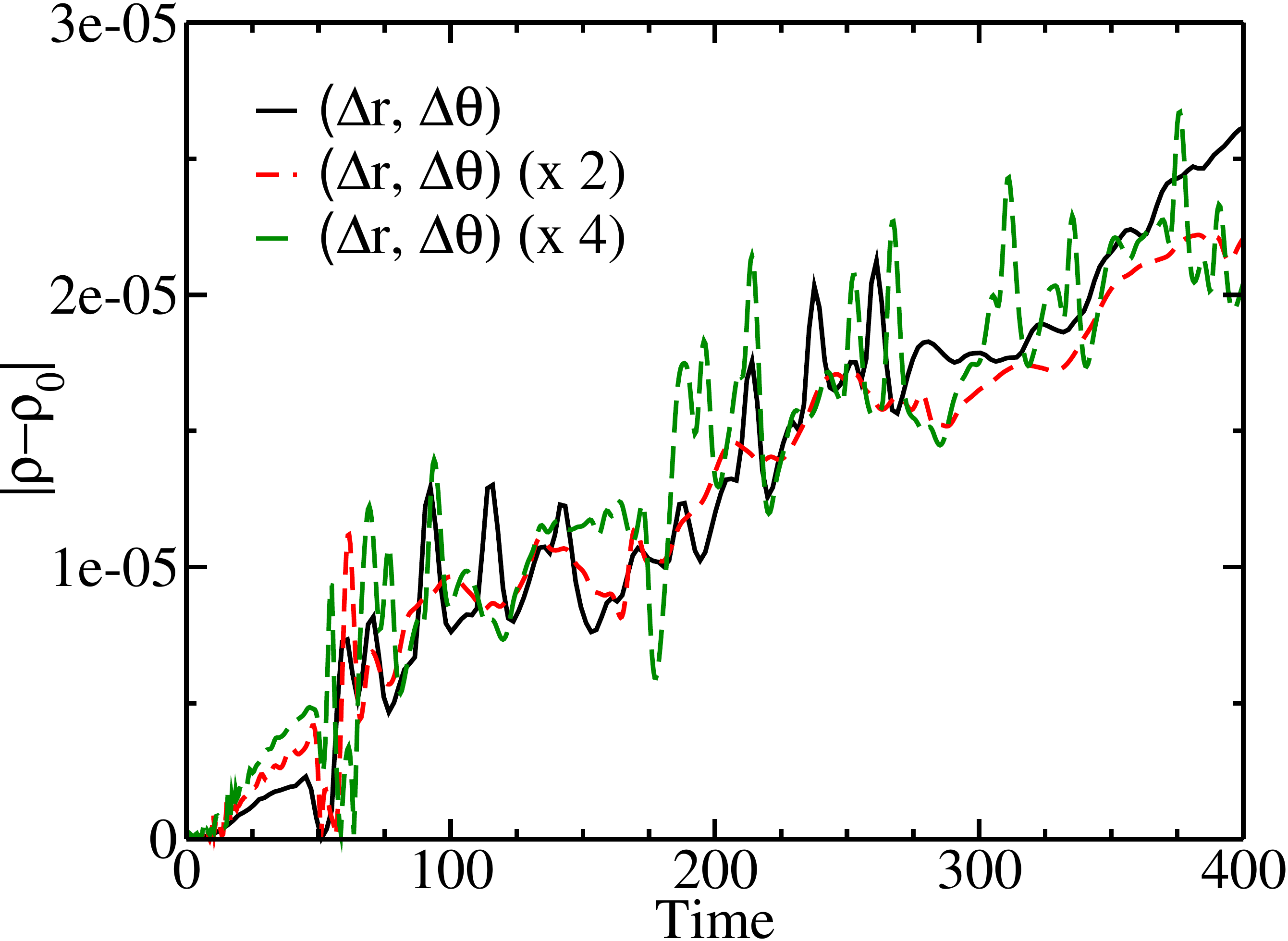}
\includegraphics[width=0.47\textwidth]{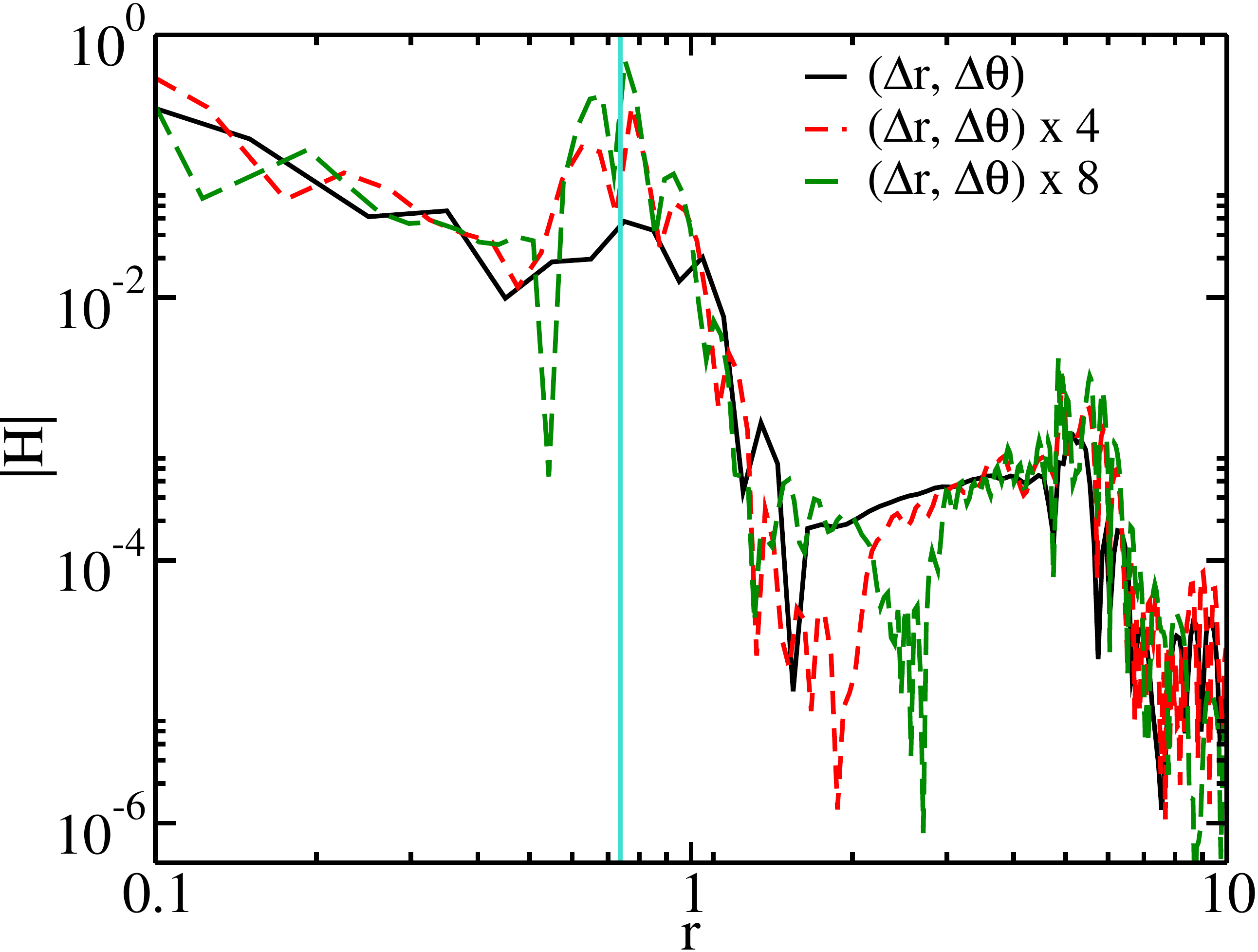}
\caption{(Top panel): L1-norm of the difference between the evolved
  rest-mass density and the initial one as a function of time for
  model BU2. Three different resolutions have been employed. In the
  legends, we show the minimum radial grid spacings of each of them
  ($\Delta r, \,\Delta \theta) =\lbrace(0.05,\pi/16), (0.071,\pi/12),
  (0.10,\pi/8)\rbrace$. The results corresponding to the finer
  resolutions are multiplied by the factors of 4 and 2 to show clearly
  the second order convergence of the method. As in all evolution
  plots, time is given in ``code units'', corresponding to
  $G=c=M_{\odot}=1$. (Bottom panel): Radial profile of the Hamiltonian
  constraint for model D1 with $\rho_{1}=1.5\times10^{-3}$ for three
  different resolutions ($\Delta r, \,\Delta \theta)
  =\lbrace(0.035,\pi/44), (0.05,\pi/32), (0.10,\pi/16)\rbrace$
  rescaled to second order convergence. All models have been evolved
  for a dimensionless time $t=50$. The snapshot corresponds to the
  dashed curve in the top right panel in Fig.~\ref{fg:NC2}.
  The vertical cyan line signals the location of the shock
    wave in model D1 with $\rho_1=1.5\times 10^{-3}$
    (Tab.\,\ref{table2}). Around the shock location is where the
    largest (absolute value) violations of the Hamiltonian constraint
    occur in our models.}
\label{fg:NC0}
\end{center}
\end{figure}

 \begin{table}
\begin{center}
\vspace{.3cm}
\caption{Uniformly rotating neutron star models with $\gamma$ = 2 and $\kappa$ = 100. From left to right the columns report the name of the model, the central density $\rho_{\rm c}$ in code units and in CGS units, the ratio of polar-to-equatorial coordinate radii $r_{\rm p}/r_{\rm e}$, the gravitational mass $M_{\rm G}$, and the circumferential equatorial radius $R_{\rm e}$.} 
\begin{tabular}{cccccc}
\hline \hline 
Model&$\rho_{\rm c}$&$\rho_{\rm c}$  &$r_{\rm p}/r_{\rm e}$&$M_{\rm G}$&$R_{\rm e}$ \\
& [code units] & [g\,cm$^{-3}$] & & $M_\odot$& [km]\\
\hline
D1&$3.280\times 10^{-3}$ & $2.046 \times 10^{15}$ & 0.95&1.665&11.5\\
D4&$3.116\times 10^{-3}$ & $1.944 \times 10^{15}$ & 0.65&1.861&14.4\\
BU2&$1.280\times 10^{-3}$ & $7.984 \times 10^{14}$ & 0.90& 1.466&15.0\\
\hline \hline
\label{T3}
\end{tabular}
\end{center}
\end{table}

\begin{table}
\begin{center}
\vspace{.3cm}
\caption{Parameters of the GGL-EoS used in the rotating neutron star collapse simulations.} 
\begin{tabular}{ccccc}
\hline \hline 
$\gamma_{0}$&$\gamma_{1}$&$\sigma/\rho_{1}$&$\rho_{1}$&$\rho_{1}$ \\
& & & [code units] & [g\,cm$^{-3}$] \\
\hline
4/3&1.9&1.10&$1.5\times 10^{-3}$& 9.356 $\times 10^{14}$\\
4/3&1.9&1.10&$1.7\times 10^{-3}$& 1.060 $\times 10^{15}$\\
4/3&1.9&1.10 / 1.15 / 1.20 / 1.50&$2.1\times 10^{-3}$& 1.310 $\times 10^{15}$\\
4/3&1.9&1.10&$2.5\times 10^{-3}$& 1.559 $\times 10^{15}$\\
\hline \hline
\label{table2}
\end{tabular}
\end{center}
\end{table}

\begin{table}
\begin{center}
\vspace{.3cm}
\caption{Central properties of various models used in the rotating neutron star collapse simulations. $\gamma_0=4/3$ and $\gamma_1=1.9$ for all models.} 
\begin{tabular}{ccccc}
\hline \hline 
Model& $\sigma/\rho_{1}$&$\rho_{1}$& $\rho_{c}/\rho_1$ & $p_{\rm c}/\rho_1$\\
\hline
D1&1.10&$1.5\times 10^{-3}$& 2.187 & 0.366\\
D1&1.10&$1.7\times 10^{-3}$& 1.929 & 0.387\\
D1&1.10&$2.1\times 10^{-3}$& 1.562 & 0.394\\
D1&1.10&$2.5\times 10^{-3}$& 1.312 & 0.368\\
D4&1.10&$1.5\times 10^{-3}$& 2.077 & 0.356\\
D4&1.10&$1.7\times 10^{-3}$& 1.833 & 0.373\\
D4&1.10&$2.1\times 10^{-3}$& 1.484 & 0.370\\
D4&1.10&$2.5\times 10^{-3}$& 1.246 & 0.339\\
\hline \hline
\label{table2a}
\end{tabular}
\end{center}
\end{table}

\begin{figure}
\begin{center}
\includegraphics[width=0.47\textwidth]{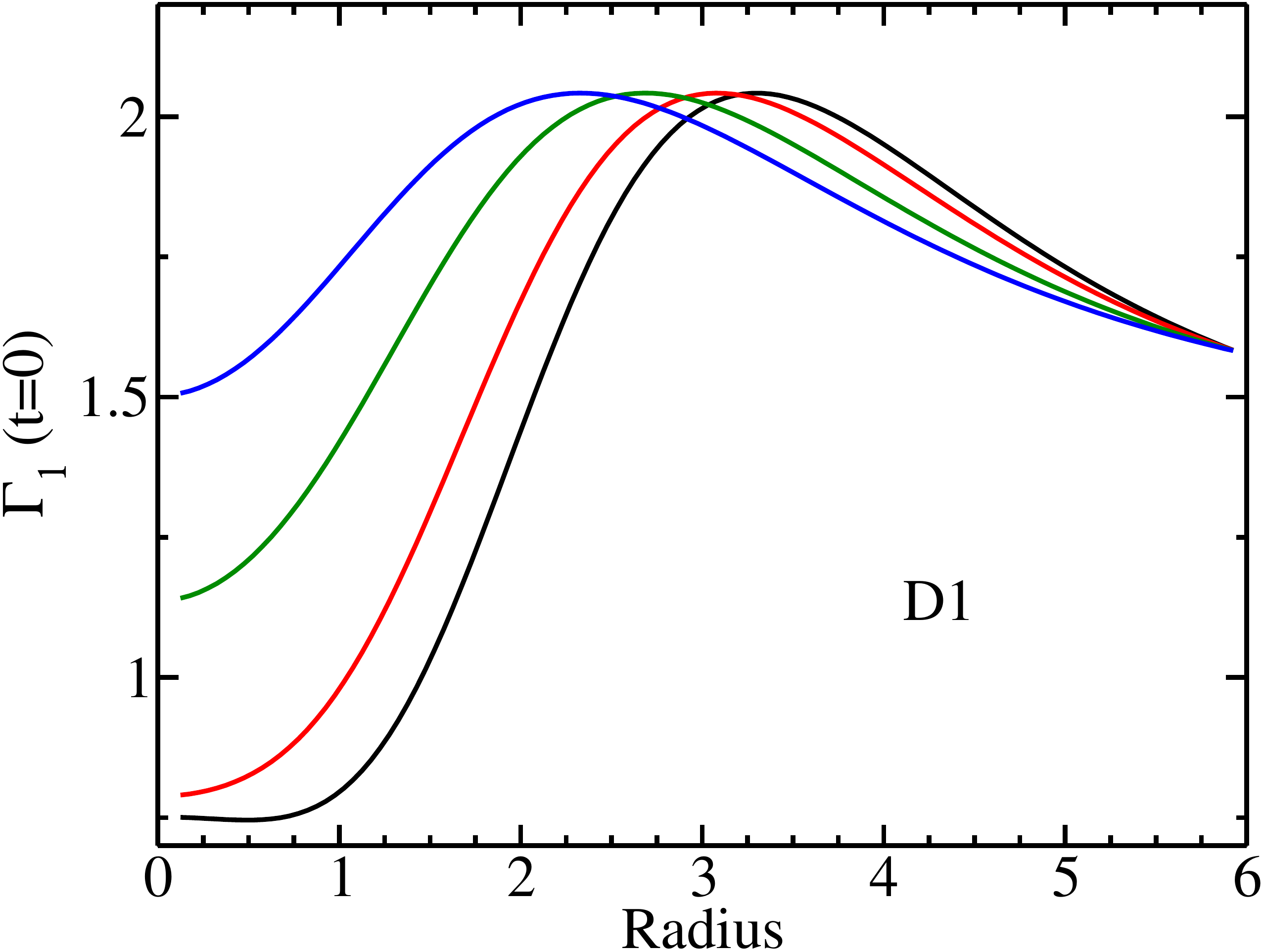}
\includegraphics[width=0.47\textwidth]{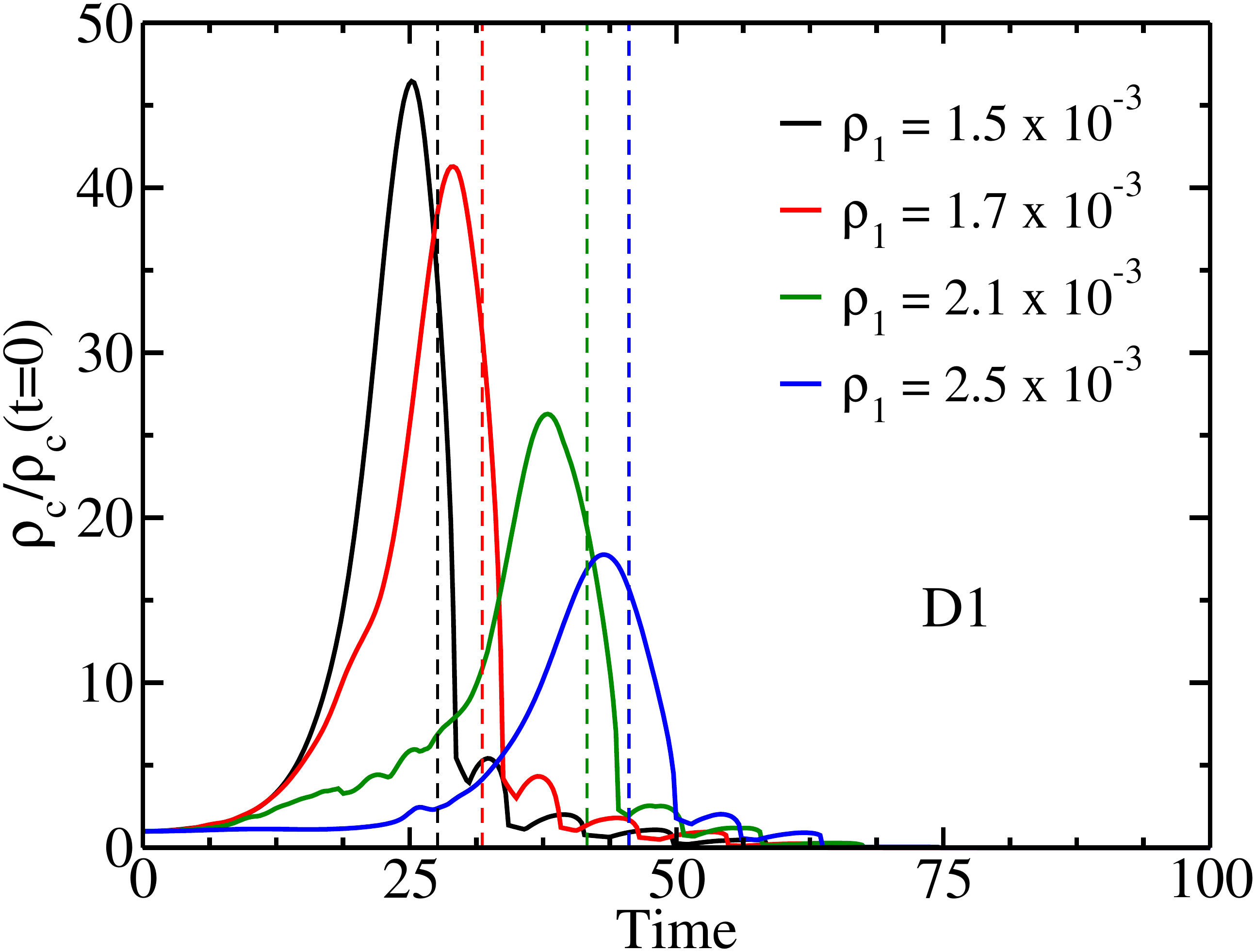}
\caption{{\it Top panel}: Initial radial distribution of $\Gamma_1$
  along the equatorial plane for model D1. {\it Bottom panel}: Time
  evolution of the central rest-mass density $\rho$ for model D1. The
  vertical dashed lines indicate the time at which the apparent
  horizon forms for each value of $\rho_1$. The legend is the same for
  the two panels.}
\label{fg:NC1}
\end{center}
\end{figure}

\begin{figure} 
\begin{center} 
\includegraphics[width=0.47\textwidth]{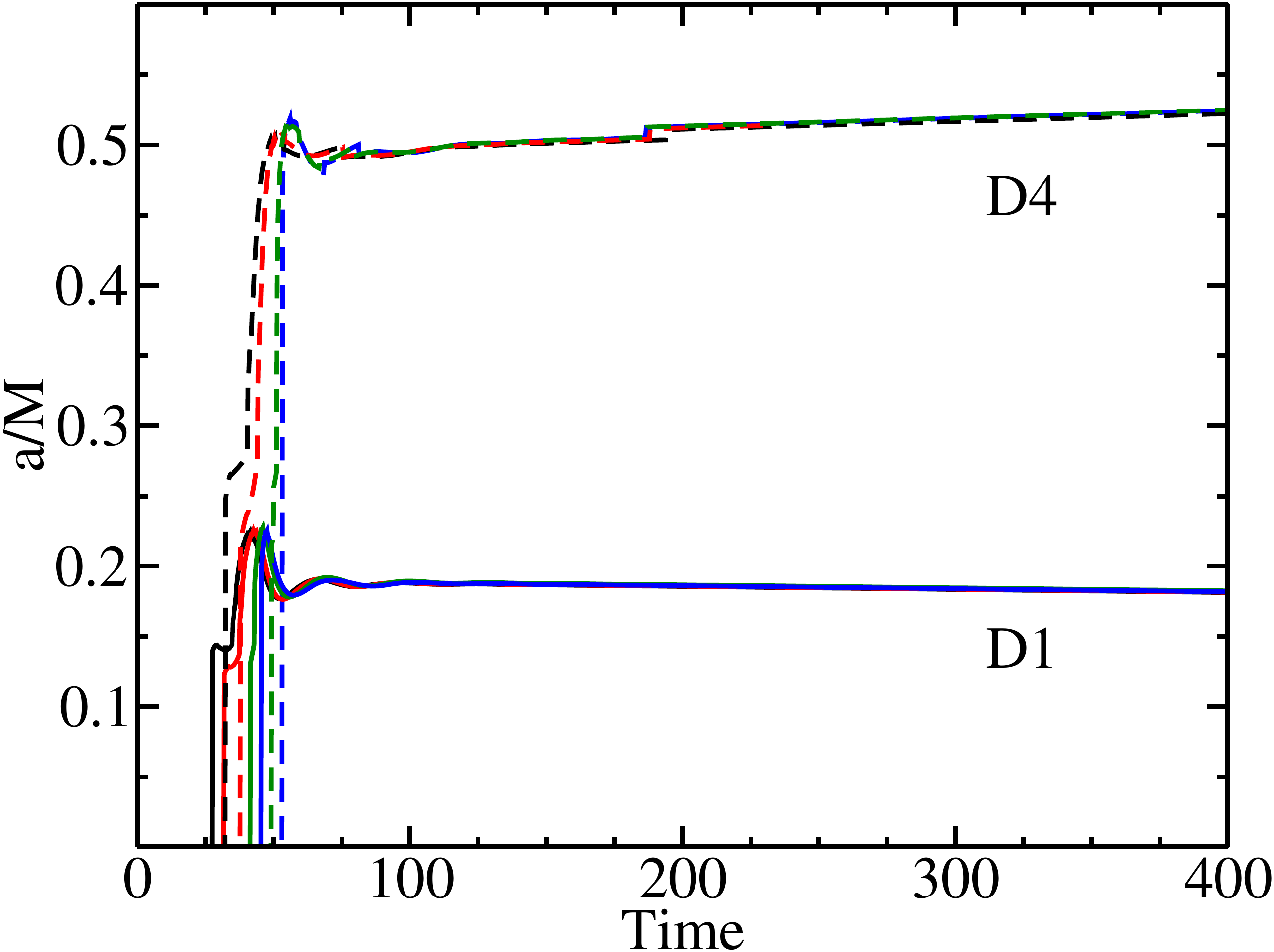} 
\caption{Spins of the final BHs after the collapse of model D1 (solid
  lines) and model D4 (dashed lines). Each set of four curves
  corresponds to the four values of $\rho_1$ in the same colour code
  as in Fig.~\ref{fg:NC1}.}
\label{fg:NC1-spin}
\end{center}
\end{figure}

\begin{figure*}
\begin{center}
\includegraphics[width=0.47\textwidth]{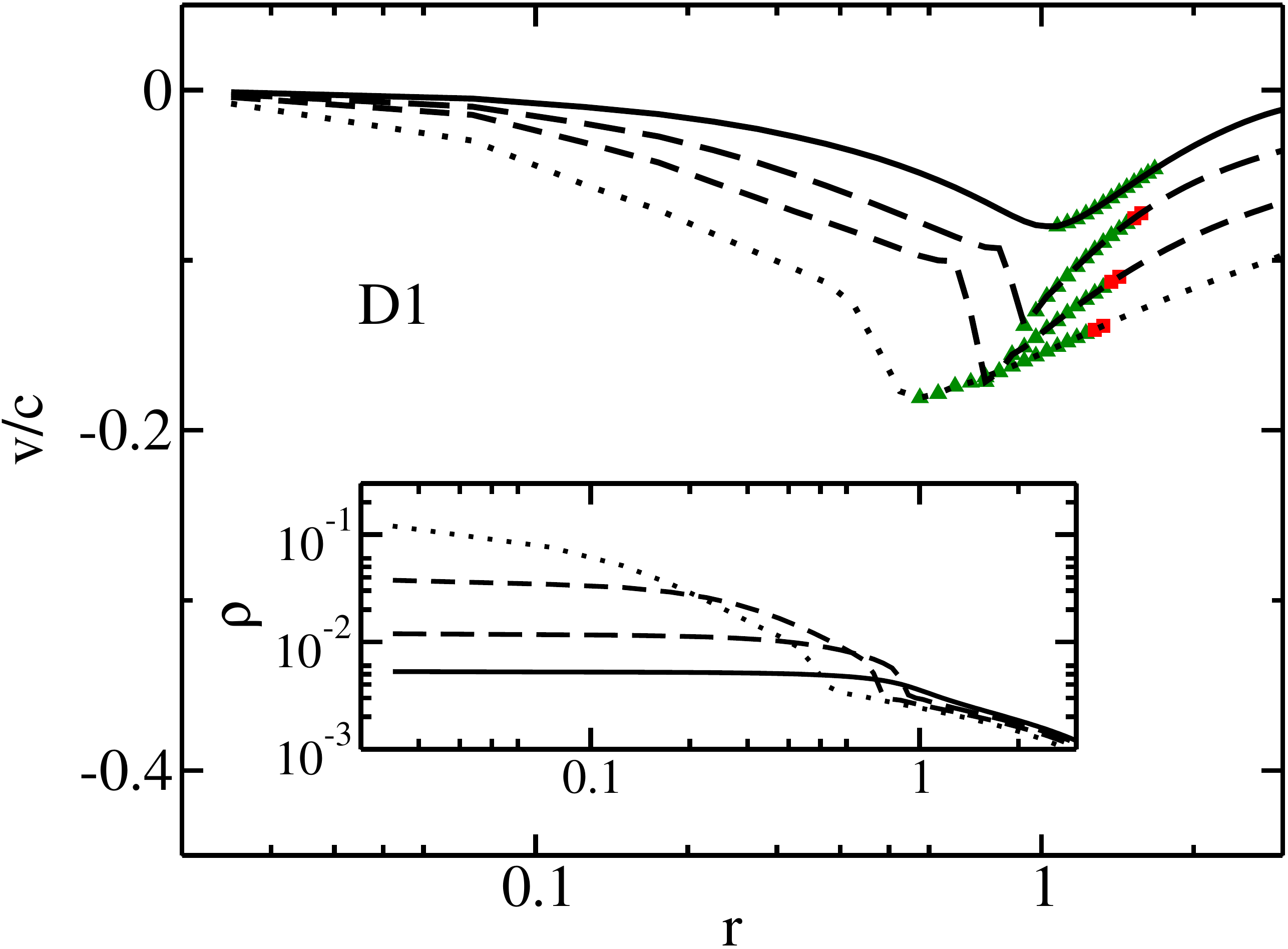} 
\includegraphics[width=0.47\textwidth]{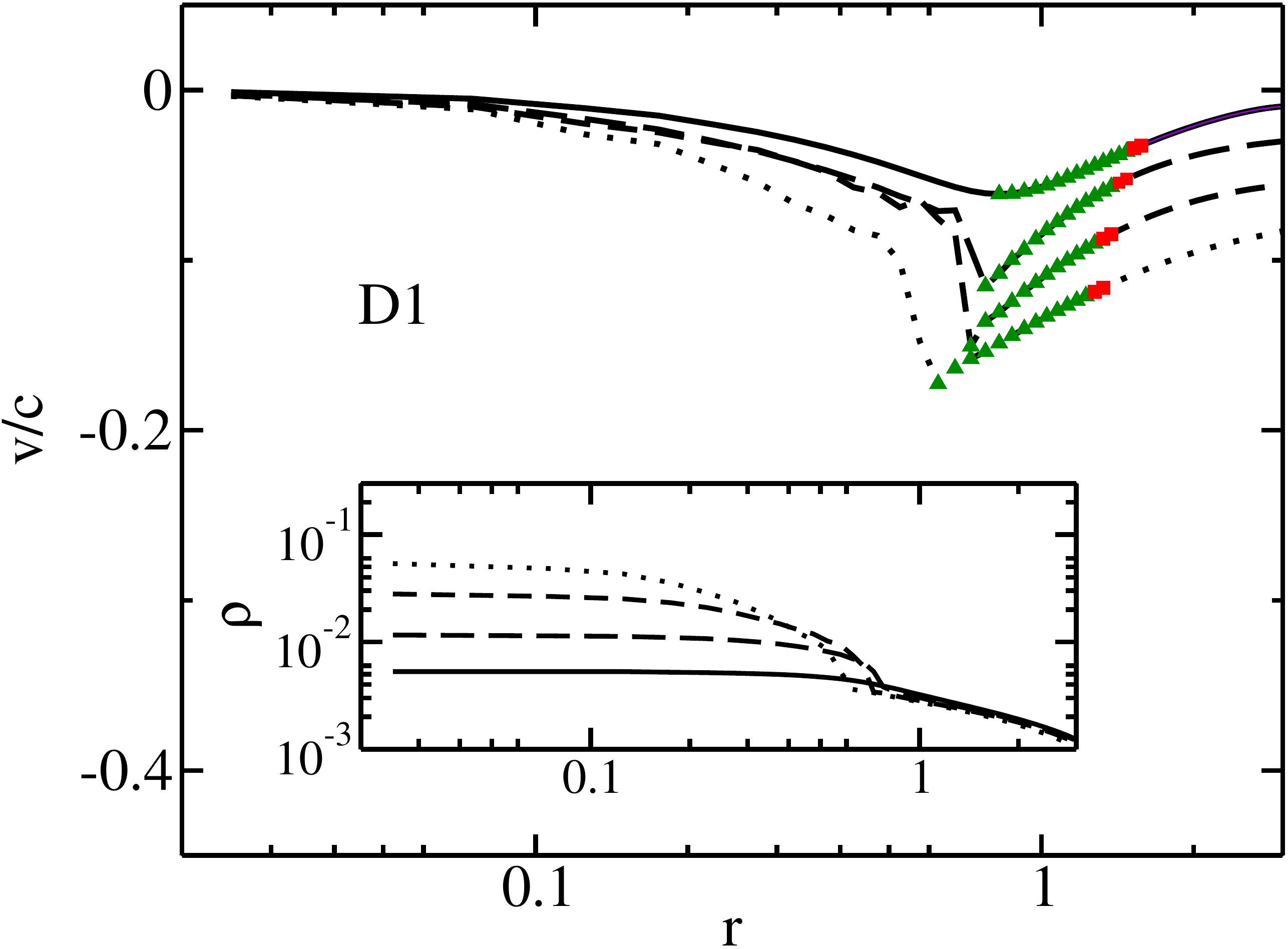}\\ 
\includegraphics[width=0.47\textwidth]{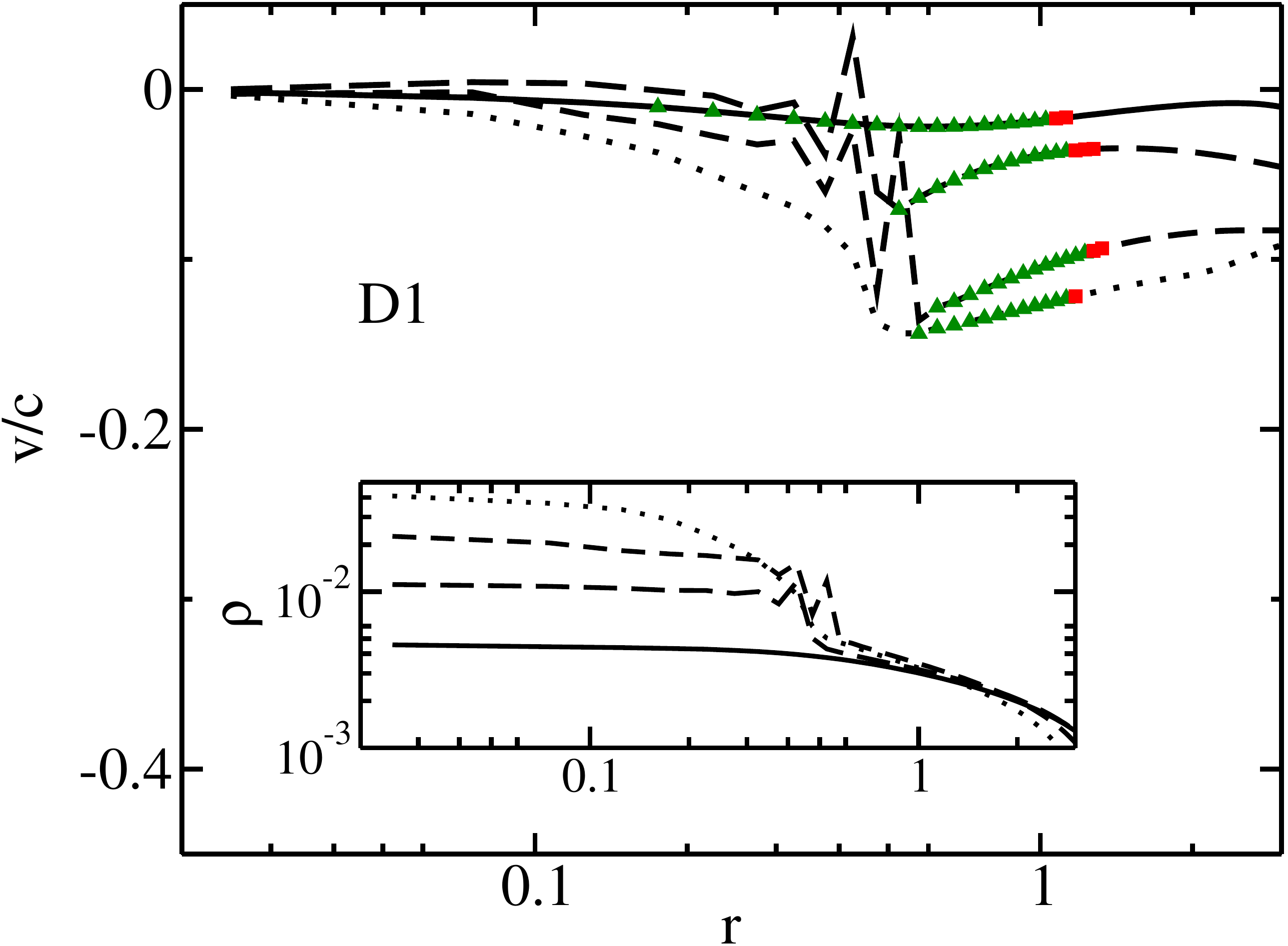} 
\includegraphics[width=0.47\textwidth]{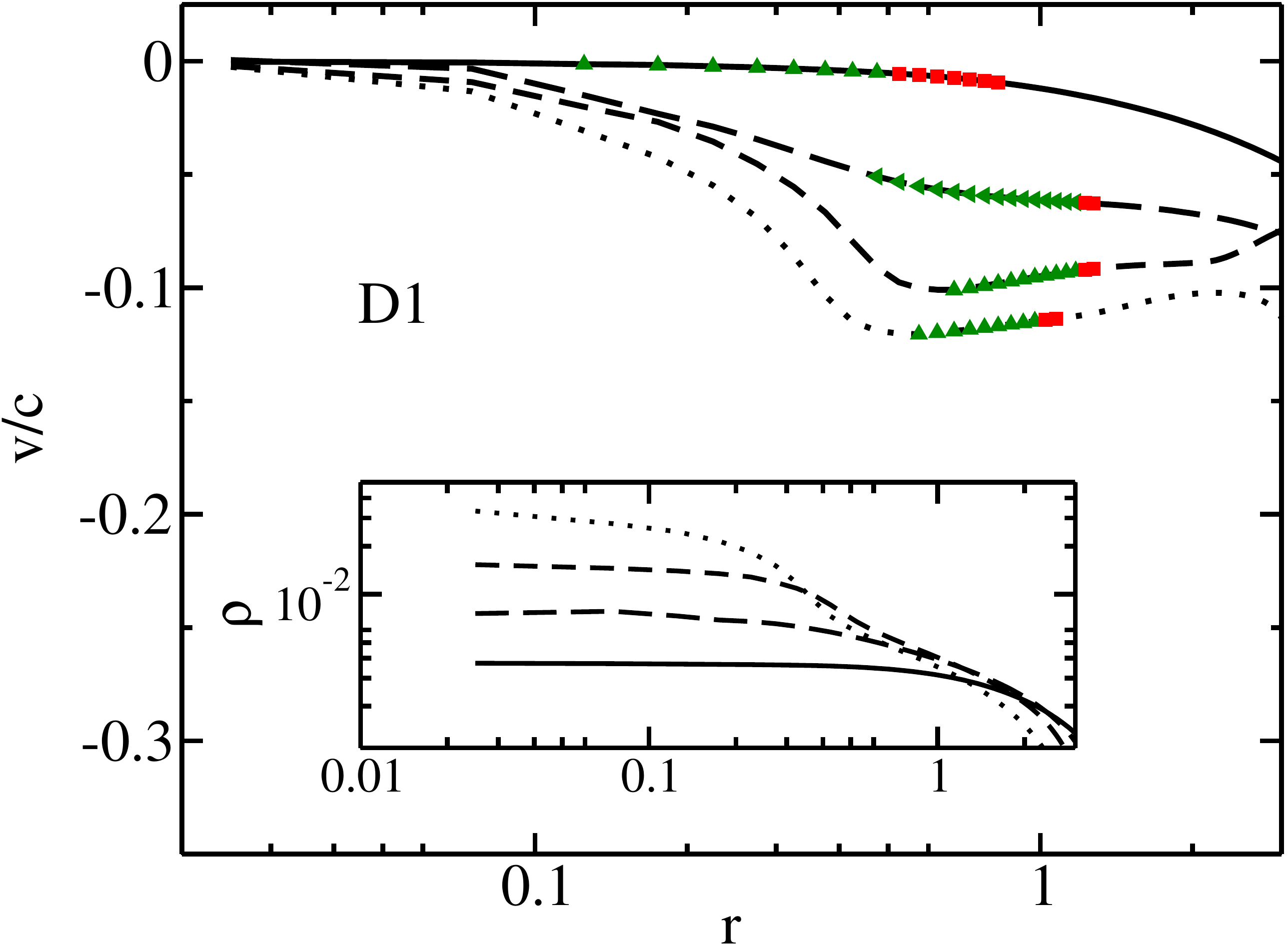} 
\caption{Radial profile of the velocity at the equatorial plane for
  model D1 at different times of the evolution and for different
  values of the $\rho_1$ parameter of the GGL-EoS. Within each panel,
  the time increases according to the following line style ordering:
  solid, long-dashed, dashed and dotted. {\it Top-left panel}:
  $\rho_{1} = 1.5\times 10^{-3}$, {\it Top-right panel}: $\rho_{1} =
  1.7\times 10^{-3}$, {\it Bottom-left panel}: $\rho_{1}= 2.1\times
  10^{-3}$, {\it Bottom-right panel}: $\rho_{1} = 2.5\times
  10^{-3}$. The insets show the corresponding radial profile of the
  rest-mass density at the same evolution times as the
  velocity. Green triangles locate the computational cells
    where the $\Gclas<0$ and $\Grel<0$, while red squares are drawn
    for cells where $\Gclas>0$ and $\Grel<0$.}
\label{fg:NC2}
\end{center}
\end{figure*}

\begin{figure*}
\begin{center}
\includegraphics[width=0.47\textwidth]{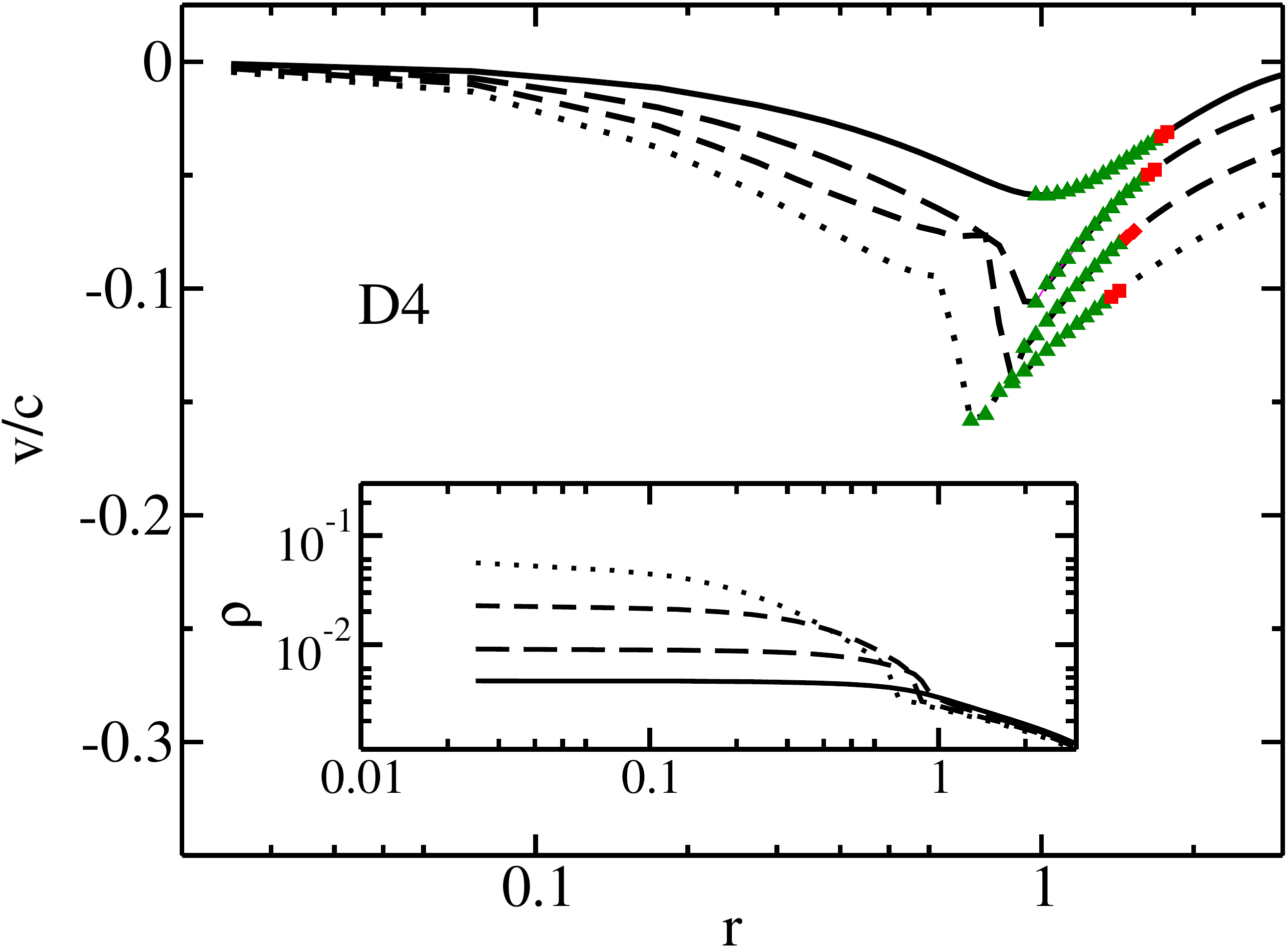} 
\includegraphics[width=0.47\textwidth]{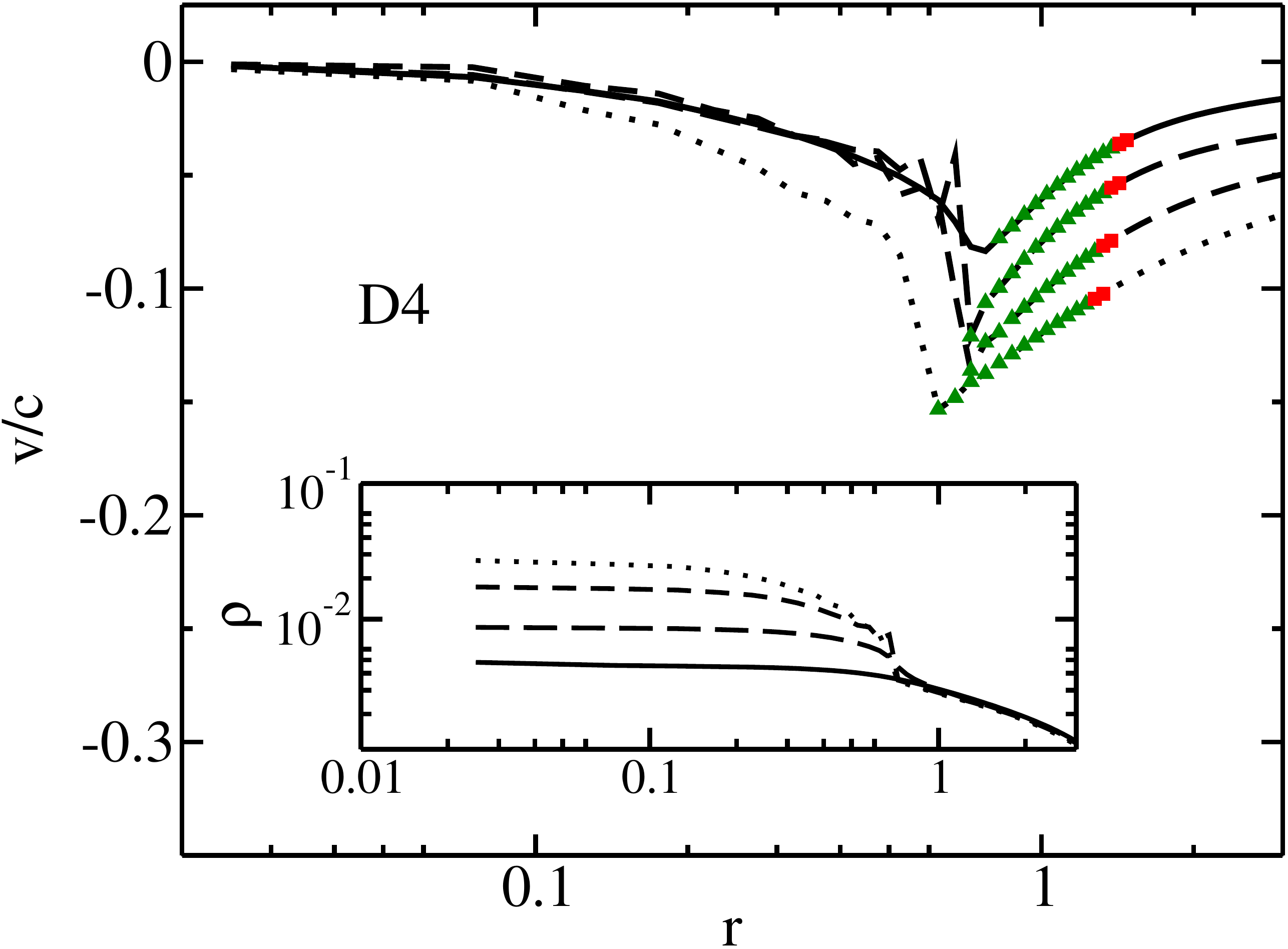}\\ 
\includegraphics[width=0.47\textwidth]{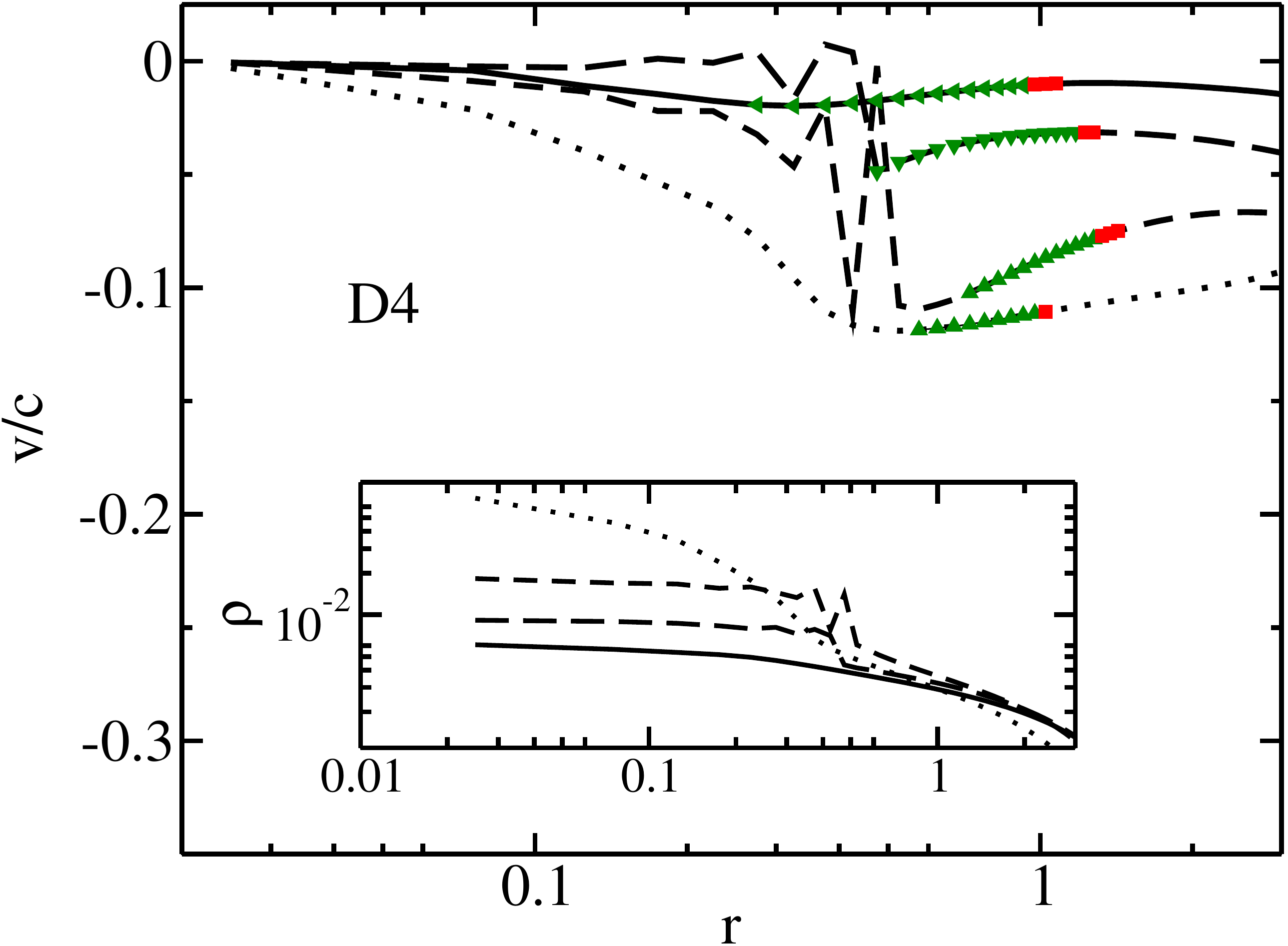} 
\includegraphics[width=0.47\textwidth]{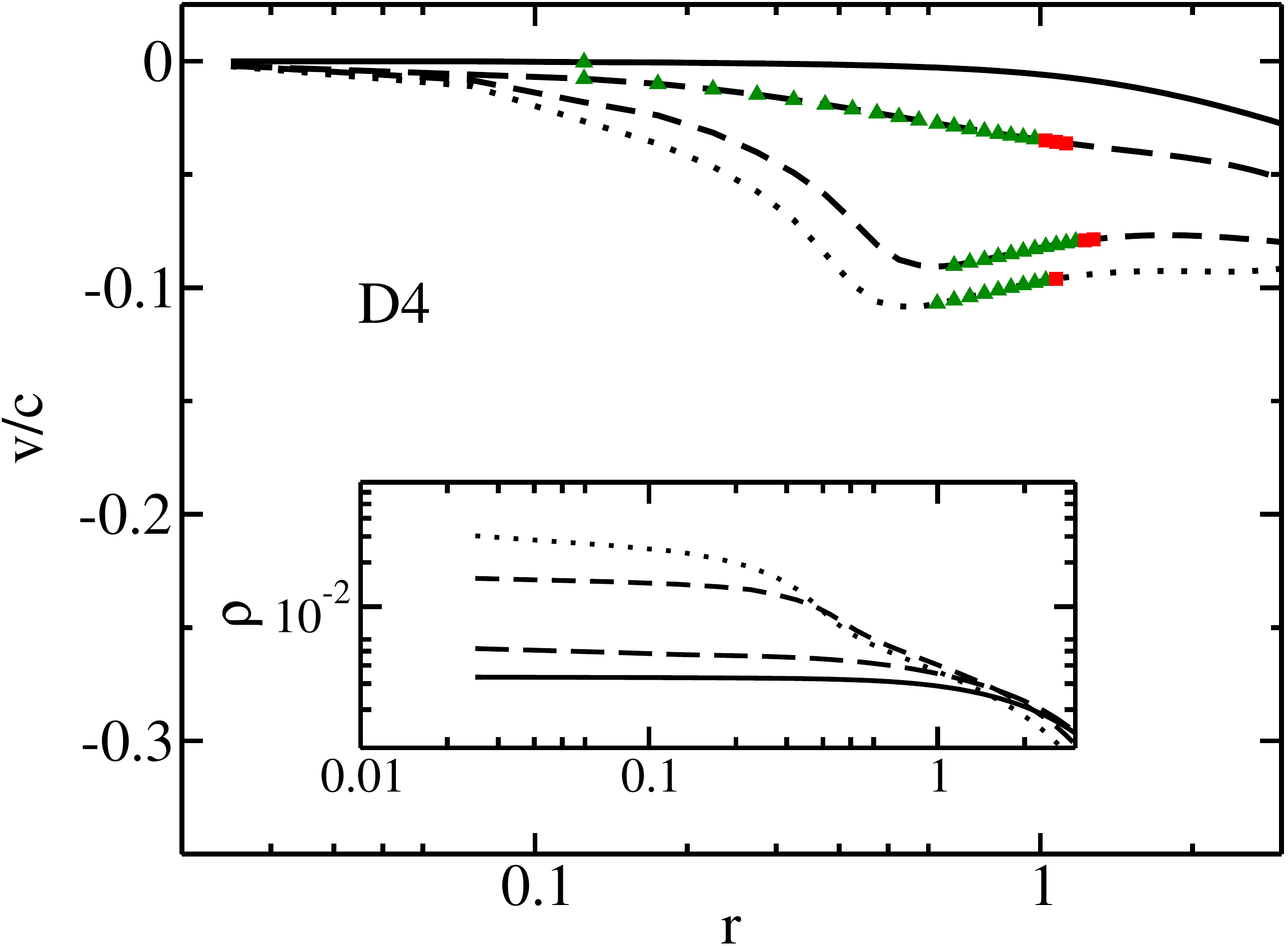} 
\caption{Same as Fig.~\ref{fg:NC2} but for model D4.}
\label{fg:D4}
\end{center}
\end{figure*}

\begin{figure}
\begin{center}
\includegraphics[width=0.45\textwidth]{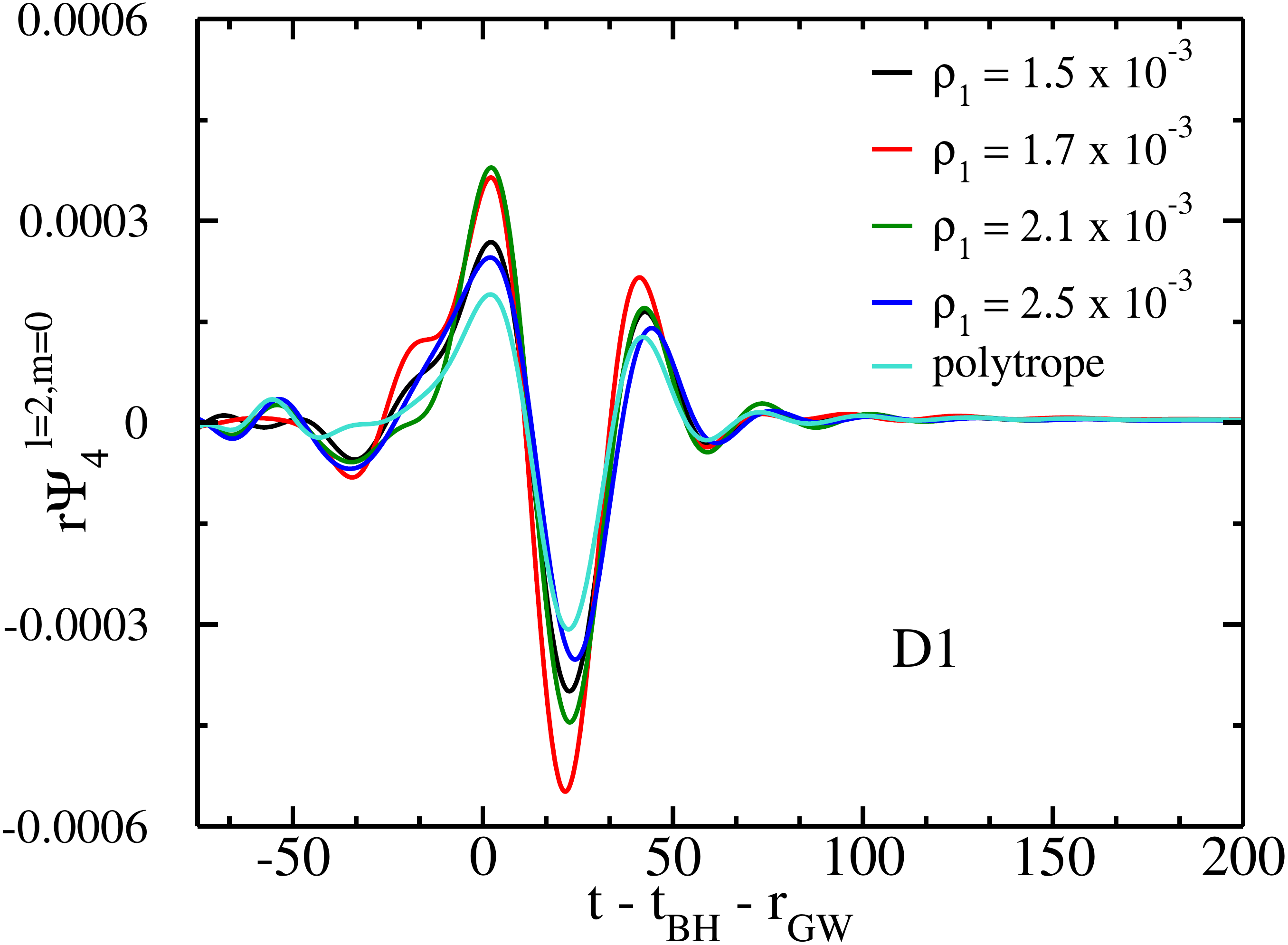} \\
\includegraphics[width=0.45\textwidth]{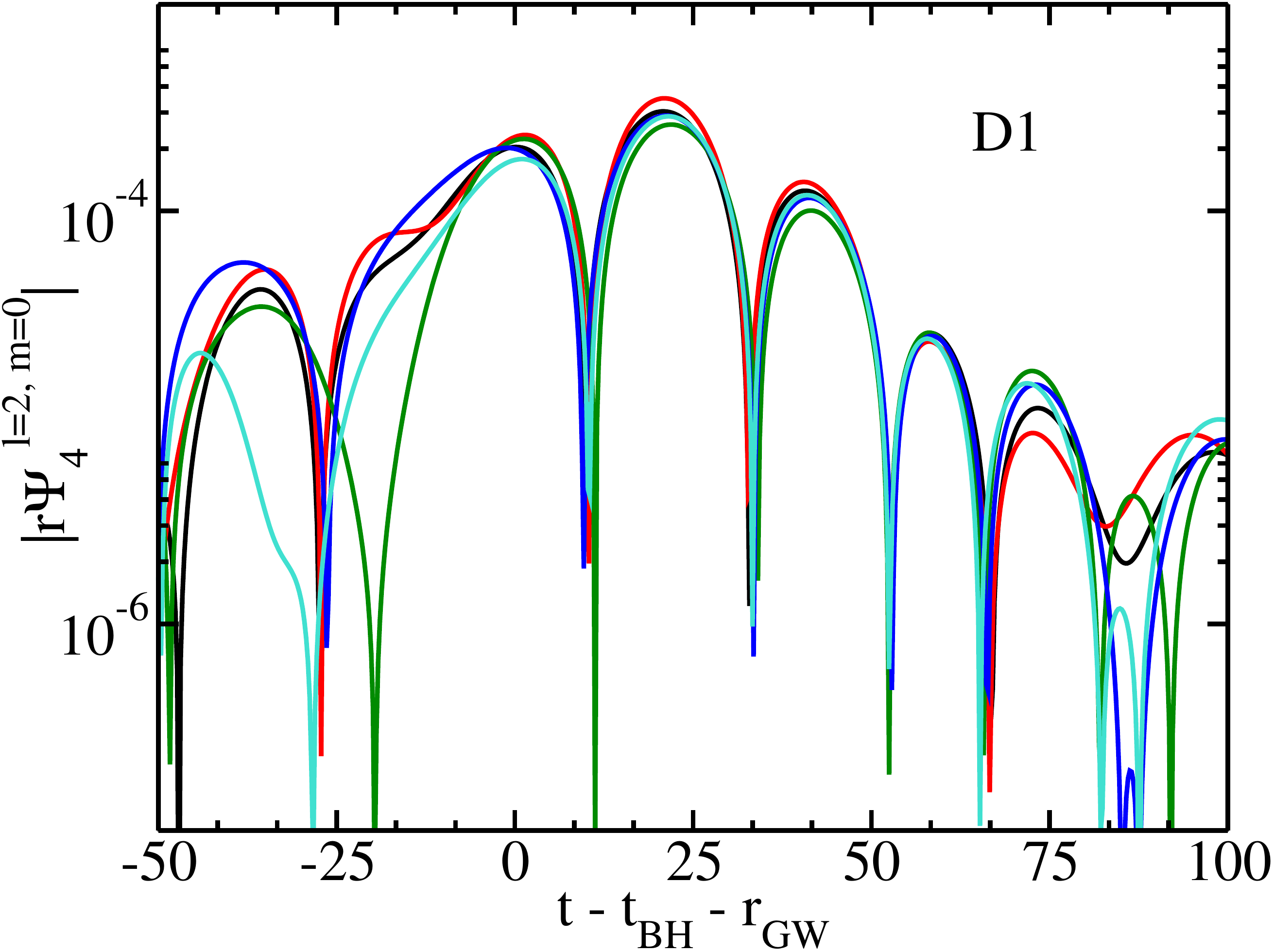}\\ 
\includegraphics[width=0.45\textwidth]{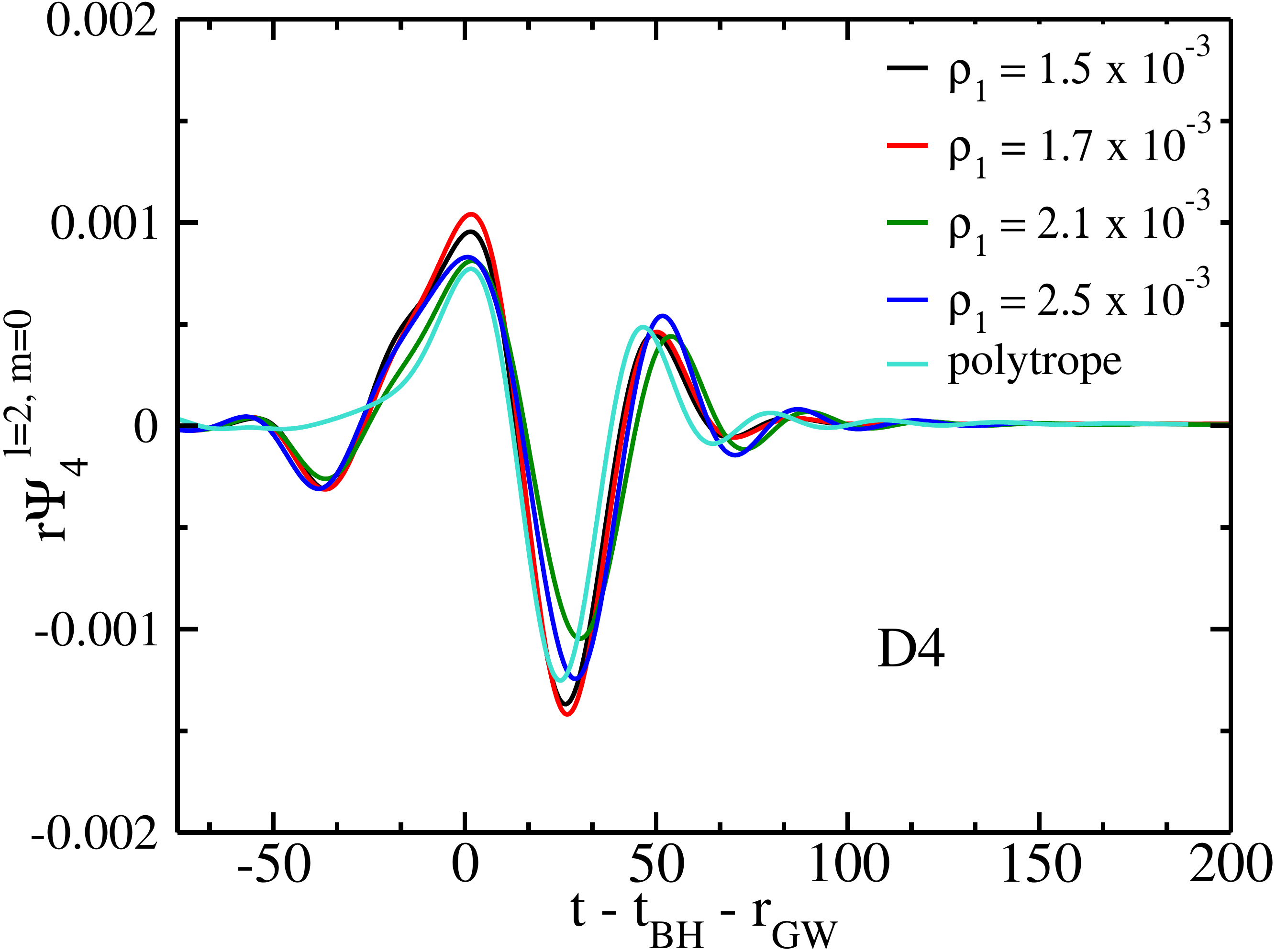} 
\caption{Real part of the ($l=2,\,m=0$) $\Psi_{4}$ mode extracted at
  $r_{\rm GW}=200$ for two different RNS models. {\it Top panel}:
  model D1. {\it Middle panel}: log scale of the rescaled
  gravitational waves for the different values of $\rho_{1}$. {\it
    Bottom panel}: model D4. The time axis is given in code units. To
  convert it to CGS, the reader must multiply the values by $\simeq
  4.926\times 10^{-6}$. For a better comparison, each model time is
  shifted by its own time of collapse ($t_{\rm BH}$), defined as
  the instant when an apparent horizon forms. }
\label{fg:NC3}
\end{center}
\end{figure}

The gravitational radiation produced during the collapse of the
neutron stars is computed using the Newman-Penrose
formalism~\citep{Newman1962}. More precisely, we compute the so-called
Newman-Penrose scalar $\Psi_{4}$, defined by
$ \Psi_4 \equiv C_{\alpha \beta \gamma \delta} \,n^{\alpha}
\bar{m}^{\beta} n^{\gamma} \bar{m}^{\delta}$, where
$C_{\alpha \beta \gamma \delta}$ is the conformal Weyl tensor
associated with the spacetime metric $g_{\alpha \beta}$ and $n$,
$\bar{m}$ are part of a null-tetrad. We use the definition of the
electric and magnetic parts of the Weyl tensor, $E_{ij}$ and $B_{ij}$,
as a function of the 3+1 variables evolved by the code, to rewrite the
Weyl $\Psi_{4}$ scalar as $\Psi_{4}=Q_{ij}\,\bar m^{i}\bar m^{j}$ with
$Q_{ij}\equiv E_{ij}-B_{ij}$. We then compute the $l=2$, $m=0$
multipole (which is the dominant mode since the collapse is
essentially axisymmetric) from
\begin{eqnarray}
  \Psi_4(t,~\theta,~\phi) &=& \sum_{\ell, m} \Psi^{\ell m}_4(t) {}_{-2}Y_{\ell m}(\theta, \phi),
      \\
  \Psi^{\ell m}_4(t) &=& \int \Psi_4(t,\theta,\phi) \bar{{}_{-2}Y_{\ell m}}(\theta,\phi)
      d\Omega\,.
\end{eqnarray}
where ${}_{-2}Y_{\ell m}$ are the $(s=-2)$ spin-weighted spherical
harmonics \citep[see, e.g.][]{Thorne_1980RvMP...52..299}.

In order to test the convergence and the gravitational-wave extraction properties of our code we first evolve the {\it stable} rotating neutron star model BU2 in Table~\ref{T3}~\citep{stergioulas2004non}. Following~\citet{font2002three}, we perturb the velocity of the initial model according to
\begin{eqnarray}
u_{\theta}(t=0)=0.02\,\sin\left(\frac{\pi r}{R_{\rm e}}\right)\,\sin\theta\,\cos\theta\, ,
\end{eqnarray}
where $R_{\rm e}$ is the circumferential equatorial radius.

The top panel of Fig.\,\ref{fg:NC0} shows the time evolution of the
L1-norm of the difference between the evolved rest-mass density and
its initial value computed for all the grid points inside the
star. The three different curves correspond to three different
resolutions, and have been conveniently rescaled to show second-order
convergence (see Fig.\,\ref{fg:NC0} caption), as expected. We extract
the gravitational wave emitted in the evolution of this perturbed
model and compute the frequencies of the fundamental quadrupole $(l =
2)$ mode, $^{2}f=1.65\pm0.20$\,kHz, and its first overtone,
$^{2}p_{1}=3.65\pm0.20$\,kHz.
  These values are in good agreement with the results obtained in
\cite{dimmelmeier2006non}.

\begin{figure*}
\begin{center}
\includegraphics[width=0.33\textwidth]{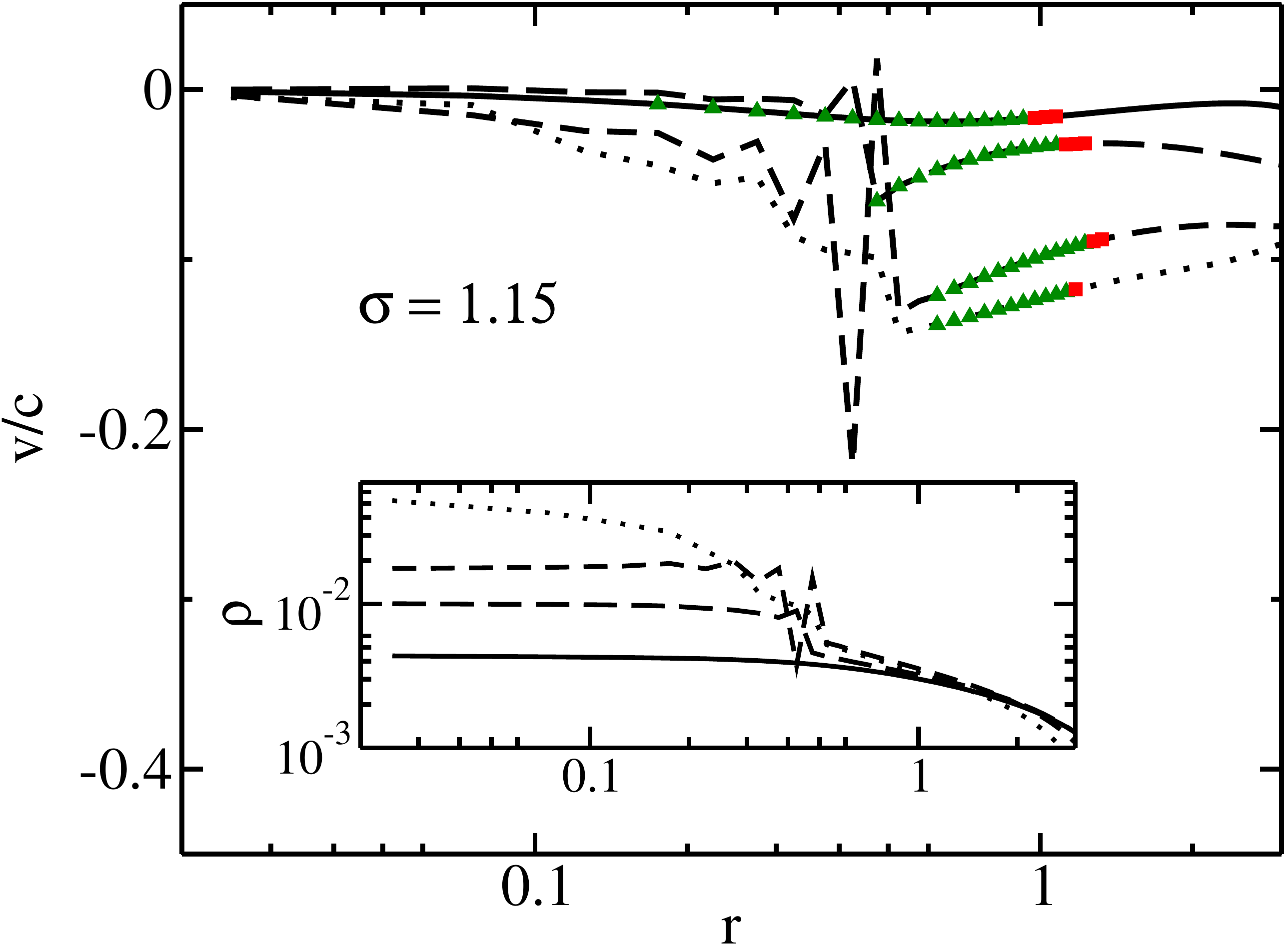} 
\includegraphics[width=0.33\textwidth]{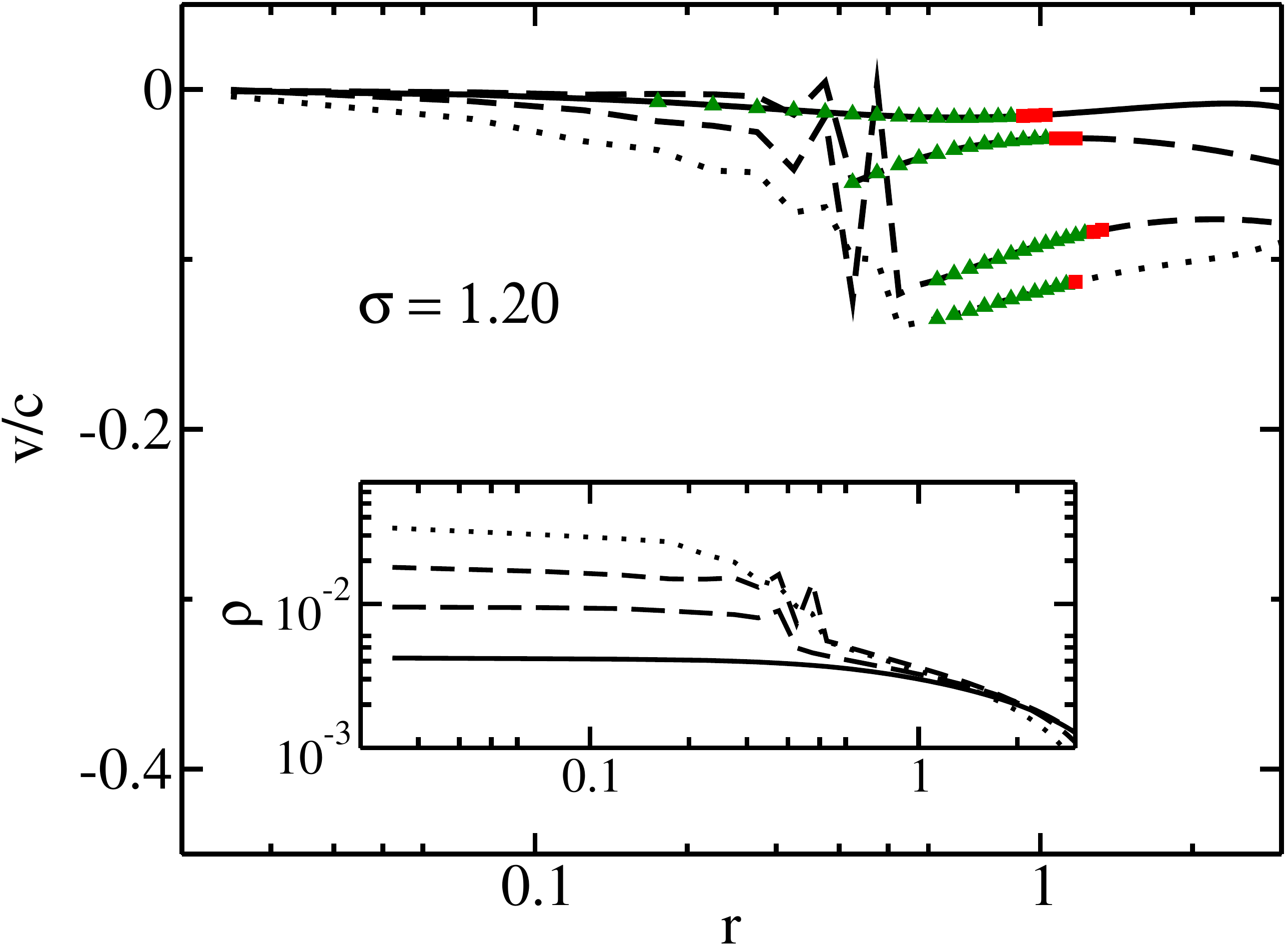} 
\includegraphics[width=0.33\textwidth]{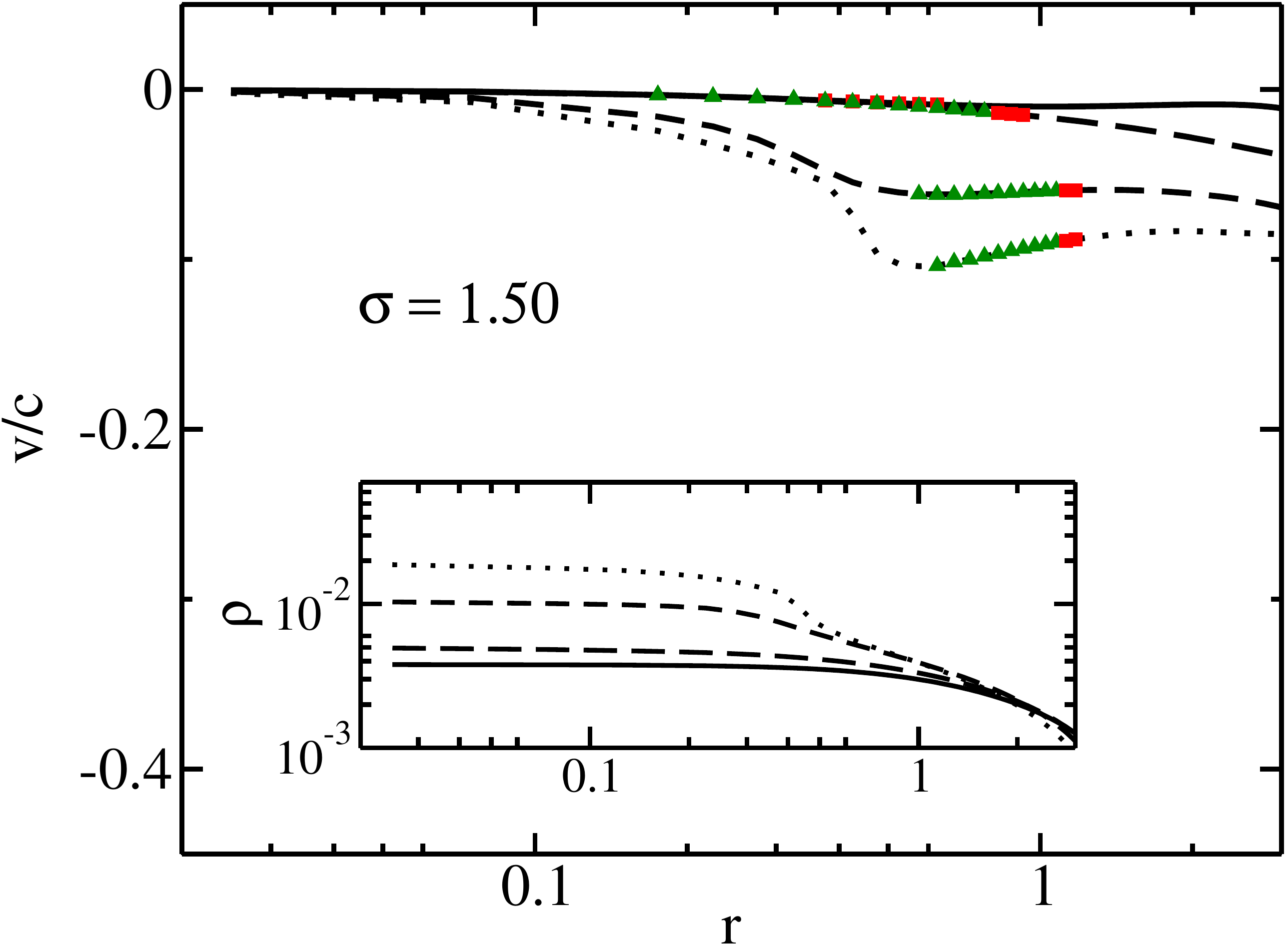}\\ 
\includegraphics[width=0.33\textwidth]{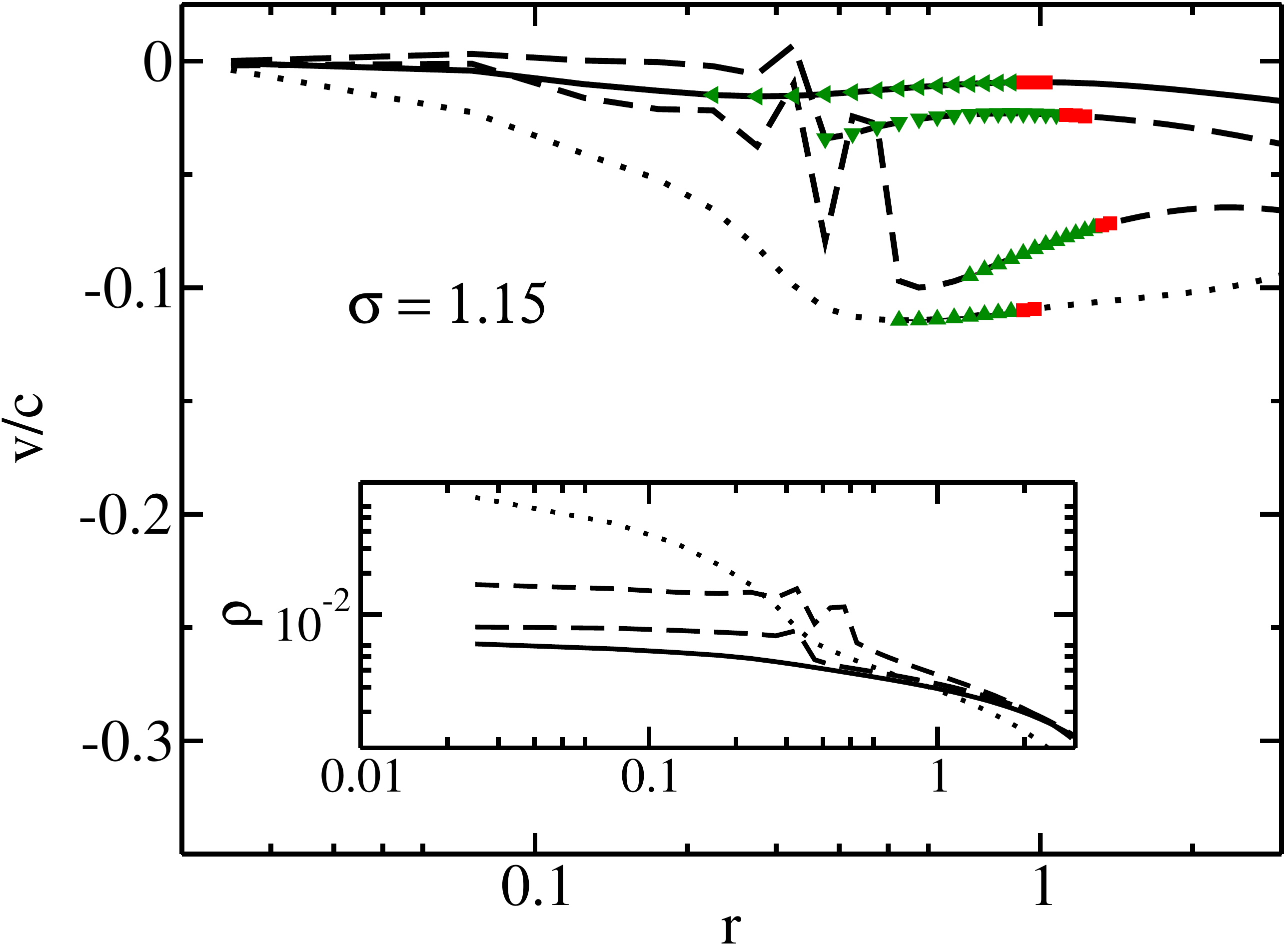} 
\includegraphics[width=0.33\textwidth]{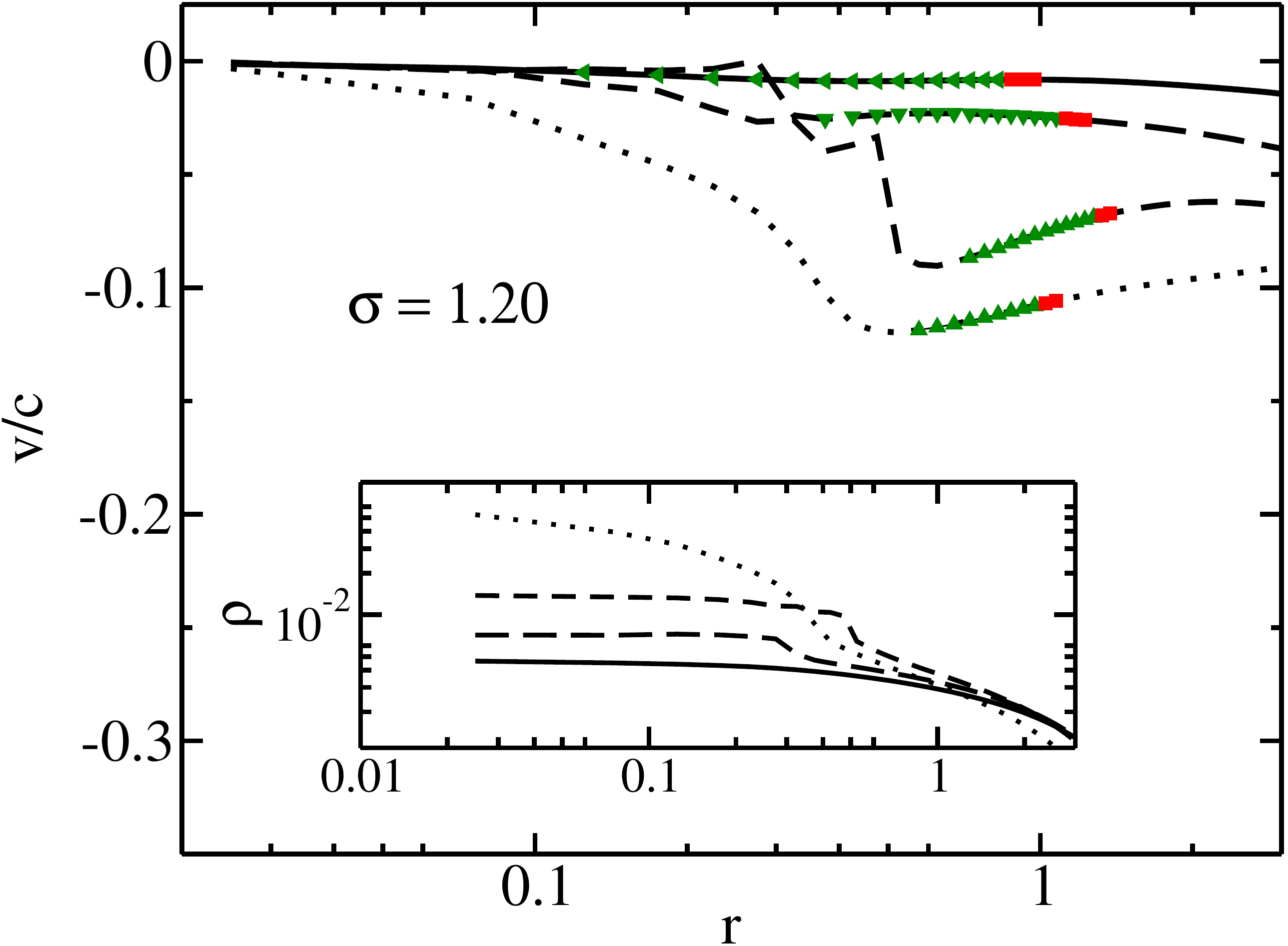} 
\includegraphics[width=0.33\textwidth]{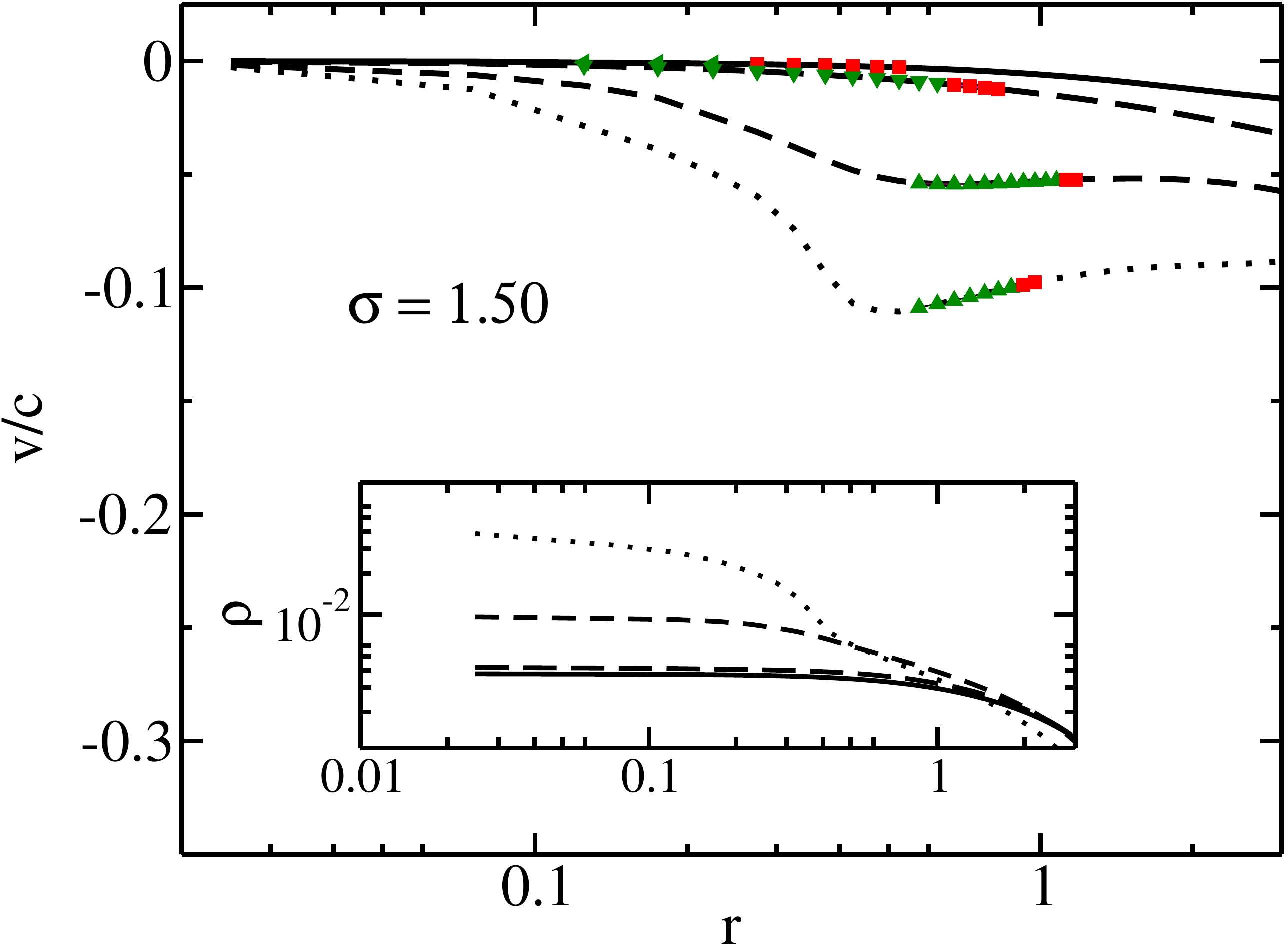} 
\caption{Detail of the radial velocity for different times of models ({\it top panels}) D1 and ({\it bottom panels}) D4 with $\rho_{1}=2.1\times 10^{-3}$: {\it Left panels}: $\sigma$ = 1.15, {\it Middle panels}: $\sigma$ = 1.20, {\it Right panels}: $\sigma$ = 1.50.}
\label{fg:NC4}
\end{center}
\end{figure*}

\begin{figure}
\begin{center}
\includegraphics[width=0.47\textwidth]{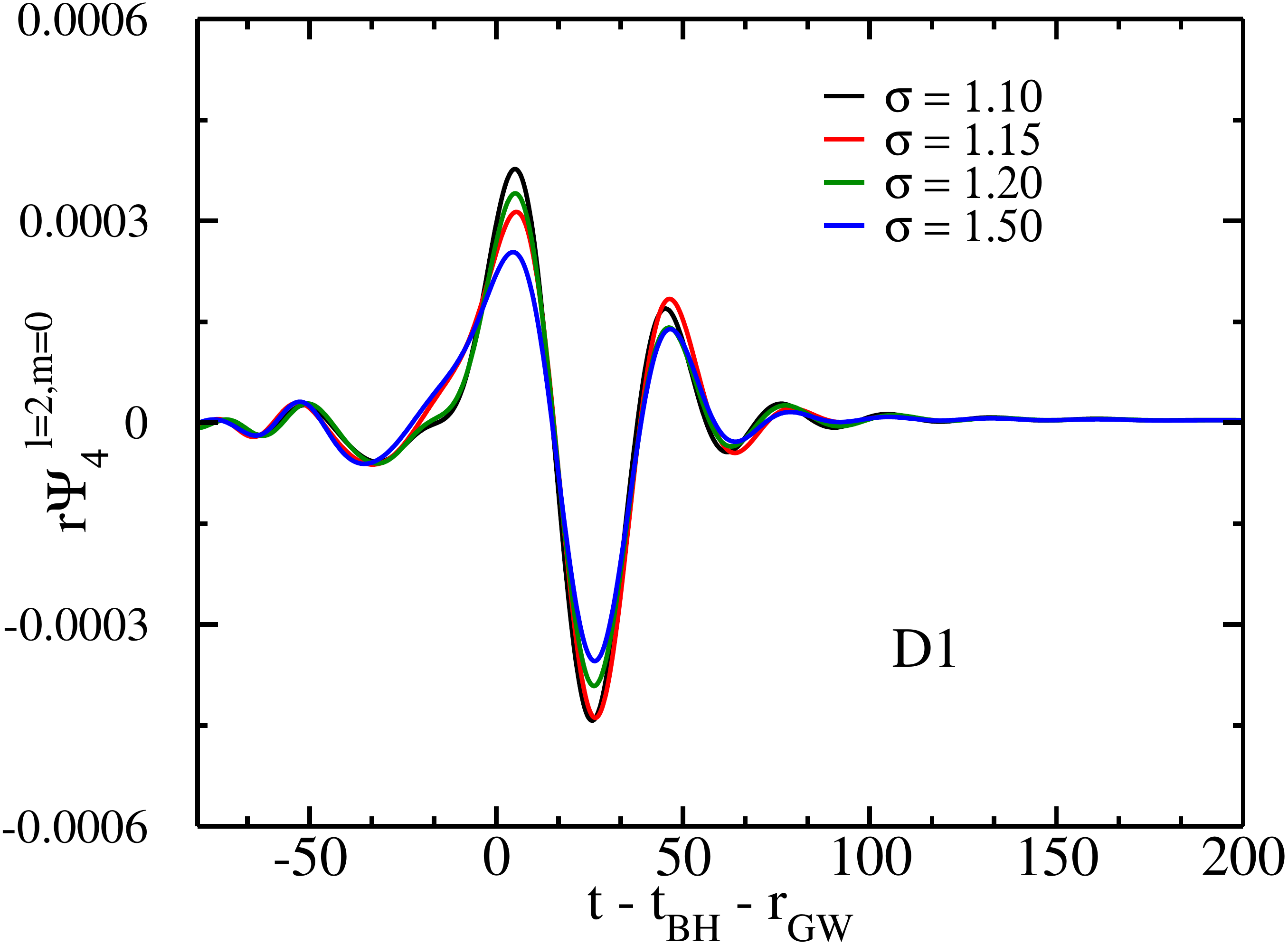} 
\includegraphics[width=0.47\textwidth]{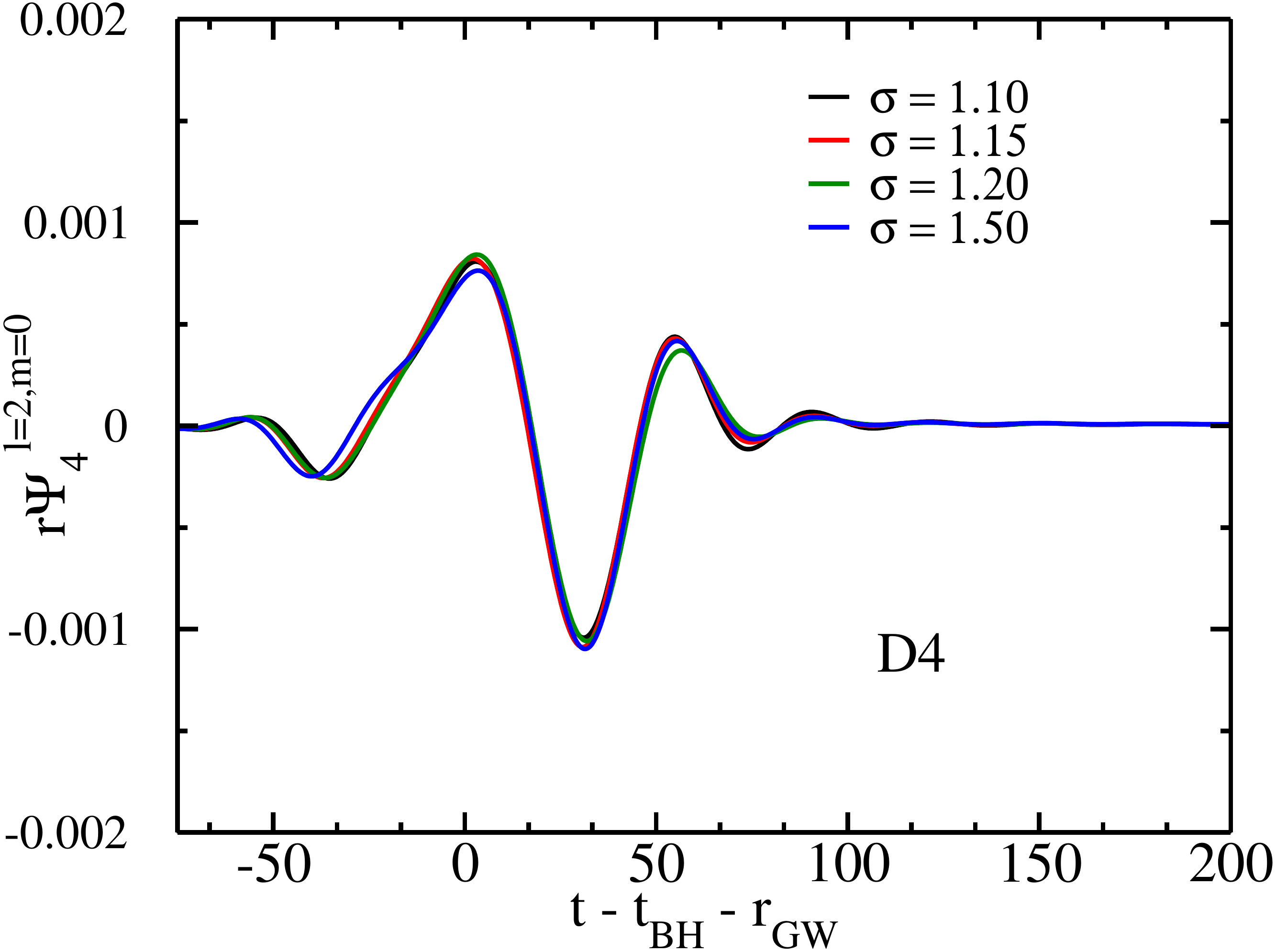} 
\caption{Real part of the ($l=2,\,m=0$) $\Psi_{4}$ mode for the model D1 (top) and D4 (bottom) with $\rho_{1}=2.1\times 10^{-3}$.}
\label{fg:NC5}
\end{center}
\end{figure}

We next consider the two rotating neutron star models described in
Table\,\ref{T3} using different values of the parameters of our
phenomenological GGL-EoS. The particular parameters are reported in
Table\,\ref{table2}.  Our simulations are performed in
equatorial-plane symmetry. They use a logarithmic radial grid that
extends from the origin to $r_{\text{max}} = 600$ and has the finest
resolution close to the origin, namely $\Delta r = 0.05$ ($\simeq
74\,$m). The angular grid is equally spaced and employs a resolution
of $\Delta \theta=\pi/32$.  These values for $\Delta r$ and $\Delta
\theta$ have been chosen after a suitable convergence test, but this
time comparing the radial distribution of the Hamiltonian constraint
at a time where a (bounce) shock has developed in our models (see
below). We chose this time since the largest violations of the
Hamiltonian constraint are expected to happen in the vicinity of
shocks.  In the bottom panel of Fig.\,\ref{fg:NC0}, we plot the radial
profile of the Hamiltonian constraint for model D1 with
$\rho_{1}=1.5\times10^{-3}$. The three different angular resolutions
are rescaled to highlight the second-order convergence, almost
everywhere, except in the region $0.1\lesssim r \lesssim 1$.  We note that the radial
resolution of our best resolved models in 2D is a bit better than that
of our 1D models of Sec.\,\ref{section:Collapse-1d}. However, the 1D
models have been computed with a higher spatial and temporal order of
accuracy, so that they effectively feature a better resolution. In
spite of these small differences, as we shall see (below) the dynamics
of two dimensional models with a rough counterpart in the previous
spherically symmetric cases is qualitatively the same.

We start by fixing the value of the Gaussian width to $\sigma=1.1$ and
study the effects of varying the parameter $\rho_1$. In the top panel
of Figure \ref{fg:NC1} we show the initial radial profile of
$\Gamma_1$ along the equator for model D1 and for the different values
of $\rho_{1}$ we are considering.  For later reference, we point out
that the set of models D1 with
$\rho_1=\{0.9356,1.06,1.31,1.559\}\times10^{15}\,$gr\,cm$^{-3}$
(Table\,\ref{table2}) can be regarded as 2D rotating counterparts of
models P-1.9G1, P-1.9G2, P-1.9G3 and P-1.9G4 of
Sec.\,\ref{section:Collapse-1d} (see Table~\ref{Tab:1d-models}). The
non-convex region of the EoS becomes -- in radius -- larger as $\rho_1$ becomes
smaller, as can be seen from the larger region of non-monotonicity of
$\Gamma_1$ in the top panel of Fig.\,\ref{fg:NC1}. The time evolution
of the central density of model D1 for the four different values of
$\rho_{1}$ is shown in the bottom panel. Note that the radius and the
time is given in these two panels in code units. The time evolution
shows that the smaller the value of $\rho_{1}$, the faster the
collapse takes place.  This happens because $\Gamma_1$ (and also
$\gamma$) is significantly smaller near the central regions of the
star as $\rho_1$ is reduced (cf.~top panel of Fig.\,\ref{fg:NC1}) and,
therefore, the pressure becomes smaller. The time of the formation of
the apparent horizon of the BH is indicated in the figure by the
vertical dashed lines.  We note that BH formation time for models of
the series D1 and different values of $\rho_1$ are about a
  factor two longer than the values found for models P-1.9G1 to
P-1.9G4. The BH formation times in the D1 series range from
$t_{\rm BH}\simeq 27$ to $t_{\rm BH}\simeq 45$ code units, or
  equivalently, $t_{\rm BH}\simeq 0.13\,$ms to $t_{\rm BH}\simeq
  0.22\,$ms. We attribute the small differences to the rotation
  present in the 2D models rather than to the approximate treatment of
  the general relativistic gravitational potential in the
  \textsc{Aenus} code.

The final outcome is in
all cases a rotating Kerr BH whose spin parameter is plotted in
Fig.\,\ref{fg:NC1-spin}. This figure shows that for all the unstable
models, the final value of the BH spin is fairly independent of
$\rho_{1}$. The spin is computed using the expression
\begin{eqnarray}
\frac{a}{M_{\rm hor}}=\sqrt{1-(-1.55+2.55\,C_{r})^{2}}\,,
\label{BHspin}
\end{eqnarray}
where $C_{r}$ is the ratio of polar-to-equatorial proper circumference and $M_{\rm hor}$ is the mass of the horizon, which coincides with $M$ when the spacetime has become axisymmetric and stationary. This expression has an accuracy of $\sim2.5\%$~\citep{brandt1995evolution,baiotti2005three}. The values for the spin and for the irreducible mass with our GGL-EoS differ with those obtained employing a polytropic EoS~\citep{baiotti2005three} by less than 1\%.

In Figs.~\ref{fg:NC2} and \ref{fg:D4} we plot the radial profiles of
the velocity of the fluid and of the rest-mass density (shown in the
insets) for models D1 and D4, respectively. The profiles are plotted
at the equatorial plane ($\theta=\pi/2$). The different curves
indicate different times during the evolution. The four panels in each
of the two figures correspond to the four values of $\rho_1$, as
indicated in the caption of Fig.~\ref{fg:NC2}. We note that for convex
EoS, as a polytrope or a gamma-law, the dynamics of the collapse
proceeds smoothly towards the formation of a BH, as discussed
in~\citet{Font2002a,Baiotti2005} and we have show in
  Sec.\,\ref{section:Collapse-1d}. The larger the centrifugal support
of the initial model, the more it takes for the model to collapse. As
shown in~\cite{Baiotti2005}, the collapse of the rapidly-rotating
model D4 goes through a short-lived centrifugal hang-up when the
stellar surface slows its inward motion and stalls, although
ultimately it shrinks to a volume smaller than that of the
radially-increasing event horizon that forms at the central
regions. During the evolution of these models, a shock develops at the
edge between the homologous inner core of the star and the outer core,
which falls supersonically. Consistent with the dynamics
  observed in the 1D models of Sec.\,\ref{section:Collapse-1d}, this
shock is eventually engulfed by the growing BH that forms as a result
of the collapse. For the nearly-spherical D1 model this process is
much faster than for the rapidly-rotating model D4. We have also
performed the evolutions using an ideal gas EoS, in order to
qualitatively compare our findings on the dynamics and on the
gravitational-wave emission with the results from these previous
works.

It is important to highlight that the formation of the former shock is
entirely due to the non-convex dynamics. In the case with
$\rho_{1}=2.5\times 10^{-3}$ (bottom-right panels of
Figs.\,\ref{fg:NC2} and \ref{fg:D4}), there is no such shock because
for that value of $\rho_1$ the sound speed in the non-convex region is
(much) larger than that of other models with smaller values of
$\rho_1$. This fact prevents reaching a supersonic regime in the
convex region and avoids the formation of the shock.  We also point
out that for the case with $\rho_{1}=2.1\times 10^{-3}$ (bottom-left
panels of Figs.\,\ref{fg:NC2} and \ref{fg:D4}), and contrary to the
two cases displayed in the top panels of both figures, the shock
propagates outwards. Furthermore, the flow speed ahead of the shock
location is slightly positive. This is due to the {\em borderline}
behaviour of this model, which develops a tiny supersonic region right
to the inner radial boundary where the classical fundamental
derivative is negative (green triangles in Figs.\,\ref{fg:NC2} and
\ref{fg:D4}). This supersonic region persists for a relatively short
time an along its inner boundary is where the shock forms. We note
that the behaviour described for the models D1 with
$\rho_1=2.1\times 10^{-3}$ and $\rho_1=2.5\times 10^{-3}$ bears
qualitative similarities with the 1D models P-1-9G3 and P-1.9G4,
respectively. In model P-1.9G3, we also observe a tiny radial outwards
displacement of the shock and the shock formation is significantly
delayed with respect to other models of the same series in the case of
model P-1.9G4. Thus, we conclude that there is a gross qualitative
agreement between the 2D models D1 and their non-rotating counterparts
in 1D. The small quantitative differences are almost exclusively
induced by the rotation of the former models.

Figure~\ref{fg:NC3} displays the gravitational-wave signals $\Psi_{4}^{20}$ for models D1 and D4 and for all values of $\rho_{1}$. For the sake of comparison, the three panels of this figure also include additional curves which correspond to a polytropic (convex) EoS. The waveforms are extracted at a radius $r_{\rm GW}=200$. For convex EoS, gravitational waveforms of the collapse of these two models have been reported before by~\cite{Giacomazzo2012}. The signal is of the burst-type, i.e.~it is characterized by an exponential increase of the amplitude and by a short-duration burst at the moment of BH formation (which coincides with the largest positive peak, see~\cite{dietrich2015simulations}\footnote{Notice that~\cite{Giacomazzo2012} associate the first negative peak to the moment of BH formation due to a global sign difference in the expression of $\Psi_{4}$ they use compared to ours and~\cite{dietrich2015simulations}.}) followed by the subsequent quasinormal mode ringdown of the BH. Our comparison with the results of~\cite{Giacomazzo2012} for convex EoS shows good agreement in the waveform morphology and amplitude, particularly for model D1 (for model D4 we obtain a few times larger amplitude; note the difference in the vertical scales between the upper and lower panels of Fig.~\ref{fg:NC3}). 

The non-convex dynamics leaves an imprint in the gravitational
waveforms produced during the process. The smallest amplitudes at the
moment of BH formation are obtained for the polytropic EoS, specially
in the case of model D1 (top panel of Fig.~\ref{fg:NC3}). For the
GGL-EoS, the frequency of the various signals is quite close to each
other, while their amplitudes are different depending on the value of
$\rho_{1}$. This is more apparent for model D1 than for model D4. In
the case of model D1 in particular, the largest gravitational-wave
amplitude is obtained for $\rho_1=1.7\times 10^{-3}$ (red curve in the
top panel of Fig.~\ref{fg:NC3}). The maximum amplitude is about twice
that attained in the polytropic case. For model D4 the maximum
amplitude is also achieved for the same value of $\rho_1$ but the
differences among the various simulated models are not as apparent as
for model D1. This means that the faster the rotation of the initial
neutron star, the smaller the imprint the loss of convexity leaves on
the gravitational-wave signal after the BH has been formed.

The radially outwards propagation of the shock in model D1 with
$\rho_1=2.1\times 10^{-3}$ (Fig.~\ref{fg:NC2}) translates in slightly
higher gravitational-wave amplitudes at the time of collapse, but
slightly smaller in the instants preceding the BH formation. The speed
of this outgoing shock is smaller than the speed at which the BH
horizon grows and eventually all neutron star matter will be inside of
the BH. This can be inferred from the middle plot of Fig.~\ref{fg:NC3}
which displays the (absolute value of) the gravitational waveforms of
model D1 in logarithmic scale. We note that the curves in this figure
have been conveniently shifted in time in order to synchronise the
time of BH formation. Later, after the amplitude reaches a minimum,
all of the infalling matter has been captured by the BH, whose area
stops growing. Irrespective of the thermodynamical details of the
collapse, encoded in our phenomenological EoS by the different values
of the $\rho_1$ parameter, the final Kerr BH must have the same mass
and angular momentum, as implied by the fact that all four
gravitational-wave signals have the same frequency and exponential
decay, associated with the distinctive quasinormal mode ringdown
signal of a BH. We observe, however, that the largest discrepancies
among different models happen in the pre-collapse phase
($t-t_{\rm BH}-r_{\rm GW}<0$ in Fig.~\ref{fg:NC3}). There, we see that
the fingerprint of convexity loss in the course of the collapse is an
increasing spectral power and amplitude in the pre-collapse phase of
model D1 compared to a polytropic model (which would be representative
of a collapse developed with a fully convex EoS; cyan line in the top
and central panels of Fig.~\ref{fg:NC3}).

The waveforms shown in the bottom panel of Fig.~\ref{fg:NC3} correspond to model D4. In this case, the maximum amplitudes of the burst signals are significantly larger than in the case of model D1, and for all values of $\rho_1$, the reason being the increased deviation from spherical symmetry of this rapidly-rotating model. While model D4 displays more similar gravitational waveforms for all values of $\rho_{1}$ than model D1, when comparing with the polytropic EoS there is still a visible change in frequency associated with the non-convexity properties of the GGL-EoS. This is particularly evident in the first part of the signal associated with the collapsing phase before the BHs form.

Additionally, we also study the effects of varying the width $\sigma$
of the Gaussian used in the definition of the GGL-EoS, fixing
$\rho_{1}=2.1\times 10^{-3}$. We analyze the dynamics of the collapse
for four different values of $\sigma$, namely
$\lbrace1.10,1.15,1.20,1.50\rbrace$. The results for models D1 and D4
are displayed in Fig.~\ref{fg:NC4}, which depicts the radial profile
of the fluid velocity at the equatorial plane. 
As we have shown before, for this value of $\rho_1$ the shock located
in the region $0.2<r<0.9$ attains a slightly positive speed if
$\sigma\lesssim 1.20$. The jumps at the latter shock become gradually
smaller when $\sigma$ increases from 1.10 (see Fig.~\ref{fg:NC2}) to
1.20. For $\sigma=1.50$ the shock is no longer visible and the
dynamics resembles that of a convex EoS. This trend is the same for
both models, i.e.~it does not depend on the initial rotation of the
unstable neutron star.

The corresponding gravitational waveforms are shown in
Fig.~\ref{fg:NC5}. The waveforms look remarkably similar irrespective
of the value of $\sigma$, with minor differences in the peak
amplitudes among all models. As in the cases previously analyzed, the
waveforms of the most rapidly rotating models D4 are less sensitive to
the changes in $\sigma$ than in models D1.

\section{Summary and outlook}
\label{section:summary}

A number of microphysical EoSs of high-density matter contain regions
in which the thermodynamics may be non-convex. These EoSs, commonly
used in a tabular form, may develop non-convex thermodynamics either
as a result of first-order PTs (regardless of whether they are
congruential or non-congruential), or non-consistent treatment of the
matter constituents (non-relativistic instead of relativistic), or
specific parameter sets in the RMF theoretical framework. In the first
group we find EoSs where transitions from nuclear hadronic matter into
quark-gluon plasma or into matter phases containing exotic particles
(e.g. hyperons) are included employing suitable Gibbs
constructions. The second group gathers EoSs in which baryons are
treated as non-relativistic particles. A prototype example of the
latter group is the LS220 EoS. To the third group belong EoSs which
include the NL3 parameter set in the RMF treatment. However, other
parameterizations of the RMF (e.g. FSU2) are convex in the classical
sense (i.e. $\Gclas>0$) even at high number densities.\footnote{The
  negative values observed for $\Gclas$ in the \texttt{GSHen(FSU1)}
  case at high number densities are likely numerical artifacts (see
  App.\,\ref{sec:numerical_artifacts}).}
The NL3 RMF parameterization yields a clean and genuinely relativistic
situation, namely, the large magnitude of the sound speed drives
negative values only of the relativistic fundamental derivative, but
the classical fundamental derivative remains positive (i.e.  $\Grel<0$
and $\Gclas>0$) for sufficiently large number densities
($n\gtrsim 0.8\,\text{fm}^{-3}$. This is clearly observed in the
\texttt{HS(NL3)} EoS at both $s=0.5$ and $s=2.5$, as well as in the
\texttt{GSHen(NL3)} at $s=0.5$ (at higher entropies per baryon we do
not have available thermodynamical data to confirm this point, but
clearly both equations should behave qualitatively in the same way at
sufficiently large number densities and entropies). In light of the
latest developments for the constituents of the merger GW170817
\citep{Abbott3,Abbot_2018PhRvL.121p1101}, we point out that the NL3
parameterization may to be too stiff, giving in particular too high
values for the tidal deformability, the neutron star radii and the
slope of the symmetry energy if one assumes low spin priors for the
merging objects \citep[see, e.g.][]{Malik_2018PhRvC..98c5804}.
In any instance, a good number of the studied microphysical EoSs
display a sensitive reduction of the relativistic fundamental
derivative as the baryon number density grows above
$n\gtrsim 1$\,fm$^{-3}$. In that regime
($n\gtrsim 1\,\text{fm}^{-3}; \,\, \Grel\gtrsim 0$), even small scale
oscillations of numerical origin, namely due to the discretization of
high-order derivatives across coexistence boundaries in PTs, may be
enough to \emph{drive} ($\Grel\lesssim 0$). Since small scale
oscillations in the evaluation of high-order derivatives are
\emph{hardly avoidable} in tabular representations of dense-matter EoS
(broadly used in computational astrophysics), and since the EoS at
number densities above $1\,\text{fm}^{-3}$ is poorly constrained, we
warn that physical, but most likely \emph{numerical}, non-convex
thermodynamics may develop in that regime. We note, however, that the
convexity across first-order PTs may be numerically recovered. Some
times (but not in all cases), the singularities exhibited by the Gibbs
(or Helmholtz) free energy are removable
singularities. Thermodynamical consistency requires that the Gibbs
free energy be a jointly concave function. This requirement may be
enforced convolving the Gibbs free energy with a non-negative
smoothing function, which mollifies the singularities at phase
transitions \citep[c.f.][]{Menikoff1989}. The physical and
mathematical conditions required to undertake such convexity recovery
are beyond the scope of this paper, but may be the subject of a future
work.

Unfortunately, most available microphysical EoSs are only tabulated up
to baryon number densities $n\lesssim 3$\,fm$^{-3}$, making it
difficult to assess whether convexity will be lost at high enough
baryon number density. Hopefully the present work will spark an
interest in this question, by pointing the phenomenological
consequences that such a non-convex regime would have. Adding to these
arguments, we point out the non-monotonic behaviour of the sound speed
in dense matter found by \cite{Bedaque2015}, which is a strong hint on
the non-convex character of matter at high densities. Remarkably,
\cite{Bedaque2015} found that the more abrupt the sound speed changes
with density (from its values at $n=2n_0$ to the asymptotic value
$1/\sqrt{3}$) the larger are the maximum masses of neutron stars they
can build within their model (near $2M_\odot$).
This non-monotonic behaviour of the sound speed may occur (depending
on the EoS) at baryon number densities within reach of the maximum
values of the number density predicted for \emph{ordinary} neutron
stars (namely, $5-8$ times the nuclear saturation density for most EoS
of dense matter) as well as in hybrid stars containing a quark phase
\citep[see
e.g.,][]{Bonanno_2009PhysRevC.79.045801,Alford_2013PhysRevD.88.083013}. Indeed,
we have shown and example of a hadronic EoS that contains the
transition to quark matter in the above mentioned density range (the
case of \texttt{CMF($\Lambda$B)}), which displays a significant
decrease of the sound speed and satisfies the existing astrophysical
and experimental constraints.
Connecting \cite{Bedaque2015} results with ours could suggest that
neutron stars or hybrid stars with masses above $\sim 2M_\odot$ (if
hyperons are included as possible degrees of freedom this limit may be
a bit smaller; see below) may have undergone a phase during their
formation, either at bounce or on longer (post-bounce) time scales
where thermodynamics could have been non-convex. This possibility
strongly depends on whether the non-monotonic behaviour of the sound
speed also drives a negative fundamental derivative, i.e. it depends
on the EoS as well as on other additional dynamic effects as,
e.g. whether the stellar core is strongly rotating. The reason for it
is that it is necessary to significantly exceed nuclear saturation
density in order to reach the regime in which non-monotonicity of the
sound speed (and hence, a possible convexity loss) may happen. During
the dynamical phase of stellar collapse, the maximum number densities
are reached just at bounce and these can be $\sim (2-3) \times n_0$
\citep[e.g.][]{Dimmelmeier_2008PhysRevD.78.064056}. Later, on longer
time scales, the density of the proto-neutron star increases as it
contracts and cools down
\cite[e.g.][]{Sumiyoshi_2005ApJ...629..922,Suwa_2014doi:10.1093/pasj/pst030},
though the central density only experiences very little increments on
time scales of $\sim 20\,$s \citep{Fischer_2010A&A...517A..80} except
if matter also contains hyperons. In this case the central number
density may increase (within less than 100\,s post bounce) and reach
values $n \simeq 0.85\,\text{fm}^{-3}$ for a proto-neutron star with
mass $\simeq 1.79\,M_\odot$ \citep[see Fig.\,22
in][]{Pons_1999ApJ...513..780}. Thus, if hyperons are present it is
much more likely to eventually reach a non-convex region (in some EoS)
in the post-bounce phase than right at bounce. 

Indeed, the above mentioned ultra-high densities are of interest when
the final fate of the collapse of stellar cores or binary neutron star
mergers is the formation of a BH. This is, for instance, the case of
metastable proto-neutron stars having a hyperon phase in the core and
baryonic masses $\gtrsim 1.8 M_\odot$. Unless fine-tuned parameters of
the hyperon-hyperon interaction are considered, these configurations
undergo a BH collapse on time escales of $\lesssim 100\,$s, during
which they may build a central number density
$\simeq 3-4\,\text{fm}^{-3}$ \citep{Pons1999}, improving the prospects
of finding a non-monotonic behaviour of the sound speed and a
potential convexity loss. An extension of the available microphysical
EoSs beyond the current upper boundaries in baryon number density is
needed to thoroughly explore any potential non-convex regime happening
before the formation of the apparent horizon.

 In this paper we have presented a numerical study of the structure, dynamics and gravitational-wave signature of compact stellar configurations described by a BZT fluid. Missing the appropriate extensions of tabular microphysical EoSs to explore the ultra-high density regime, we have resorted to a simple, phenomenological, non-convex EoS, which mimics some of the qualitative properties that microphysical EoSs possess. This ideal-gas-like EoS holds a density-dependent adiabatic index (or similarly, a non-monotonic sound speed dependence with density) and a causal behavior within a broad range of EoS parameters. The reason behind our simplistic choice of such a toy-model EoS has been to provide a proof-of-concept of the peculiarities associated with non-convex EoS before attempting further work employing state-of-the-art, microphysical EoSs. 

 We have studied the dynamics triggered by the non-convexity of the
 EoS analyzing three different situations. First, the equilibrium
 structure of stable compact stars. Second, the collapse of
 spherically symmetric neutron stars to BHs. Third, the dynamics of
 unstable and uniformly-rotating neutron stars that collapse
 gravitationally to BHs on a dynamical timescale.  The numerical
 simulations have been performed with two different codes, which
 guarantees the numerical robustness of our results.  We have used the
 most basic HLL solver to prevent a breach in our simulations that may
 happen between adjacent numerical zones across which the fundamental
 derivative changes sign as it is the case of the (S)GGL-EoS. For the
 fluid flow system of equations closed with a non-convex EoS, it has
 been demonstrated that if the approximate Riemann solver provides a
 sufficient amount of numerical viscosity to allow the formation of
 compound waves, the resulting numerical method is stable
 \cite[see][]{Argrow:1996,Guardone2002,Voss2005,Cinnella2006,Serna_Marquina:2014}. In
 particular, the HLL approximate Riemann solver satisfies the above
 requirements in relativistic fluid dynamics \citep{Ibanez:2017TUBOS}.

 The numerical simulations of collapsing stars have shown the
 appearance of non-convex dynamics. Remarkably, the non-convexity of
 the dynamics does not result in compound waves (e.g. rarefaction
 shocks or compressive rarefactions). This result is somewhat
 unexpected in view of the fact that our models produce BZT fluids,
 which may develop anomalous dynamics \citep[see,
 e.g.][]{Ibanez:2017TUBOS}. Instead, the new dynamics produced by the
 non-convexity of our phenomenological EoS stems from the
 non-monotonic dependence of the sound speed with density. As a
 result, regions where the sound speed decreases significantly form in
 the course of the collapse. In these regions the infalling matter
 becomes \emph{suddenly} supersonic and a shock forms. This shock is
 not expansive as one may guess, since it is produced as a result of
 the development of a non-convex region in the collapsing
 core. Noteworthy, all shock structures developed during the infalling
 phase are engulfed by the nascent BH. This result holds independent
 of whether the initial neutron star is rotating or it is spherically
 symmetric. To our knowledge, the behaviour we have found in our
 models has some precedent even using a microphysical
 EoS. Calculations of collapsing proto-neutron stars with a kaon
 condensate also showed the formation of an accretion (compression)
 shock in \cite{Pons1999}. That feature was attributed to the fact
 that $dp/dn=0$ in the region where the Maxwell construction for the
 PT was used. \cite{Pons1999} argued that a
 different treatment of the PT
 (e.g. employing a Gibbs rather than a Maxwell construction) would
 have prevented the formation of discontinuities, keeping finite the
 compressibility. While this conclusion is correct, we also note that
 the treatment of PTs in nuclear matter is
 an active field of research. So far, there is no global consensus in
 the Nuclear Physics community on, e.g. how to properly treat the
 transition from inhomogeneous to homogeneous nuclear
 matter. Therefore, we signal in this paper the potential consequences
 of a convexity loss in the dynamics due to the loss of convexity
 especially in first-order PTs.

 The existence of regions where the fundamental derivatives are
 negative is imprinted on the gravitational-wave signals associated
 with the infalling phase. Furthermore, the increased amplitude of the
 gravitational waves in the phase immediately preceding BH formation
 might be the only signature of a non-convex dynamics, unless a
 successful SN explosion is driven as a result of the released latent
 heat of a first-order PT \citep[this is the case of the hadron-quark
 PT in,
 e.g.][]{Sagert:2009PhRvL.102h1101,Fischer_2018NatAs...2..980}. If the
 SN fails, electromagnetic signals of this phase are not foreseen
 since the system is extremely optically thick to radiation in the
 regime in which convexity is lost. Likewise, neutrino emission is
 unimportant inasmuch as neutrinos are fully trapped inside the
 collapsing neutron star. However, neutrinos may act as a source of
 physical viscosity in the system
 \citep{Guilet15,Guilet17}. Therefore, they may smooth out the shocks
 developed in the limits of the non-convex regions formed in the
 course of the collapse, and hence, they may wash out any prominent
 effect of the convexity loss in the course of the collapse
 dynamics. A future study using actual microphysical EoS from nuclear
 physics and a suitable neutrino transport is opportune and will be
 presented elsewhere. In a different context that we have addressed
 here, we point out that \cite{Most_2018arXiv180703684M} have already
 found a systematic dephasing of the GW emission after the merger of
 two neutron stars, which may produce a qualitatively distinct
 signature in the post-merger GW signal and spectrum. These authors
 further conclude that the inclusion of a first-order PT to quark
 matter significantly accelerates the collapse to BH of the
 post-merger remnant and changes the ringdown GW frequencies. We point
 out that, since the transition to quark matter of first-order kind in
 the variant of the CMF EoS that \cite{Most_2018arXiv180703684M} have
 used (corresponding with the \texttt{CMF($\Lambda$B)} EoS), the
 convexity shall be lost in their merger models (as we have shown
 here). Therefore, the strong impact of the GW signature that they
 find, is also an indirect trace of the convexity loss at densities a
 few times larger than the nuclear saturation density. Furthermore, we
 note the qualitative resemblance of their results with ours: a
 significant modification of the GW emission is found after the,
 essentially, free-fall collapse of an unstable neutron star remnant.

 As a final note, we want to convey the idea that a finer tabulation
 of nuclear matter EoSs is probably adequate when the fundamental
 derivative displays large variations, specially, when these
 variations drive negative values of $\Gclas$. This means that, for
 applications in Computational Astrophysics it is probably worth
 including additional tabular points in situations where $\Gclas
 <0$. This means mapping with more tabular points thermodynamical
 states near the boundaries of regions where Maxwell or Gibbs
 constructions are built to deal with first-order PTs.

\appendix

\section{Numerical artifacts in the evaluation of fundamental derivatives}
\label{sec:numerical_artifacts}

We have investigated whether the differences between distinct variants
of the LS220 EoSs (including cases with $\Lambda$ hyperons) may arise
due to the distinct tabulation resolution. For the LS220 EoS, the
three-dimensional table in $(T, n, Y_q)$ obtained from the CompOSE
webpage, where $n$ and $Y_q$ are the baryon number density and the
charge fraction, respectively, contains $(163,164,51)$ points
logarithmically allocated along the $T$- and $n$-directions of the
table and linearly collocated in the $Y_q$-direction, covering the
ranges $0.1\,\text{MeV}\lesssim T \lesssim 182\,\text{MeV}$,
$5.2\times 10^{-8}\,\text{fm}^{-3}\lesssim n \lesssim
12\,\text{fm}^{-3}$ and $0.03\lesssim Y_q \lesssim 0.5$. For the HS
EoS, the tables contain $(81,326,60)$ points to cover the intervals
$0.1\,\text{MeV}\lesssim T \lesssim 158\,\text{MeV}$,
$10^{-12}\,\text{fm}^{-3} \lesssim n \lesssim 10\,\,\text{fm}^{-3}$,
$0.01 \lesssim Y_q \lesssim 0.6$ and, as a result, the temperature
resolution is about twice better in the LS220 tables than in the HS
ones, while the baryon number density resolution is only $\sim 20\%$
better in the HS tables than in the LS220 case. In order to compute
the fundamental derivative using numerical derivatives along the
tabular directions we employ the expression
\begin{equation}
  \Gclas= \displaystyle{ 1 +
    \left.\frac{\partial \ln{\cscla}}{\partial \ln{n}} \right|_{T,Y_q} +
    \frac{\beta_{V}}{n c_V} \left.\frac{\partial \ln{\cscla}}{\partial \ln{T}} \right|_{n,Y_q}},
\label{eq:G4}
\end{equation}
where, $\beta_V$ is the tension coefficient at contant volume
\begin{equation}
  \beta_V= \displaystyle{\left.\frac{\partial p}{\partial T} \right|_{n,Y_q}},
\label{eq:betaV}
\end{equation}
and $c_V$ is the specific heat capacity at constant volume
\begin{equation}
  c_V= \displaystyle{\frac{T}n\left.\frac{\partial s}{\partial T} \right|_{n,Y_q}}.
\label{eq:cV}
\end{equation}

For the calculation of the derivatives involved in $\Gclas$ and
$\Grel$, the resolution in number density is more important than in
the other two directions. Thus, a priori, we may guess that a finer
number density resolution across the PT should provide
smoother results and, if the negativity of the fundamental derivatives
would come from a purely numerical origin, the expectation would also
be that the fundamental derivative would remain positive for finer
number-density discretizations too. However, this is not the case and
the HS EoS displays a more oscillatory behaviour of the fundamental
derivatives.

Restricting our attention to the LS220 EoS, we have compared the
fundamental derivatives obtained with different tabulations
(i.e. different tabular nodal points) and numerical computation of the
thermodynamic derivatives. For that, we have employed two tabular
versions of the LS220 EoS built by \cite{OConnor:2010} in addition to
the two variants obtained from the CompOSE web page shown before. The
table dubbed \texttt{LS220hr} \citep{OConnor:2010} possesses a
resolution in number density and charge fraction that is roughly the
same as in the CompOSE tables ($19.5$\,points per decade for the
number density and 50 uniform points for $Y_q$), while the temperature
resolution ($\simeq 38$\,points per decade) is slightly worse than in
the CompOSE tables ($\simeq 50$\,points per decade). The other LS220
variant (tagged \texttt{LS220lr}; \citealt{OConnor:2010}) has a poorer
resolution in temperature ($\simeq 30\,$points per decade) and number
density ($\simeq 18\,$points per decade) than the \texttt{LS220hr}
table. It is evident from Fig.\,\ref{fig:funder3} that $\Grel <0$
through the PT in the tabular versions of the LS220 EoS which do not
include hyperons (\texttt{LS220}, \texttt{LS220hr} and
\texttt{LS220lr}). All these variants of the LS220 EoS have been
broadly used in actual calculations of stellar core collapse and
supernovae
\citep[e.g.][]{Couch_2013ApJ...778L...7,Couch_2014ApJ...785..123} and,
necessarily, these computations have accessed the regime where the
transition from non-uniform to uniform nuclear matter is
located. Thus, it is very likely that state-of-the-art models of
supernovae have included regions of the thermodynamics phase space
which are non-convex (due to the particular realization of the PT
under consideration). We note that since the tabulation of the tables
\texttt{LS220hr} and \texttt{LS220lr} is done as a function of
$(T,n,Y_q)$, we do not show in Fig.\,\ref{fig:funder3} the
relativistic fundamental derivative along an isentrope (as was done in
Figs.\,\ref{fig:Compose-funder1} and \ref{fig:Compose-funder2}), but
along a curve of constant temperature and charge fraction. This
explains why the \texttt{LS220($\Lambda$)} EoS does not display
$\Grel<0$ in this case, while it does it for the isentrope $s=2.5$
(Fig.\,\ref{fig:Compose-funder2}).

Regardless of the discretization of the tables, which may induce
spurious changes of sign of the fundamental derivatives, across the
PT, \cite{Hempel:2010NuPhA.837..210} stated that
discontinuities of the second derivatives of the Helmholtz free energy
result from the enforced Maxwell construction, which we remind is the
same as used in the LS220 EoS. This is the physical root of the large
amplitude oscillations and changes of sign of the fundamental
derivative in both the HS EoS and the LS EoS without the incorporation
of hyperons. In contrast, the \texttt{LS220($\Lambda$)} EoS, was
constructed by \cite{Oertel:2012PhRvC..85e5806}, who took special care
in fixing a number of pathologies of the PT under
consideration. The result are positively defined fundamental
derivatives across the PT in the latter EoS and the
conditions considered here. Remarkably, the EoS of
\cite{Oertel:2012PhRvC..85e5806} employs also a Maxwell construction
to deal with both the PT from inhomogeneous to
homogeneous nuclear matter as well as the transition to the
hyperon phase.

Another place where numerical artifacts (associated to the calculation
of high-order derivatives) may exist is close to the tabular
boundaries. This seems to be the case in the GSHen EoS, which displays
different behaviors depending on the RMF parameterization at low
entropies. While the original FSUGold or FSU1 parameterization
included in the \texttt{GSHen(FSU1)} EoS shows large amplitude
oscillations, where both the relativistic and classical fundamental
derivatives become negative at high density
($n \gtrsim 1\,\text{fm}^{-3}$) and low entropies per baryon
(Fig.\,\ref{fig:Compose-funder1}), the FSU2 and NL3 parameter sets
(corresponding to \texttt{GSHen(FSU2)} and \texttt{GSHen(NL3)},
respectively) are classically convex (i.e. $\Gclas >0$) up to the
highest baryon density at which they are tabulated, i.e.
$n_{\rm max}\simeq 1.5\,\text{fm}^{-3}$.  As we have anticipated
above, the convexity loss of the \texttt{GSHen(FSU1)} EoS is likely
due to numerical artifacts in the computation of high-order
thermodynamic derivatives near the table boundaries. Large amplitude
oscillations at high number densities are not observed at higher
entropies per baryon ($s=2.5$) because the CompOSE tables of the GSHen
EoSs do not contain tabular points at sufficiently large temperature
to compute values of the thermodynamic quantities along the isentrope
$s=2.5$ for $n\gtrsim 0.6\,\text{fm}^{-3}$
(Fig.\,\ref{fig:Compose-funder2}; upper row).

%
\begin{figure}
\centering
\includegraphics[width=\columnwidth]{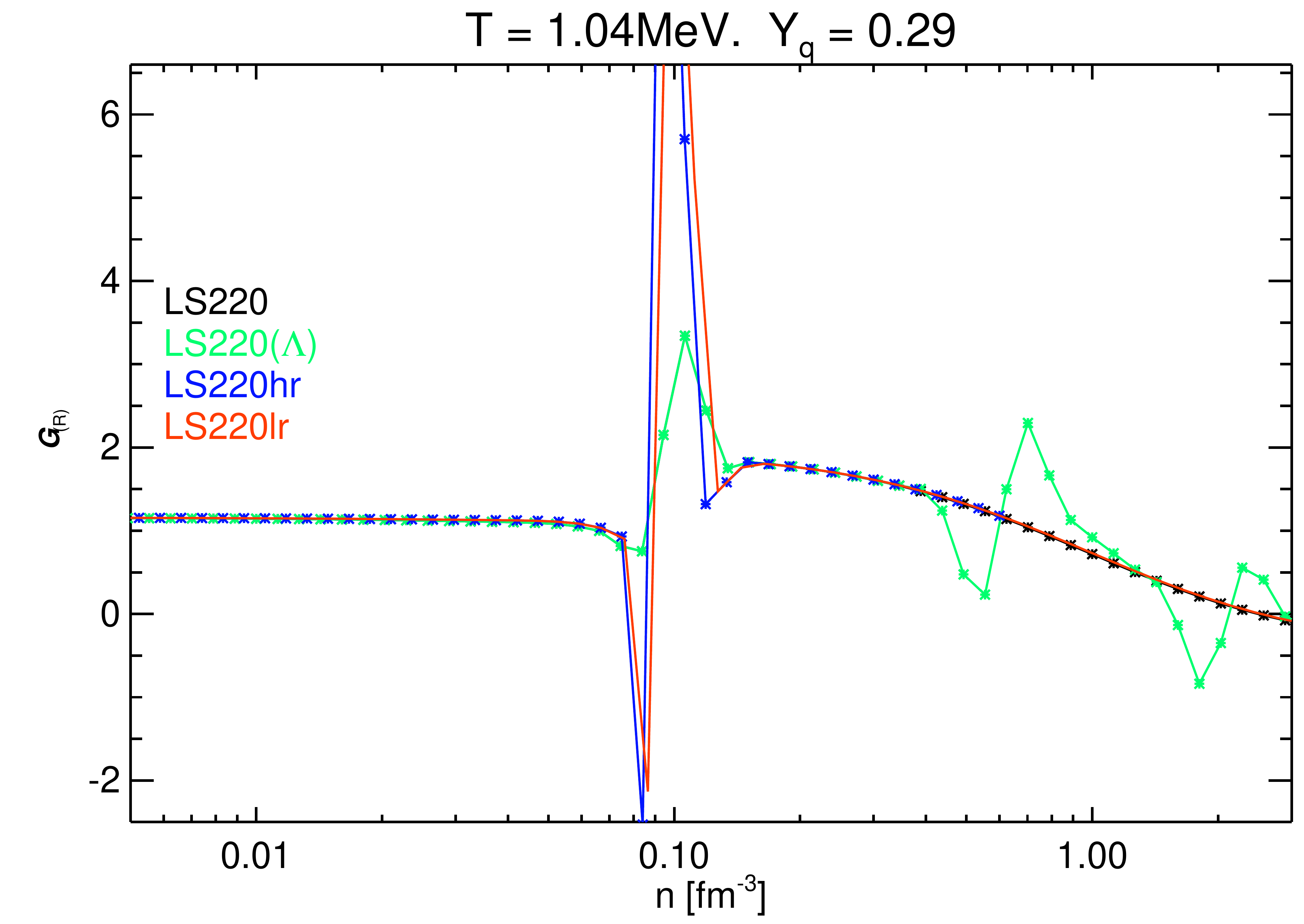}
\caption{Comparison of different tabular versions of the LS220 EoS.
  We fix the same temperature and charge fraction as in
  Fig.\,\ref{fig:Compose-funder1} and consider the two versions of the
  LS220 EoS obtained from the CompOSE web page (\texttt{LS220} and
  \texttt{LS220($\Lambda$)}) in addition to the \emph{high}- and
  \emph{low-resolution} tables from
  \texttt{https://stellarcollapse.org/equationofstate}, respectively
  labeled \texttt{LS220hr} and \texttt{LS220lr}. We note that the plot
  of the relativistic fundamental derivative corresponding to the
  \texttt{LS220hr} falls on top of the one corresponding to the
  \texttt{LS220} EoS in the number density range where both tables
  overlap.}
    %
\label{fig:funder3}
\end{figure}

\section{Analysis of the `P\lowercase{oly}T\lowercase{h}' EoS}
\label{appB}

A simple way to mimic the complex thermodynamical 
processes taking place inside a collapsing stellar core in simulations of hydrodynamical 
supernovae and in the formation of compact objects,  
considers an EoS for which the pressure has two components, namely a polytropic 
component (the cold one, $p_{\rm c}$), and an ideal-gas component which incorporates the
thermal effects $p_{\rm t}$. This EoS, that we call 'PolyTh' reads as 
\citep[see, e.g.][]{Maione2016} 
\begin{equation}
p = p_{\rm c} + p_{\rm t}
\, \, \, \,, 
\, \, \, \, 
p_{\rm c} = K \, \rho^{\Gamma_{\rm c}}
\, \, \, \,, 
\, \, \, \, 
p_{\rm t} = (\Gamma_{\rm t} - 1) \rho \epsilon_{\rm t}\,,
\end{equation}
where
\begin{equation}
\epsilon_{\rm t} 
\, \, =  \, \, 
\epsilon - \epsilon_{\rm c} 
\, \, \, \,, 
\, \, \, \, 
\epsilon_{\rm c} 
\, \, =  \, \, 
\displaystyle{
\epsilon_0 
\, \, +  \, \, 
\frac{K}{\Gamma_{\rm c} -1} \, \rho^{\Gamma_{\rm c} - 1} \,.
}
\label{ad}
\end{equation}
The total specific internal energy $\epsilon$, or its thermal component $\epsilon_{\rm t}$, can be
considered as an independent thermodynamical variable, being the cold component $\epsilon_{\rm c}$ 
a function of density given by Eq.~(\ref{ad}). In practice, we take $\epsilon_0$, 
in Eq.~(\ref{ad}), equal to zero. 
The free parameters of the PolyTh-EoS are $K, \Gamma_{\rm c}$ and $ \Gamma_{\rm t}$. 

Let us define
\begin{equation}  
a_{\alpha}^2 := \Gamma_{\alpha} \,\left( \displaystyle{\frac{p_{\alpha}}{\rho}} \right) 
\, = \,
\Gamma_{\alpha} \, (\Gamma_{\alpha} -1) \, \epsilon_{\alpha}
\end{equation}
where $\alpha\,(={\rm c}, {\rm t})$ stands, respectively, for the cold and thermal components of
pressure. Hence, the classical definition of the local speed of sound can be written
\begin{equation}  
\cscla^2  = a_{\rm c}^2 + a_{\rm t}^2\,.
\label{cs2cla2}
\end{equation}
The specific enthalpy is given by
\begin{equation}
h := 1 + \epsilon + \displaystyle{\frac{p}{\rho}} =
1+
\Gamma_{\rm c} \, \epsilon_{\rm c} 
\, +  \, 
\Gamma_{\rm t} \, \epsilon_{\rm t} \,,
\end{equation}
or, alternatively,
\begin{equation}
h = 
\displaystyle{
1+
\frac{a_{\rm c}^2}{\Gamma_{\rm c} -1} 
\, +  \, 
\frac{a_{\rm t}^2}{\Gamma_{\rm t} -1} \,.
}
\label{h}
\end{equation}
The relativistic definition of the speed of sound is related to the classical one according to: 
\begin{equation}                                                                  
\csrel^2 
\,\,  =\,\,  
h^{-1} \,\, \cscla^2
\,\,  =\,\,  
\displaystyle{
\frac{\Gamma_{\rm c} \, (\Gamma_{\rm c} -1) \, \epsilon_{\rm c}  + \Gamma_{\rm t} \, (\Gamma_{\rm t} -1) \, \epsilon_{\rm t}}
{ 1+ \Gamma_{\rm c} \, \epsilon_{\rm c} + \Gamma_{\rm t} \, \epsilon_{\rm t} }
}\,.
\label{cs2rel}
\end{equation}                                                       
%
 From this equation we obtain the following constraint that the PolyTh EoS has to satisfy in order to be causal:
 \begin{equation}
 \displaystyle{
 c_{s_{(R)}}^2 \le  1
 \,\,\Longrightarrow \,\,
\Gamma_c (\Gamma_c - 2) \epsilon_c + \Gamma_t (\Gamma_t - 2) \epsilon_t  \le 1
 }\,.
 \label{causal}
 \end{equation}       
Hence, assuming that both $\epsilon_c$ and $\epsilon_t$ are non-negative, 
a sufficient condition for causality is
 \begin{equation}
\Gamma_c \le 2  \,\,\,\,\, {\rm and} \,\,\,\,\, \Gamma_t \le 2 
 \end{equation}       
There are a critical values of 
$\epsilon_{\rm c}^{\rm crit}$ and $\epsilon_{\rm t}^{\rm crit}$, at
the stationary point of $\csrel^2$
\begin{equation}
\displaystyle{
\left. \frac{\partial \csrel^2}{\partial \epsilon_{\alpha}} \right|_{\epsilon_{\beta}} 
\,\,
= 
\,\,
0
\,\,
\Longrightarrow
\,\,
\epsilon_\alpha^{\rm crit}
\,\,
= 
\,\,
\frac{\Gamma_{\alpha} - 1}{\Gamma_{\beta} ( \Gamma_{\beta} - \Gamma_{\alpha})}
} \,\,\,\,\,\,\,\, (\alpha \ne \beta)\,.
\label{eps_crit}
\end{equation}
By definition, the adiabatic exponent, $\Gamma_1$, is
\begin{equation}
\Gamma_1 
=
\displaystyle{
\left(\frac{\rho}{p}\right)
\,\,
\cscla^2
}
=
\displaystyle{
\left(\frac{\rho}{p}\right)
\,\,
(a_{\rm c}^2 + a_{\rm t}^2)
}\,,
\end{equation}
or, alternatively
\begin{equation}
\Gamma_1 
=
\,\,
\Gamma_{\rm c} \, \beta + \Gamma_{\rm t} \, (1-\beta)\,,
\label{ad_exp}
\end{equation}
where $\beta := p_{\rm c}/p$. According to Eq.~(\ref{ad_exp}), $\Gamma_1$ can be considered as
just the average of the cold and thermal `gammas' weighted with their relative components of pressure.

 The classical fundamental derivative, $\Gclas$, for the PolyTh EoS is
\begin{equation}
\Gclas
= 
\displaystyle{
\frac{1}{2}
\,
(1+ \tilde{\Gamma})
}\,
\label{FDcla}
\end{equation}
where
\begin{equation}
\tilde{\Gamma}
= 
\displaystyle{
\frac{
\Gamma_{\rm c} \, a_{\rm c}^2 + 
\Gamma_{\rm t} \, a_{\rm t}^2
}{
a_{\rm c}^2 + a_{\rm t}^2
}
}\,,
\end{equation}
or, alternatively
\begin{equation}
\tilde{\Gamma}
= 
\displaystyle{
\frac{
\Gamma_{\rm c}^2 \, \beta + \Gamma_{\rm t}^2 \, (1-\beta)
}{
\Gamma_1
}
}\,,
\end{equation}
which can be interpreted as the ratio between the mean of both $\Gamma_{\rm c}^2$ and  $\Gamma_{\rm t}^2$ and the
adiabatic exponent $\Gamma_1$. By construction, the quantity $\tilde{\Gamma}$ varies between the values 
of $\Gamma_{\rm t}$ and $\Gamma_{\rm c}$.
 
The relativistic fundamental derivative, $\Grel$, for the PolyTh EoS is
\begin{equation}
\Grel
\,\,= \,\, 
\Gclas
\,
-
\,
\frac{3}{2} 
\,
\csrel^2 
\,\,= \,\, 
\displaystyle{
\frac{1}{2}
\,
( 1+ \tilde{\Gamma} - 3 \, \csrel^2 )
}\,.
\label{FDrel}
\end{equation}
Some comments are in order:

%
%
1) From Eq.~(\ref{FDrel}), one concludes that the PolyTh EoS can develop, due to
relativistic effects,  non-convex regions there where the following relationships are satisfied:
\begin{equation}
\displaystyle{
\frac{1+ \tilde{\Gamma}}{3} 
}
\,\,\le \,\, 
\csrel^2 
\,\,\le \,\, 1
\,\,\,\,\, {\rm and} \,\,\,\,\,
1 \,\,\le \,\, 
 \tilde{\Gamma}
\,\,\le \,\,2 \,.
\label{PolConvex}
\end{equation}
where the lower bound on $ \tilde{\Gamma} $ comes from its definition, assuming that: 
$\Gamma_{\alpha} \ge 1 \,\,\,\forall \alpha=c,t$.

2) The analysis of the particular cases $\beta=1$ and $\beta=0$ can shed light on the previous conclusion.
These cases are easily covered by taking 
$\tilde{\Gamma} = \Gamma_{\rm c}$
and $\tilde{\Gamma} = \Gamma_{\rm t}$, respectively, 
in Eqs.~(\ref{FDcla}), (\ref{FDrel}) and (\ref{PolConvex}).
Let us consider, e.g.~$\beta=1$. From Eqs.~(\ref{cs2rel}), (\ref{cs2cla2}) and (\ref{h}) we obtain
\begin{eqnarray}
i) \,\,\lim_{\epsilon_{\rm c} \rightarrow \infty} \csrel^2 &=& \Gamma_{\rm c} \,-\, 1\,,
\\
ii) \,\,\lim_{\epsilon_{\rm c} \rightarrow \infty} \Grel
&=& 2 \, - \, \Gamma_{\rm c} \,,
\end{eqnarray}
and, therefore, the thermodynamics is convex for a causal EoS, if and only if $1 \le \Gamma_{\rm c} \le 2$, 
as it happens for an ideal-gas EoS.

3) The above two comments help us to give the conditions to be satisfied by the PolyTh EoS 
in order to be both causal and convex:  
\begin{equation}
\csrel^2 
\,\,\le \,\,
\tilde{\Gamma} - 1
\,\,\,\,\, {\rm and} \,\,\,\,\,
1 \,\,\le \,\, 
 \tilde{\Gamma}
\,\,\le \,\,2 \,.
\end{equation}
%

As an example, let us complete the analysis by taking for the PolyTh EoS one of the set of parameters used in the binary neutron star merger simulations of~\citet{Maione2016}, namely $\Gamma_{\rm c} =3.005$ and $\Gamma_{\rm t} =1.8$. We take
$\epsilon_{\rm c}$ and $\epsilon_{\rm t}$  as the independent thermodynamical variables. Figure~\ref{fig:cs2r} shows the relativistic speed of sound, $\csrel^2$, defined in Eq.~(\ref{cs2rel}). It is an increasing function, in both $\epsilon_{\rm c}$ and $\epsilon_{\rm t}$, up to some value of $\epsilon_{\rm c}^{\rm crit}$ given by Eq.~(\ref{eps_crit}). In our example, this value is $\epsilon_{\rm c}^{\rm crit} = 0.22$. For $\epsilon_{\rm c} \ge \epsilon_{\rm c}^{\rm crit}$ (depending on $\epsilon_{\rm t}$) the PolyTh EoS becomes non-causal. On the other hand, Figure \ref{fig:cs2r} also shows the relativistic fundamental derivative, $\Grel$, given by Eq.~(\ref{FDrel}). It is a decreasing function in both $\epsilon_{\rm c}$ and $\epsilon_{\rm t}$. For $\epsilon_c \ge \epsilon_c^{\rm crit}$ (depending on $\epsilon_t$) the PolyTh EoS becomes non-convex.

As a summary, from the above example and from our previous analysis, we conclude 
that the PolyTh EoS is convex in those regions of the space of parameters in which it is causal. The
non-convex regions are associated with the non-causal ones and, therefore, the corresponding subset
of parameters has no physical meaning.

 \begin{figure}
  \centering
    \includegraphics[width=\columnwidth]{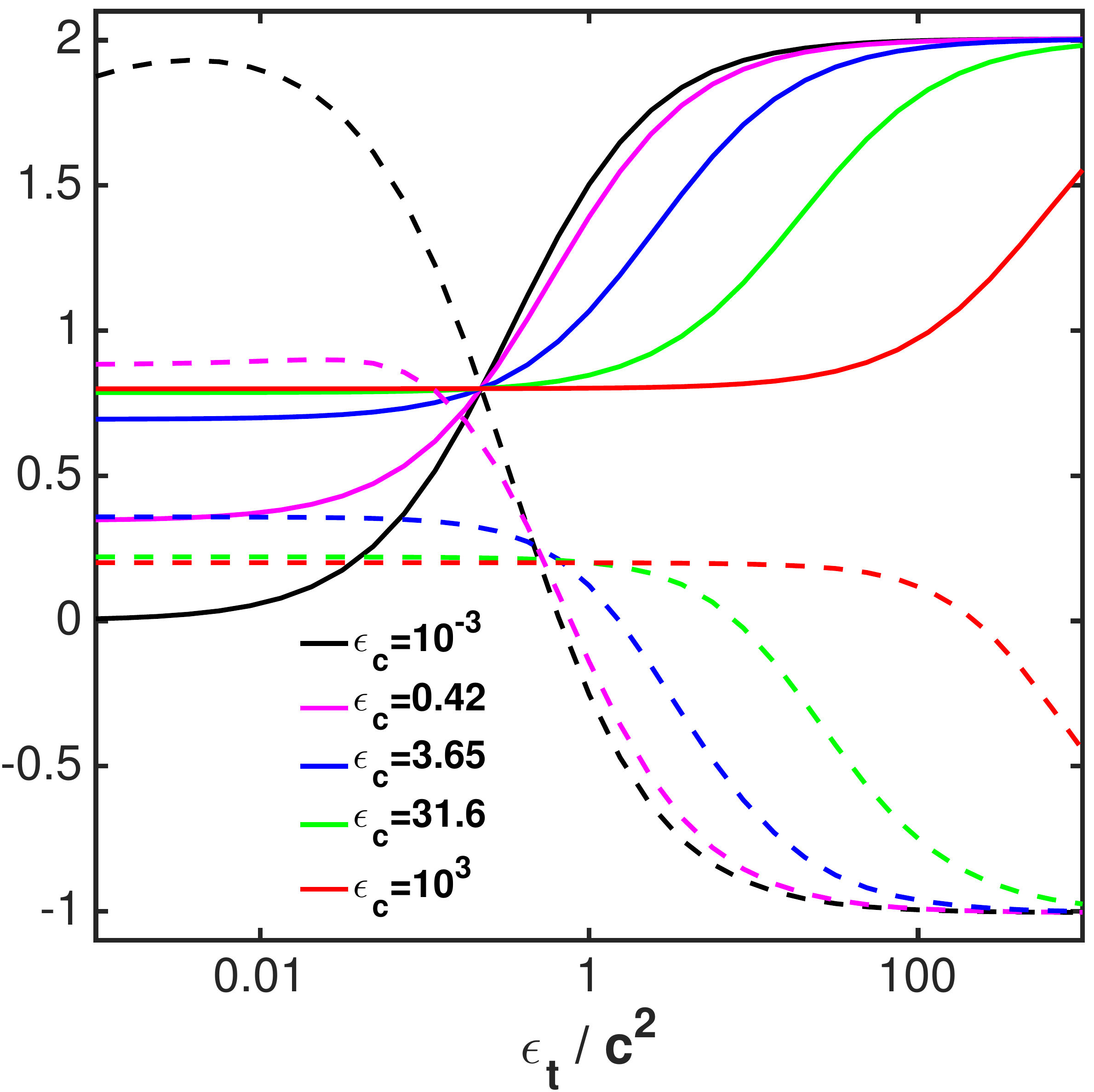}
  \caption{Relativistic fundamental derivative $\Grel$ in Eq.~(\ref{FDrel}) (dashed lines) and relativistic speed of sound $\csrel$ in Eq.~(\ref{cs2rel}) (solid lines),
versus $\epsilon_{\rm t}$, for different values of $\epsilon_{\rm c}$. We use the PolyTh EoS with $\Gamma_{\rm c} =3.005$ and  $\Gamma_{\rm t} =1.8$.}
  \label{fig:cs2r}
\end{figure}
%

%
%
%
 \section*{Acknowledgements}
%
 We gratefully acknowledge the anonymous referee for the carefull
 examination of our work, which has greatly improved as a result of
 his/her feedback. We thank V. Dexheimer and S. Schramm for allowing
 us to use their implementation of the \texttt{CMF($\Lambda$B)} EoS,
 since it is not included in the CompOSE database so far. We also
 kindly acknowledge the help of the CompOSE development team and,
 especially of M. Oertel.  Research supported by the Spanish Ministry
 of Economy and Competitiveness (MINECO) through grants
 AYA2015-66899-C2-1-P, and MTM2014-56218-C2-2-P, and by the
 Generalitat Valenciana (PROMETEOII-2014-069, ACIF/2015/216). MAA
 acknowledges support from the European Research Council (grant
 CAMAP-259276). NSG and JAF acknowledge support from the European
 Union's Horizon 2020 research and innovation programme under the
 H2020-MSCA-RISE-2017 Grant No. FunFiCO-777740. We thank the PHAROS
 COST Action (CA16214) and the GWverse COST Action (CA16104) for
 partial support. The computations were performed under grants
 AECT-2017-3-0007 and AECT-2018-1-0010 of the Spanish Supercomputing
 Network on the \textit{MareNostrum} cluster of the Barcelona
 Supercomputing Centre, and on the clusters \textit{Tirant} and
 \textit{Lluisvives} of the Servei d'Inform\`atica of the University
 of Valencia.


\bibliographystyle{mnras} 
\bibliography{referencesNCC}

\begin{thebibliography}{}
\makeatletter
\relax
\def\mn@urlcharsother{\let\do\@makeother \do\$\do\&\do\#\do\^\do\_\do\%\do\~}
\def\mn@doi{\begingroup\mn@urlcharsother \@ifnextchar [ {\mn@doi@}
  {\mn@doi@[]}}
\def\mn@doi@[#1]#2{\def\@tempa{#1}\ifx\@tempa\@empty \href
  {http://dx.doi.org/#2} {doi:#2}\else \href {http://dx.doi.org/#2} {#1}\fi
  \endgroup}
\def\mn@eprint#1#2{\mn@eprint@#1:#2::\@nil}
\def\mn@eprint@arXiv#1{\href {http://arxiv.org/abs/#1} {{\tt arXiv:#1}}}
\def\mn@eprint@dblp#1{\href {http://dblp.uni-trier.de/rec/bibtex/#1.xml}
  {dblp:#1}}
\def\mn@eprint@#1:#2:#3:#4\@nil{\def\@tempa {#1}\def\@tempb {#2}\def\@tempc
  {#3}\ifx \@tempc \@empty \let \@tempc \@tempb \let \@tempb \@tempa \fi \ifx
  \@tempb \@empty \def\@tempb {arXiv}\fi \@ifundefined
  {mn@eprint@\@tempb}{\@tempb:\@tempc}{\expandafter \expandafter \csname
  mn@eprint@\@tempb\endcsname \expandafter{\@tempc}}}

\bibitem[\protect\citeauthoryear{{Abbott} et~al.,}{{Abbott}
  et~al.}{2016a}]{Abbott1}
{Abbott} B.~P.,  et~al., 2016a, \mn@doi [Physical Review Letters]
  {10.1103/PhysRevLett.116.061102}, \href
  {http://adsabs.harvard.edu/abs/2016PhRvL.116f1102A} {116, 061102}

\bibitem[\protect\citeauthoryear{{Abbott} et~al.,}{{Abbott}
  et~al.}{2016b}]{Abbott2}
{Abbott} B.~P.,  et~al., 2016b, \mn@doi [Physical Review Letters]
  {10.1103/PhysRevLett.116.241103}, \href
  {http://adsabs.harvard.edu/abs/2016PhRvL.116x1103A} {116, 241103}

\bibitem[\protect\citeauthoryear{{Abbott} et~al.,}{{Abbott}
  et~al.}{2017a}]{Abbott:2017vtc}
{Abbott} B.~P.,  et~al., 2017a, \mn@doi [\prl]
  {10.1103/PhysRevLett.118.221101}, 118, 221101

\bibitem[\protect\citeauthoryear{{Abbott} et~al.,}{{Abbott}
  et~al.}{2017b}]{GW170814-prl}
{Abbott} B.~P.,  et~al., 2017b, \mn@doi [\prl]
  {10.1103/PhysRevLett.119.141101}, \href
  {http://adsabs.harvard.edu/abs/2017PhRvL.119n1101A} {119, 141101}

\bibitem[\protect\citeauthoryear{{Abbott} et~al.,}{{Abbott}
  et~al.}{2017c}]{Abbott:2017a}
{Abbott} B.~P.,  et~al., 2017c, \mn@doi [Phys. Rev. Lett.]
  {10.1103/PhysRevLett.119.161101}, \href
  {http://adsabs.harvard.edu/abs/2017PhRvL.119p1101A} {119, 161101}

\bibitem[\protect\citeauthoryear{{Abbott} et~al.,}{{Abbott}
  et~al.}{2017d}]{Abbott_2017ApJ...848L..12}
{Abbott} B.~P.,  et~al., 2017d, \mn@doi [\apjl] {10.3847/2041-8213/aa91c9},
  \href {http://adsabs.harvard.edu/abs/2017ApJ...848L..12A} {848, L12}

\bibitem[\protect\citeauthoryear{{Abbott} et~al.,}{{Abbott}
  et~al.}{2017e}]{Abbott3}
{Abbott} B.~P.,  et~al., 2017e, \mn@doi [\apjl] {10.3847/2041-8213/aa920c},
  \href {http://adsabs.harvard.edu/abs/2017ApJ...848L..13A} {848, L13}

\bibitem[\protect\citeauthoryear{{Abbott} et~al.,}{{Abbott}
  et~al.}{2017f}]{GW170608}
{Abbott} B.~P.,  et~al., 2017f, \mn@doi [\apjl] {10.3847/2041-8213/aa9f0c},
  \href {http://adsabs.harvard.edu/abs/2017ApJ...851L..35A} {851, L35}

\bibitem[\protect\citeauthoryear{{Abbott} et~al.,}{{Abbott}
  et~al.}{2018}]{Abbot_2018PhRvL.121p1101}
{Abbott} B.~P.,  et~al., 2018, \mn@doi [Physical Review Letters]
  {10.1103/PhysRevLett.121.161101}, \href
  {http://cdsads.u-strasbg.fr/abs/2018PhRvL.121p1101A} {121, 161101}

\bibitem[\protect\citeauthoryear{{Abdikamalov}, {Dimmelmeier}, {Rezzolla}  \&
  {Miller}}{{Abdikamalov} et~al.}{2009}]{Abdikamalov2009}
{Abdikamalov} E.~B.,  {Dimmelmeier} H.,  {Rezzolla} L.,   {Miller} J.~C.,
  2009, \mn@doi [\mnras] {10.1111/j.1365-2966.2008.14056.x}, \href
  {http://adsabs.harvard.edu/abs/2009MNRAS.392...52A} {392, 52}

\bibitem[\protect\citeauthoryear{{Alford}, {Blaschke}, {Drago}, {Kl{\"a}hn},
  {Pagliara}  \& {Schaffner-Bielich}}{{Alford}
  et~al.}{2007}]{Alford_2007Natur.445E...7}
{Alford} M.,  {Blaschke} D.,  {Drago} A.,  {Kl{\"a}hn} T.,  {Pagliara} G.,
  {Schaffner-Bielich} J.,  2007, \mn@doi [\nat] {10.1038/nature05582}, \href
  {http://adsabs.harvard.edu/abs/2007Natur.445E...7A} {445, 7}

\bibitem[\protect\citeauthoryear{Alford, Han  \& Prakash}{Alford
  et~al.}{2013}]{Alford_2013PhysRevD.88.083013}
Alford M.~G.,  Han S.,   Prakash M.,  2013, \mn@doi [Phys. Rev. D]
  {10.1103/PhysRevD.88.083013}, 88, 083013

\bibitem[\protect\citeauthoryear{{Annala}, {Gorda}, {Kurkela}  \&
  {Vuorinen}}{{Annala} et~al.}{2018}]{Annala_2018PhRvL.120q2703}
{Annala} E.,  {Gorda} T.,  {Kurkela} A.,   {Vuorinen} A.,  2018, \mn@doi
  [Physical Review Letters] {10.1103/PhysRevLett.120.172703}, \href
  {http://adsabs.harvard.edu/abs/2018PhRvL.120q2703A} {120, 172703}

\bibitem[\protect\citeauthoryear{{Antoniadis} et~al.,}{{Antoniadis}
  et~al.}{2013}]{Antoniadis2013}
{Antoniadis} J.,  et~al., 2013, \mn@doi [Science] {10.1126/science.1233232},
  \href {http://adsabs.harvard.edu/abs/2013Sci...340..448A} {340, 448}

\bibitem[\protect\citeauthoryear{{Aoki}, {Endr{\H o}di}, {Fodor}, {Katz}  \&
  {Szab{\'o}}}{{Aoki} et~al.}{2006}]{Aoki2006}
{Aoki} Y.,  {Endr{\H o}di} G.,  {Fodor} Z.,  {Katz} S.~D.,   {Szab{\'o}} K.~K.,
   2006, \mn@doi [\nat] {10.1038/nature05120}, \href
  {http://adsabs.harvard.edu/abs/2006Natur.443..675A} {443, 675}

\bibitem[\protect\citeauthoryear{Argrow}{Argrow}{1996}]{Argrow:1996}
Argrow B.~M.,  1996, \mn@doi [Shock Waves] {10.1007/BF02511381}, 6, 241

\bibitem[\protect\citeauthoryear{Baiotti, Hawke, Montero, L\"{o}ffler,
  Rezzolla, Stergioulas, Font  \& Seidel}{Baiotti et~al.}{2005a}]{Baiotti2005}
Baiotti L.,  Hawke I.,  Montero P.~J.,  L\"{o}ffler F.,  Rezzolla L.,
  Stergioulas N.,  Font J.~A.,   Seidel E.,  2005a, \mn@doi [\prd]
  {10.1103/PhysRevD.71.024035}, 71, 024035

\bibitem[\protect\citeauthoryear{Baiotti, Hawke, Montero, L{\"o}ffler,
  Rezzolla, Stergioulas, Font  \& Seidel}{Baiotti
  et~al.}{2005b}]{baiotti2005three}
Baiotti L.,  Hawke I.,  Montero P.~J.,  L{\"o}ffler F.,  Rezzolla L.,
  Stergioulas N.,  Font J.~A.,   Seidel E.,  2005b, Physical Review D, 71,
  024035

\bibitem[\protect\citeauthoryear{{Baiotti}, {Hawke}  \& {Rezzolla}}{{Baiotti}
  et~al.}{2007}]{Baiotti_etal2007}
{Baiotti} L.,  {Hawke} I.,   {Rezzolla} L.,  2007, \mn@doi [Classical and
  Quantum Gravity] {10.1088/0264-9381/24/12/S13}, \href
  {http://adsabs.harvard.edu/abs/2007CQGra..24S.187B} {24, S187}

\bibitem[\protect\citeauthoryear{{Balberg} \& {Gal}}{{Balberg} \&
  {Gal}}{1997}]{Balberg:1997NuPhA.625..435}
{Balberg} S.,  {Gal} A.,  1997, \mn@doi [Nuclear Physics A]
  {10.1016/S0375-9474(97)81465-0}, \href
  {http://adsabs.harvard.edu/abs/1997NuPhA.625..435B} {625, 435}

\bibitem[\protect\citeauthoryear{{Banik}, {Hempel}  \& {Bandyopadhyay}}{{Banik}
  et~al.}{2014}]{Banik:2014ApJS..214...22}
{Banik} S.,  {Hempel} M.,   {Bandyopadhyay} D.,  2014, \mn@doi [\apjs]
  {10.1088/0067-0049/214/2/22}, \href
  {http://adsabs.harvard.edu/abs/2014ApJS..214...22B} {214, 22}

\bibitem[\protect\citeauthoryear{{Banyuls}, {Font}, {Ib{\'a}{\~n}ez},
  {Mart{\'{\i}}}  \& {Miralles}}{{Banyuls} et~al.}{1997}]{Banyuls1997}
{Banyuls} F.,  {Font} J.~A.,  {Ib{\'a}{\~n}ez} J.~M.~l.,  {Mart{\'{\i}}}
  J.~M.~l.,   {Miralles} J.~A.,  1997, \apj, 476, 221

\bibitem[\protect\citeauthoryear{{Bardeen}, {Thorne}  \& {Meltzer}}{{Bardeen}
  et~al.}{1966}]{Bardeen1966}
{Bardeen} J.~M.,  {Thorne} K.~S.,   {Meltzer} D.~W.,  1966, \mn@doi [\apj]
  {10.1086/148791}, \href {http://adsabs.harvard.edu/abs/1966ApJ...145..505B}
  {145, 505}

\bibitem[\protect\citeauthoryear{Baumgarte \& Shapiro}{Baumgarte \&
  Shapiro}{1998}]{Baumgarte1998}
Baumgarte T.~W.,  Shapiro S.~L.,  1998, \mn@doi [\prd]
  {10.1103/PhysRevD.59.024007}, 59, 024007

\bibitem[\protect\citeauthoryear{{Baumgarte}, {Montero}, {Cordero-Carri{\'o}n}
  \& {M{\"u}ller}}{{Baumgarte} et~al.}{2013}]{Baumgarte2013}
{Baumgarte} T.~W.,  {Montero} P.~J.,  {Cordero-Carri{\'o}n} I.,   {M{\"u}ller}
  E.,  2013, \mn@doi [\prd] {10.1103/PhysRevD.87.044026}, \href
  {http://adsabs.harvard.edu/abs/2013PhRvD..87d4026B} {87, 044026}

\bibitem[\protect\citeauthoryear{{Bauswein}, {Janka}  \& {Oechslin}}{{Bauswein}
  et~al.}{2010}]{Bauswein2010}
{Bauswein} A.,  {Janka} H.-T.,   {Oechslin} R.,  2010, \mn@doi [\prd]
  {10.1103/PhysRevD.82.084043}, \href
  {http://adsabs.harvard.edu/abs/2010PhRvD..82h4043B} {82, 084043}

\bibitem[\protect\citeauthoryear{{Bazavov} et~al.,}{{Bazavov}
  et~al.}{2014}]{Bazavov2014}
{Bazavov} A.,  et~al., 2014, \mn@doi [\prd] {10.1103/PhysRevD.90.094503}, \href
  {http://adsabs.harvard.edu/abs/2014PhRvD..90i4503B} {90, 094503}

\bibitem[\protect\citeauthoryear{{Bedaque} \& {Steiner}}{{Bedaque} \&
  {Steiner}}{2015}]{Bedaque2015}
{Bedaque} P.,  {Steiner} A.~W.,  2015, \mn@doi [Physical Review Letters]
  {10.1103/PhysRevLett.114.031103}, \href
  {http://adsabs.harvard.edu/abs/2015PhRvL.114c1103B} {114, 031103}

\bibitem[\protect\citeauthoryear{{Bednarek}, {Haensel}, {Zdunik}, {Bejger}  \&
  {Ma{\'n}ka}}{{Bednarek} et~al.}{2012}]{Bednarek2012}
{Bednarek} I.,  {Haensel} P.,  {Zdunik} J.~L.,  {Bejger} M.,   {Ma{\'n}ka} R.,
  2012, \mn@doi [\aap] {10.1051/0004-6361/201118560}, \href
  {http://adsabs.harvard.edu/abs/2012A%26A...543A.157B} {543, A157}

\bibitem[\protect\citeauthoryear{{Bejger}, {Dimmelmeier}, {Haensel}  \&
  {Zdunik}}{{Bejger} et~al.}{2012}]{Bejger2012}
{Bejger} M.,  {Dimmelmeier} H.,  {Haensel} P.,   {Zdunik} J.~L.,  2012, in
  {Chamseddine} A.~H.,  ed., Twelfth Marcel Grossmann Meeting on General
  Relativity. pp 785--787, \mn@doi{10.1142/9789814374552_0056}

\bibitem[\protect\citeauthoryear{{Bejger}, {Blaschke}, {Haensel}, {Zdunik}  \&
  {Fortin}}{{Bejger} et~al.}{2017}]{Bejger2017}
{Bejger} M.,  {Blaschke} D.,  {Haensel} P.,  {Zdunik} J.~L.,   {Fortin} M.,
  2017, \mn@doi [\aap] {10.1051/0004-6361/201629580}, \href
  {http://adsabs.harvard.edu/abs/2017A%26A...600A..39B} {600, A39}

\bibitem[\protect\citeauthoryear{Bethe}{Bethe}{1942}]{Bethe1942}
Bethe H.,  1942, The theory of shock waves for an arbitrary equation of state,
  Tech. Paper 545, Office of Scientific Research and Development

\bibitem[\protect\citeauthoryear{Bonanno \& Drago}{Bonanno \&
  Drago}{2009}]{Bonanno_2009PhysRevC.79.045801}
Bonanno L.,  Drago A.,  2009, \mn@doi [Phys. Rev. C]
  {10.1103/PhysRevC.79.045801}, 79, 045801

\bibitem[\protect\citeauthoryear{{Bors{\'a}nyi}, {Fodor}, {Hoelbling}, {Katz},
  {Krieg}  \& {Szab{\'o}}}{{Bors{\'a}nyi} et~al.}{2014}]{Borsanyi2014}
{Bors{\'a}nyi} S.,  {Fodor} Z.,  {Hoelbling} C.,  {Katz} S.~D.,  {Krieg} S.,
  {Szab{\'o}} K.~K.,  2014, \mn@doi [Physics Letters B]
  {10.1016/j.physletb.2014.01.007}, \href
  {http://adsabs.harvard.edu/abs/2014PhLB..730...99B} {730, 99}

\bibitem[\protect\citeauthoryear{Brandt \& Seidel}{Brandt \&
  Seidel}{1995}]{brandt1995evolution}
Brandt S.~R.,  Seidel E.,  1995, Physical Review D, 52, 856

\bibitem[\protect\citeauthoryear{{Buballa} et~al.,}{{Buballa}
  et~al.}{2014}]{Buballa2014}
{Buballa} M.,  et~al., 2014, \mn@doi [Journal of Physics G Nuclear Physics]
  {10.1088/0954-3899/41/12/123001}, \href
  {http://adsabs.harvard.edu/abs/2014JPhG...41l3001B} {41, 123001}

\bibitem[\protect\citeauthoryear{Callen}{Callen}{1985}]{Callen_1985thermodynamics}
Callen H.,  1985, Thermodynamics and an Introduction to Thermostatistics.
Wiley

\bibitem[\protect\citeauthoryear{Camenzind}{Camenzind}{2007}]{Camenzind2007}
Camenzind M.,  2007, Compact objects in astrophysics: white dwarfs, neutron
  stars, and black holes.
Springer, Berlin, Germany

\bibitem[\protect\citeauthoryear{Chandrasekhar}{Chandrasekhar}{1939}]{Chandra1939}
Chandrasekhar S.,  1939, An Introduction to the Stellar Structure.
~ Vol. 1, University of Chicago Press, Chicago, USA

\bibitem[\protect\citeauthoryear{{Cinnella}}{{Cinnella}}{2008}]{Cinnella2008}
{Cinnella} P.,  2008, \mn@doi [Physics of Fluids] {10.1063/1.2907212}, \href
  {http://adsabs.harvard.edu/abs/2008PhFl...20d6103C} {20, 046103}

\bibitem[\protect\citeauthoryear{{Cinnella} \& {Corre}}{{Cinnella} \&
  {Corre}}{2006}]{Cinnella2006}
{Cinnella} P.,  {Corre} C.,  2006, in APS Division of Fluid Dynamics Meeting
  Abstracts.

\bibitem[\protect\citeauthoryear{{Cordero-Carri{\'o}n} \&
  {Cerd{\'a}-Dur{\'a}n}}{{Cordero-Carri{\'o}n} \&
  {Cerd{\'a}-Dur{\'a}n}}{2012}]{Isabel:2012arx}
{Cordero-Carri{\'o}n} I.,  {Cerd{\'a}-Dur{\'a}n} P.,  2012, preprint, \href
  {http://adsabs.harvard.edu/abs/2012arXiv1211.5930C} {} (\mn@eprint {arXiv}
  {1211.5930})

\bibitem[\protect\citeauthoryear{Cordero-Carri{\'o}n \&
  Cerd{\'a}-Dur{\'a}n}{Cordero-Carri{\'o}n \&
  Cerd{\'a}-Dur{\'a}n}{2014}]{Casas:2014}
Cordero-Carri{\'o}n I.,  Cerd{\'a}-Dur{\'a}n P.,  2014, Advances in
  Differential Equations and Applications.
SEMA SIMAI Springer Series Vol. 4, Springer International Publishing
  Switzerland, Switzerland

\bibitem[\protect\citeauthoryear{{Couch} \& {O'Connor}}{{Couch} \&
  {O'Connor}}{2014}]{Couch_2014ApJ...785..123}
{Couch} S.~M.,  {O'Connor} E.~P.,  2014, \mn@doi [\apj]
  {10.1088/0004-637X/785/2/123}, \href
  {http://adsabs.harvard.edu/abs/2014ApJ...785..123C} {785, 123}

\bibitem[\protect\citeauthoryear{{Couch} \& {Ott}}{{Couch} \&
  {Ott}}{2013}]{Couch_2013ApJ...778L...7}
{Couch} S.~M.,  {Ott} C.~D.,  2013, \mn@doi [\apjl]
  {10.1088/2041-8205/778/1/L7}, \href
  {http://adsabs.harvard.edu/abs/2013ApJ...778L...7C} {778, L7}

\bibitem[\protect\citeauthoryear{{De}, {Finstad}, {Lattimer}, {Brown}, {Berger}
   \& {Biwer}}{{De} et~al.}{2018}]{De_2018PhRvL.121i1102}
{De} S.,  {Finstad} D.,  {Lattimer} J.~M.,  {Brown} D.~A.,  {Berger} E.,
  {Biwer} C.~M.,  2018, \mn@doi [Physical Review Letters]
  {10.1103/PhysRevLett.121.091102}, \href
  {http://cdsads.u-strasbg.fr/abs/2018PhRvL.121i1102D} {121, 091102}

\bibitem[\protect\citeauthoryear{{Demorest}, {Pennucci}, {Ransom}, {Roberts}
  \& {Hessels}}{{Demorest} et~al.}{2010}]{Demorest2010}
{Demorest} P.~B.,  {Pennucci} T.,  {Ransom} S.~M.,  {Roberts} M.~S.~E.,
  {Hessels} J.~W.~T.,  2010, \mn@doi [\nat] {10.1038/nature09466}, \href
  {http://adsabs.harvard.edu/abs/2010Natur.467.1081D} {467, 1081}

\bibitem[\protect\citeauthoryear{{Dexheimer} \& {Schramm}}{{Dexheimer} \&
  {Schramm}}{2008}]{Dexheimer08}
{Dexheimer} V.,  {Schramm} S.,  2008, \mn@doi [\apj] {10.1086/589735}, \href
  {http://adsabs.harvard.edu/abs/2008ApJ...683..943D} {683, 943}

\bibitem[\protect\citeauthoryear{{Dexheimer}, {Negreiros}  \&
  {Schramm}}{{Dexheimer} et~al.}{2015}]{Dexheimer15}
{Dexheimer} V.,  {Negreiros} R.,   {Schramm} S.,  2015, \mn@doi [\prc]
  {10.1103/PhysRevC.92.012801}, \href
  {http://adsabs.harvard.edu/abs/2015PhRvC..92a2801D} {92, 012801}

\bibitem[\protect\citeauthoryear{{Diaz Alonso}}{{Diaz
  Alonso}}{1985}]{Diaz-Alonso1985}
{Diaz Alonso} J.,  1985, \mn@doi [\prd] {10.1103/PhysRevD.31.1315}, \href
  {http://adsabs.harvard.edu/abs/1985PhRvD..31.1315D} {31, 1315}

\bibitem[\protect\citeauthoryear{Dietrich \& Bernuzzi}{Dietrich \&
  Bernuzzi}{2015}]{dietrich2015simulations}
Dietrich T.,  Bernuzzi S.,  2015, Physical Review D, 91, 044039

\bibitem[\protect\citeauthoryear{Dimmelmeier, Stergioulas  \& Font}{Dimmelmeier
  et~al.}{2006}]{dimmelmeier2006non}
Dimmelmeier H.,  Stergioulas N.,   Font J.~A.,  2006, Monthly Notices of the
  Royal Astronomical Society, 368, 1609

\bibitem[\protect\citeauthoryear{Dimmelmeier, Ott, Marek  \& Janka}{Dimmelmeier
  et~al.}{2008}]{Dimmelmeier_2008PhysRevD.78.064056}
Dimmelmeier H.,  Ott C.~D.,  Marek A.,   Janka H.-T.,  2008, \mn@doi [Phys.
  Rev. D] {10.1103/PhysRevD.78.064056}, 78, 064056

\bibitem[\protect\citeauthoryear{Dimmelmeier, Bejger, Haensel  \&
  Zdunik}{Dimmelmeier et~al.}{2009}]{Dimmelmeier2009}
Dimmelmeier H.,  Bejger M.,  Haensel P.,   Zdunik J.~L.,  2009, \mn@doi
  [Monthly Notices of the Royal Astronomical Society]
  {10.1111/j.1365-2966.2009.14891.x}, 396, 2269

\bibitem[\protect\citeauthoryear{{Drago} \& {Tambini}}{{Drago} \&
  {Tambini}}{1999}]{Drago_1999JPhG...25..971}
{Drago} A.,  {Tambini} U.,  1999, \mn@doi [Journal of Physics G Nuclear
  Physics] {10.1088/0954-3899/25/5/302}, \href
  {http://adsabs.harvard.edu/abs/1999JPhG...25..971D} {25, 971}

\bibitem[\protect\citeauthoryear{{Ducoin}, {Hasnaoui}, {Napolitani}, {Chomaz}
  \& {Gulminelli}}{{Ducoin} et~al.}{2007}]{Ducoin_2007}
{Ducoin} C.,  {Hasnaoui} K.~H.~O.,  {Napolitani} P.,  {Chomaz} P.,
  {Gulminelli} F.,  2007, \prc, \href
  {http://adsabs.harvard.edu/abs/2007PhRvC..75f5805D} {75, 065805}

\bibitem[\protect\citeauthoryear{{Fischer}, {Whitehouse}, {Mezzacappa},
  {Thielemann}  \& {Liebend{\"o}rfer}}{{Fischer}
  et~al.}{2010}]{Fischer_2010A&A...517A..80}
{Fischer} T.,  {Whitehouse} S.~C.,  {Mezzacappa} A.,  {Thielemann} F.~K.,
  {Liebend{\"o}rfer} M.,  2010, \mn@doi [\aap] {10.1051/0004-6361/200913106},
  \href {https://ui.adsabs.harvard.edu/#abs/2010A&A...517A..80F} {517, A80}

\bibitem[\protect\citeauthoryear{{Fischer} et~al.,}{{Fischer}
  et~al.}{2011}]{Fisher:2011ApJS..194...39}
{Fischer} T.,  et~al., 2011, \mn@doi [\apjs] {10.1088/0067-0049/194/2/39},
  \href {http://adsabs.harvard.edu/abs/2011ApJS..194...39F} {194, 39}

\bibitem[\protect\citeauthoryear{{Fischer} et~al.,}{{Fischer}
  et~al.}{2018}]{Fischer_2018NatAs...2..980}
{Fischer} T.,  et~al., 2018, \mn@doi [Nature Astronomy]
  {10.1038/s41550-018-0583-0}, \href
  {http://adsabs.harvard.edu/abs/2018NatAs...2..980F} {2, 980}

\bibitem[\protect\citeauthoryear{{Font} et~al.,}{{Font}
  et~al.}{2002a}]{Font2002a}
{Font} J.~A.,  et~al., 2002a, \mn@doi [\prd] {10.1103/PhysRevD.65.084024}, 65,
  084024

\bibitem[\protect\citeauthoryear{Font et~al.,}{Font
  et~al.}{2002b}]{font2002three}
Font J.~A.,  et~al., 2002b, Physical Review D, 65, 084024

\bibitem[\protect\citeauthoryear{{Giacomazzo} \& {Perna}}{{Giacomazzo} \&
  {Perna}}{2012}]{Giacomazzo2012}
{Giacomazzo} B.,  {Perna} R.,  2012, \mn@doi [\apjl]
  {10.1088/2041-8205/758/1/L8}, \href
  {http://adsabs.harvard.edu/abs/2012ApJ...758L...8G} {758, L8}

\bibitem[\protect\citeauthoryear{{Giacomazzo}, {Rezzolla}  \&
  {Stergioulas}}{{Giacomazzo} et~al.}{2011}]{Giacomazzo_etal_2011}
{Giacomazzo} B.,  {Rezzolla} L.,   {Stergioulas} N.,  2011, \mn@doi [\prd]
  {10.1103/PhysRevD.84.024022}, \href
  {http://adsabs.harvard.edu/abs/2011PhRvD..84b4022G} {84, 024022}

\bibitem[\protect\citeauthoryear{Glendenning}{Glendenning}{2000}]{Glendenning2000}
Glendenning N.~K.,  2000, Compact stars: nuclear physics, particle physics, and
  general relativity.
Springer, New York, USA

\bibitem[\protect\citeauthoryear{{Glendenning}}{{Glendenning}}{2001}]{Glendenning2001}
{Glendenning} N.~K.,  2001, \mn@doi [\physrep] {10.1016/S0370-1573(00)00080-6},
  \href {http://adsabs.harvard.edu/abs/2001PhR...342..393G} {342, 393}

\bibitem[\protect\citeauthoryear{{Glendenning} \& {Kettner}}{{Glendenning} \&
  {Kettner}}{2000}]{Glendenning2000a}
{Glendenning} N.~K.,  {Kettner} C.,  2000, \aap, \href
  {http://adsabs.harvard.edu/abs/2000A%26A...353L...9G} {353, L9}

\bibitem[\protect\citeauthoryear{{Guardone} \& {Vigevano}}{{Guardone} \&
  {Vigevano}}{2002}]{Guardone2002}
{Guardone} A.,  {Vigevano} L.,  2002, \mn@doi [Journal of Computational
  Physics] {10.1006/jcph.2001.6915}, \href
  {http://adsabs.harvard.edu/abs/2002JCoPh.175...50G} {175, 50}

\bibitem[\protect\citeauthoryear{{Guardone}, {Zamfirescu}  \&
  {Colonna}}{{Guardone} et~al.}{2010}]{Guardone2010}
{Guardone} A.,  {Zamfirescu} C.,   {Colonna} P.,  2010, \mn@doi [Journal of
  Fluid Mechanics] {10.1017/S0022112009991716}, \href
  {http://adsabs.harvard.edu/abs/2010JFM...642..127G} {642, 127}

\bibitem[\protect\citeauthoryear{{Guilet}, {M{\"u}ller}  \& {Janka}}{{Guilet}
  et~al.}{2015}]{Guilet15}
{Guilet} J.,  {M{\"u}ller} E.,   {Janka} H.-T.,  2015, \mn@doi [\mnras]
  {10.1093/mnras/stu2550}, \href
  {http://adsabs.harvard.edu/abs/2015MNRAS.447.3992G} {447, 3992}

\bibitem[\protect\citeauthoryear{{Guilet}, {M{\"u}ller}, {Janka}, {Rembiasz},
  {Obergaulinger}, {Cerd{\'a}-Dur{\'a}n}  \& {Aloy}}{{Guilet}
  et~al.}{2017}]{Guilet17}
{Guilet} J.,  {M{\"u}ller} E.,  {Janka} H.-T.,  {Rembiasz} T.,  {Obergaulinger}
  M.,  {Cerd{\'a}-Dur{\'a}n} P.,   {Aloy} M.-A.,  2017, in {Marcowith} A.,
  {Renaud} M.,  {Dubner} G.,  {Ray} A.,   {Bykov} A.,  eds,  IAU Symposium Vol.
  331, Supernova 1987A:30 years later - Cosmic Rays and Nuclei from Supernovae
  and their Aftermaths. pp 119--124 (\mn@eprint {arXiv} {1706.08733}),
  \mn@doi{10.1017/S1743921317004732}

\bibitem[\protect\citeauthoryear{{Gulminelli}, {Raduta}, {Oertel}  \&
  {Margueron}}{{Gulminelli} et~al.}{2013}]{Gulminelli:2013PhRvC..87e5809}
{Gulminelli} F.,  {Raduta} A.~R.,  {Oertel} M.,   {Margueron} J.,  2013,
  \mn@doi [\prc] {10.1103/PhysRevC.87.055809}, \href
  {http://adsabs.harvard.edu/abs/2013PhRvC..87e5809G} {87, 055809}

\bibitem[\protect\citeauthoryear{{Haensel} \& {Potekhin}}{{Haensel} \&
  {Potekhin}}{2004}]{Haensel2004}
{Haensel} P.,  {Potekhin} A.~Y.,  2004, \mn@doi [\aap]
  {10.1051/0004-6361:20041722}, \href
  {http://adsabs.harvard.edu/abs/2004A%26A...428..191H} {428, 191}

\bibitem[\protect\citeauthoryear{{Haensel}, {Levenfish}  \&
  {Yakovlev}}{{Haensel} et~al.}{2002}]{Haensel2002}
{Haensel} P.,  {Levenfish} K.~P.,   {Yakovlev} D.~G.,  2002, \mn@doi [\aap]
  {10.1051/0004-6361:20021112}, \href
  {http://adsabs.harvard.edu/abs/2002A%26A...394..213H} {394, 213}

\bibitem[\protect\citeauthoryear{{Haensel}, {Potekhin}  \&
  {Yakovlev}}{{Haensel} et~al.}{2007}]{Haensel2007}
{Haensel} P.,  {Potekhin} A.~Y.,   {Yakovlev} D.~G.,  eds, 2007, {Neutron Stars
  1: Equation of State and Structure}  Astrophysics and Space Science Library
  Vol. 326.
New York: Springer

\bibitem[\protect\citeauthoryear{{Heiselberg} \& {Hjorth-Jensen}}{{Heiselberg}
  \& {Hjorth-Jensen}}{2000}]{Heiselberg2000}
{Heiselberg} H.,  {Hjorth-Jensen} M.,  2000, \mn@doi [\physrep]
  {10.1016/S0370-1573(99)00110-6}, \href
  {http://adsabs.harvard.edu/abs/2000PhR...328..237H} {328, 237}

\bibitem[\protect\citeauthoryear{{Hempel} \& {Schaffner-Bielich}}{{Hempel} \&
  {Schaffner-Bielich}}{2010}]{Hempel:2010NuPhA.837..210}
{Hempel} M.,  {Schaffner-Bielich} J.,  2010, \mn@doi [Nuclear Physics A]
  {10.1016/j.nuclphysa.2010.02.010}, \href
  {http://adsabs.harvard.edu/abs/2010NuPhA.837..210H} {837, 210}

\bibitem[\protect\citeauthoryear{{Hempel}, {Fischer}, {Schaffner-Bielich}  \&
  {Liebend{\"o}rfer}}{{Hempel} et~al.}{2012}]{Hempel:2012ApJ...748...70}
{Hempel} M.,  {Fischer} T.,  {Schaffner-Bielich} J.,   {Liebend{\"o}rfer} M.,
  2012, \mn@doi [\apj] {10.1088/0004-637X/748/1/70}, \href
  {http://adsabs.harvard.edu/abs/2012ApJ...748...70H} {748, 70}

\bibitem[\protect\citeauthoryear{{Hempel}, {Dexheimer}, {Schramm}  \&
  {Iosilevskiy}}{{Hempel} et~al.}{2013}]{Hempel_2013PhRvC..88a4906}
{Hempel} M.,  {Dexheimer} V.,  {Schramm} S.,   {Iosilevskiy} I.,  2013, \mn@doi
  [\prc] {10.1103/PhysRevC.88.014906}, \href
  {http://adsabs.harvard.edu/abs/2013PhRvC..88a4906H} {88, 014906}

\bibitem[\protect\citeauthoryear{{Ib{\'a}{\~n}ez}, {Cordero-Carri{\'o}n},
  {Mart{\'{\i}}}  \& {Miralles}}{{Ib{\'a}{\~n}ez} et~al.}{2013}]{Ibanez2013}
{Ib{\'a}{\~n}ez} J.~M.,  {Cordero-Carri{\'o}n} I.,  {Mart{\'{\i}}} J.~M.,
  {Miralles} J.~A.,  2013, \mn@doi [Classical and Quantum Gravity]
  {10.1088/0264-9381/30/5/057002}, \href
  {http://adsabs.harvard.edu/abs/2013CQGra..30e7002I} {30, 057002}

\bibitem[\protect\citeauthoryear{{Ib{\'a}{\~n}ez}, {Marquina}, {Serna}  \&
  {Aloy}}{{Ib{\'a}{\~n}ez} et~al.}{2018}]{Ibanez:2017TUBOS}
{Ib{\'a}{\~n}ez} J.~M.,  {Marquina} A.,  {Serna} S.,   {Aloy} M.~A.,  2018,
  \mn@doi [\mnras] {10.1093/mnras/sty137}, \href
  {http://adsabs.harvard.edu/abs/2018MNRAS.476.1100I} {476, 1100}

\bibitem[\protect\citeauthoryear{{Iosilevskiy}}{{Iosilevskiy}}{2010}]{Iosilevskiy_2010arXiv1005.4186}
{Iosilevskiy} I.,  2010, Acta Phys. Polon. B (Proc. Supl.), \href
  {http://adsabs.harvard.edu/abs/2010arXiv1005.4186I} {3, 589}

\bibitem[\protect\citeauthoryear{{Janka}, {Hanke}, {H{\"u}depohl}, {Marek},
  {M{\"u}ller}  \& {Obergaulinger}}{{Janka} et~al.}{2012}]{Janka_etal:2012}
{Janka} H.-T.,  {Hanke} F.,  {H{\"u}depohl} L.,  {Marek} A.,  {M{\"u}ller} B.,
   {Obergaulinger} M.,  2012, \mn@doi [Progress of Theoretical and Experimental
  Physics] {10.1093/ptep/pts067}, \href
  {http://adsabs.harvard.edu/abs/2012PTEP.2012aA309J} {2012, 01A309}

\bibitem[\protect\citeauthoryear{{Lalazissis}, {K{\"o}nig}  \&
  {Ring}}{{Lalazissis} et~al.}{1997}]{Lalazissis:1997PhRvC..55..540}
{Lalazissis} G.~A.,  {K{\"o}nig} J.,   {Ring} P.,  1997, \mn@doi [\prc]
  {10.1103/PhysRevC.55.540}, \href
  {http://adsabs.harvard.edu/abs/1997PhRvC..55..540L} {55, 540}

\bibitem[\protect\citeauthoryear{{Lattimer} \& {Swesty}}{{Lattimer} \&
  {Swesty}}{1991}]{Lattimer:1991NuPhA.535..331}
{Lattimer} J.~M.,  {Swesty} F.~D.,  1991, \mn@doi [Nuclear Physics A]
  {10.1016/0375-9474(91)90452-C}, \href
  {http://adsabs.harvard.edu/abs/1991NuPhA.535..331L} {535, 331}

\bibitem[\protect\citeauthoryear{{Maione}, {De Pietri}, {Feo}  \&
  {L{\"o}ffler}}{{Maione} et~al.}{2016}]{Maione2016}
{Maione} F.,  {De Pietri} R.,  {Feo} A.,   {L{\"o}ffler} F.,  2016, \mn@doi
  [Classical and Quantum Gravity] {10.1088/0264-9381/33/17/175009}, \href
  {http://adsabs.harvard.edu/abs/2016CQGra..33q5009M} {33, 175009}

\bibitem[\protect\citeauthoryear{{Malik}, {Alam}, {Fortin}, {Provid{\^e}ncia},
  {Agrawal}, {Jha}, {Kumar}  \& {Patra}}{{Malik}
  et~al.}{2018}]{Malik_2018PhRvC..98c5804}
{Malik} T.,  {Alam} N.,  {Fortin} M.,  {Provid{\^e}ncia} C.,  {Agrawal} B.~K.,
  {Jha} T.~K.,  {Kumar} B.,   {Patra} S.~K.,  2018, \mn@doi [\prc]
  {10.1103/PhysRevC.98.035804}, \href
  {http://adsabs.harvard.edu/abs/2018PhRvC..98c5804M} {98, 035804}

\bibitem[\protect\citeauthoryear{{Marek}, {Dimmelmeier}, {Janka}, {M{\"u}ller}
  \& {Buras}}{{Marek} et~al.}{2006}]{Marek_etal__2006__AA__TOV-potential}
{Marek} A.,  {Dimmelmeier} H.,  {Janka} H.-T.,  {M{\"u}ller} E.,   {Buras} R.,
  2006, \mn@doi [\aap] {10.1051/0004-6361:20052840}, \href
  {http://esoads.eso.org/abs/2006A%26A...445..273M} {445, 273}

\bibitem[\protect\citeauthoryear{{Margalit} \& {Metzger}}{{Margalit} \&
  {Metzger}}{2017}]{Margalit_2017ApJ...850L..19}
{Margalit} B.,  {Metzger} B.~D.,  2017, \mn@doi [\apjl]
  {10.3847/2041-8213/aa991c}, \href
  {http://adsabs.harvard.edu/abs/2017ApJ...850L..19M} {850, L19}

\bibitem[\protect\citeauthoryear{{Marques}, {Oertel}, {Hempel}  \&
  {Novak}}{{Marques} et~al.}{2017}]{Marques_2017PhRvC..96d5806}
{Marques} M.,  {Oertel} M.,  {Hempel} M.,   {Novak} J.,  2017, \mn@doi [\prc]
  {10.1103/PhysRevC.96.045806}, \href
  {http://adsabs.harvard.edu/abs/2017PhRvC..96d5806M} {96, 045806}

\bibitem[\protect\citeauthoryear{{Mart{\'{\i}}}, {Miralles}, {Ibanez}  \& {Diaz
  Alonso}}{{Mart{\'{\i}}} et~al.}{1988}]{Marti1988}
{Mart{\'{\i}}} J.~M.,  {Miralles} J.~A.,  {Ibanez} J.~M.,   {Diaz Alonso} J.,
  1988, \mn@doi [\apj] {10.1086/166420}, \href
  {http://adsabs.harvard.edu/abs/1988ApJ...329..780M} {329, 780}

\bibitem[\protect\citeauthoryear{{Menikoff} \& {Plohr}}{{Menikoff} \&
  {Plohr}}{1989}]{Menikoff1989}
{Menikoff} R.,  {Plohr} B.~J.,  1989, \mn@doi [Reviews of Modern Physics]
  {10.1103/RevModPhys.61.75}, \href
  {http://adsabs.harvard.edu/abs/1989RvMP...61...75M} {61, 75}

\bibitem[\protect\citeauthoryear{Montero \& Cordero-Carrion}{Montero \&
  Cordero-Carrion}{2012}]{Montero:2012yr}
Montero P.~J.,  Cordero-Carrion I.,  2012, \mn@doi [Phys.Rev.]
  {10.1103/PhysRevD.85.124037}, D85, 124037

\bibitem[\protect\citeauthoryear{{Montero}, {Baumgarte}  \&
  {M{\"u}ller}}{{Montero} et~al.}{2014}]{Montero2014}
{Montero} P.~J.,  {Baumgarte} T.~W.,   {M{\"u}ller} E.,  2014, \mn@doi [\prd]
  {10.1103/PhysRevD.89.084043}, \href
  {http://adsabs.harvard.edu/abs/2014PhRvD..89h4043M} {89, 084043}

\bibitem[\protect\citeauthoryear{{Most}, {Papenfort}, {Dexheimer}, {Hanauske},
  {Schramm}, {St{\"o}cker}  \& {Rezzolla}}{{Most}
  et~al.}{2018a}]{Most_2018arXiv180703684M}
{Most} E.~R.,  {Papenfort} L.~J.,  {Dexheimer} V.,  {Hanauske} M.,  {Schramm}
  S.,  {St{\"o}cker} H.,   {Rezzolla} L.,  2018a, arXiv e-prints, \href
  {http://cdsads.u-strasbg.fr/abs/2018arXiv180703684M} {}

\bibitem[\protect\citeauthoryear{{Most}, {Weih}, {Rezzolla}  \&
  {Schaffner-Bielich}}{{Most} et~al.}{2018b}]{Most_2018PhRvL.120z1103M}
{Most} E.~R.,  {Weih} L.~R.,  {Rezzolla} L.,   {Schaffner-Bielich} J.,  2018b,
  \mn@doi [Physical Review Letters] {10.1103/PhysRevLett.120.261103}, \href
  {http://cdsads.u-strasbg.fr/abs/2018PhRvL.120z1103M} {120, 261103}

\bibitem[\protect\citeauthoryear{{M{\"u}ller} \& {Eriguchi}}{{M{\"u}ller} \&
  {Eriguchi}}{1985}]{Mueller1985}
{M{\"u}ller} E.,  {Eriguchi} Y.,  1985, \aap, \href
  {http://adsabs.harvard.edu/abs/1985A%26A...152..325M} {152, 325}

\bibitem[\protect\citeauthoryear{{Nakazato}, {Sumiyoshi}  \&
  {Yamada}}{{Nakazato} et~al.}{2008}]{Nakazato_2008PhRvD..77j3006}
{Nakazato} K.,  {Sumiyoshi} K.,   {Yamada} S.,  2008, \mn@doi [\prd]
  {10.1103/PhysRevD.77.103006}, \href
  {http://adsabs.harvard.edu/abs/2008PhRvD..77j3006N} {77, 103006}

\bibitem[\protect\citeauthoryear{{Nandi} \& {Schramm}}{{Nandi} \&
  {Schramm}}{2017}]{Nandi_2017PhRvC..95f5801}
{Nandi} R.,  {Schramm} S.,  2017, \mn@doi [\prc] {10.1103/PhysRevC.95.065801},
  \href {http://adsabs.harvard.edu/abs/2017PhRvC..95f5801N} {95, 065801}

\bibitem[\protect\citeauthoryear{{Newman} \& {Penrose}}{{Newman} \&
  {Penrose}}{1962}]{Newman1962}
{Newman} E.,  {Penrose} R.,  1962, \mn@doi [Journal of Mathematical Physics]
  {10.1063/1.1724257}, \href
  {http://adsabs.harvard.edu/abs/1962JMP.....3..566N} {3, 566}

\bibitem[\protect\citeauthoryear{O'Connor \& Ott}{O'Connor \&
  Ott}{2010}]{OConnor:2010}
O'Connor E.,  Ott C.~D.,  2010, Classical and Quantum Gravity, 27, 114103

\bibitem[\protect\citeauthoryear{{O'Connor} \& {Ott}}{{O'Connor} \&
  {Ott}}{2011}]{OConnor2011}
{O'Connor} E.,  {Ott} C.~D.,  2011, \mn@doi [\apj]
  {10.1088/0004-637X/730/2/70}, \href
  {http://adsabs.harvard.edu/abs/2011ApJ...730...70O} {730, 70}

\bibitem[\protect\citeauthoryear{{Obergaulinger} \& {Aloy}}{{Obergaulinger} \&
  {Aloy}}{2017}]{Obergaulinger_Aloy:2017}
{Obergaulinger} M.,  {Aloy} M.~{\'A}.,  2017, \mn@doi [\mnras]
  {10.1093/mnrasl/slx046}, \href
  {http://adsabs.harvard.edu/abs/2017MNRAS.469L..43O} {469, L43}

\bibitem[\protect\citeauthoryear{{Obergaulinger}, {Janka}  \&
  {Aloy}}{{Obergaulinger} et~al.}{2014}]{Obergaulinger:2014MNRAS.445.3169}
{Obergaulinger} M.,  {Janka} H.-T.,   {Aloy} M.~A.,  2014, \mn@doi [\mnras]
  {10.1093/mnras/stu1969}, \href
  {http://adsabs.harvard.edu/abs/2014MNRAS.445.3169O} {445, 3169}

\bibitem[\protect\citeauthoryear{{Obergaulinger}, {Just}  \&
  {Aloy}}{{Obergaulinger} et~al.}{2018}]{Obergaulinger_Just_Aloy:2018}
{Obergaulinger} M.,  {Just} O.,   {Aloy} M.~{\'A}.,  2018, Journal of Physics
  G, 45, 084001

\bibitem[\protect\citeauthoryear{{Oertel}, {Fantina}  \& {Novak}}{{Oertel}
  et~al.}{2012}]{Oertel:2012PhRvC..85e5806}
{Oertel} M.,  {Fantina} A.~F.,   {Novak} J.,  2012, \mn@doi [\prc]
  {10.1103/PhysRevC.85.055806}, \href
  {http://adsabs.harvard.edu/abs/2012PhRvC..85e5806O} {85, 055806}

\bibitem[\protect\citeauthoryear{{Pais}, {Newton}  \& {Stone}}{{Pais}
  et~al.}{2014}]{Pais_2014PhRvC..90f5802}
{Pais} H.,  {Newton} W.~G.,   {Stone} J.~R.,  2014, \mn@doi [\prc]
  {10.1103/PhysRevC.90.065802}, \href
  {http://adsabs.harvard.edu/abs/2014PhRvC..90f5802P} {90, 065802}

\bibitem[\protect\citeauthoryear{Peres}{Peres}{2013}]{Peres2013b}
Peres B.,  2013, PhD thesis, University of Paris VII

\bibitem[\protect\citeauthoryear{{Peres}, {Oertel}  \& {Novak}}{{Peres}
  et~al.}{2013a}]{Peres2013}
{Peres} B.,  {Oertel} M.,   {Novak} J.,  2013a, \mn@doi [\prd]
  {10.1103/PhysRevD.87.043006}, \href
  {http://adsabs.harvard.edu/abs/2013PhRvD..87d3006P} {87, 043006}

\bibitem[\protect\citeauthoryear{{Peres}, {Oertel}  \& {Novak}}{{Peres}
  et~al.}{2013b}]{Peres:2013PhRvD..87d3006}
{Peres} B.,  {Oertel} M.,   {Novak} J.,  2013b, \mn@doi [\prd]
  {10.1103/PhysRevD.87.043006}, \href
  {http://adsabs.harvard.edu/abs/2013PhRvD..87d3006P} {87, 043006}

\bibitem[\protect\citeauthoryear{Pons}{Pons}{1999}]{Pons1999}
Pons J.~A.,  1999, PhD thesis, University of Valencia

\bibitem[\protect\citeauthoryear{{Pons}, {Reddy}, {Prakash}, {Lattimer}  \&
  {Miralles}}{{Pons} et~al.}{1999}]{Pons_1999ApJ...513..780}
{Pons} J.~A.,  {Reddy} S.,  {Prakash} M.,  {Lattimer} J.~M.,   {Miralles}
  J.~A.,  1999, \mn@doi [\apj] {10.1086/306889}, \href
  {http://adsabs.harvard.edu/abs/1999ApJ...513..780P} {513, 780}

\bibitem[\protect\citeauthoryear{{Radice}, {Perego}, {Zappa}  \&
  {Bernuzzi}}{{Radice} et~al.}{2018}]{Radice_2018ApJ...852L..29}
{Radice} D.,  {Perego} A.,  {Zappa} F.,   {Bernuzzi} S.,  2018, \mn@doi [\apjl]
  {10.3847/2041-8213/aaa402}, \href
  {http://adsabs.harvard.edu/abs/2018ApJ...852L..29R} {852, L29}

\bibitem[\protect\citeauthoryear{Raduta \& Gulminelli}{Raduta \&
  Gulminelli}{2010}]{Raduta_2010PhysRevC.82.065801}
Raduta A.~R.,  Gulminelli F.,  2010, \mn@doi [Phys. Rev. C]
  {10.1103/PhysRevC.82.065801}, 82, 065801

\bibitem[\protect\citeauthoryear{{Raithel}, {{\"O}zel}  \& {Psaltis}}{{Raithel}
  et~al.}{2018}]{Raithel_2018ApJ...857L..23}
{Raithel} C.~A.,  {{\"O}zel} F.,   {Psaltis} D.,  2018, \mn@doi [\apjl]
  {10.3847/2041-8213/aabcbf}, \href
  {http://adsabs.harvard.edu/abs/2018ApJ...857L..23R} {857, L23}

\bibitem[\protect\citeauthoryear{{Sagert}, {Fischer}, {Hempel}, {Pagliara},
  {Schaffner-Bielich}, {Mezzacappa}, {Thielemann}  \&
  {Liebend{\"o}rfer}}{{Sagert} et~al.}{2009}]{Sagert:2009PhRvL.102h1101}
{Sagert} I.,  {Fischer} T.,  {Hempel} M.,  {Pagliara} G.,  {Schaffner-Bielich}
  J.,  {Mezzacappa} A.,  {Thielemann} F.-K.,   {Liebend{\"o}rfer} M.,  2009,
  \mn@doi [Physical Review Letters] {10.1103/PhysRevLett.102.081101}, \href
  {http://adsabs.harvard.edu/abs/2009PhRvL.102h1101S} {102, 081101}

\bibitem[\protect\citeauthoryear{{Sagert}, {Fischer}, {Hempel}, {Pagliara},
  {Schaffner-Bielich}, {Thielemann}  \& {Liebend{\"o}rfer}}{{Sagert}
  et~al.}{2010}]{Sagert:2010JPhG...37i4064}
{Sagert} I.,  {Fischer} T.,  {Hempel} M.,  {Pagliara} G.,  {Schaffner-Bielich}
  J.,  {Thielemann} F.-K.,   {Liebend{\"o}rfer} M.,  2010, \mn@doi [Journal of
  Physics G Nuclear Physics] {10.1088/0954-3899/37/9/094064}, \href
  {http://adsabs.harvard.edu/abs/2010JPhG...37i4064S} {37, 094064}

\bibitem[\protect\citeauthoryear{{Sanchis-Gual}, {Degollado}, {Montero}, {Font}
   \& {Mewes}}{{Sanchis-Gual} et~al.}{2015}]{Sanchis-Gual:2015sms}
{Sanchis-Gual} N.,  {Degollado} J.~C.,  {Montero} P.~J.,  {Font} J.~A.,
  {Mewes} V.,  2015, \mn@doi [\prd] {10.1103/PhysRevD.92.083001}, \href
  {http://adsabs.harvard.edu/abs/2015PhRvD..92h3001S} {92, 083001}

\bibitem[\protect\citeauthoryear{{Sanchis-Gual}, {Herdeiro}, {Radu},
  {Degollado}  \& {Font}}{{Sanchis-Gual} et~al.}{2017}]{Sanchis-Gual:2017ps}
{Sanchis-Gual} N.,  {Herdeiro} C.,  {Radu} E.,  {Degollado} J.~C.,   {Font}
  J.~A.,  2017, \mn@doi [\prd] {10.1103/PhysRevD.95.104028}, \href
  {http://adsabs.harvard.edu/abs/2017PhRvD..95j4028S} {95, 104028}

\bibitem[\protect\citeauthoryear{{Schertler}, {Greiner}, {Schaffner-Bielich}
  \& {Thoma}}{{Schertler} et~al.}{2000}]{Schertler2000}
{Schertler} K.,  {Greiner} C.,  {Schaffner-Bielich} J.,   {Thoma} M.~H.,  2000,
  \mn@doi [Nuclear Physics A] {10.1016/S0375-9474(00)00305-5}, \href
  {http://adsabs.harvard.edu/abs/2000NuPhA.677..463S} {677, 463}

\bibitem[\protect\citeauthoryear{{Sch{\"u}rhoff}, {Schramm}  \&
  {Dexheimer}}{{Sch{\"u}rhoff} et~al.}{2010}]{Schurhoff10}
{Sch{\"u}rhoff} T.,  {Schramm} S.,   {Dexheimer} V.,  2010, \mn@doi [\apjl]
  {10.1088/2041-8205/724/1/L74}, \href
  {http://adsabs.harvard.edu/abs/2010ApJ...724L..74S} {724, L74}

\bibitem[\protect\citeauthoryear{{Serna} \& {Marquina}}{{Serna} \&
  {Marquina}}{2014}]{Serna_Marquina:2014}
{Serna} S.,  {Marquina} A.,  2014, \mn@doi [Physics of Fluids]
  {10.1063/1.4851415}, \href
  {http://adsabs.harvard.edu/abs/2014PhFl...26a6101S} {26, 016101}

\bibitem[\protect\citeauthoryear{{Shapiro} \& {Teukolsky}}{{Shapiro} \&
  {Teukolsky}}{1983}]{Shapiro1983}
{Shapiro} S.~L.,  {Teukolsky} S.~A.,  1983, {Black holes, white dwarfs, and
  neutron stars: The physics of compact objects}.
Wiley-Interscience

\bibitem[\protect\citeauthoryear{{Shen}, {Toki}, {Oyamatsu}  \&
  {Sumiyoshi}}{{Shen} et~al.}{1998a}]{Shen:1998PThPh.100.1013}
{Shen} H.,  {Toki} H.,  {Oyamatsu} K.,   {Sumiyoshi} K.,  1998a, \mn@doi
  [Progress of Theoretical Physics] {10.1143/PTP.100.1013}, \href
  {http://adsabs.harvard.edu/abs/1998PThPh.100.1013S} {100, 1013}

\bibitem[\protect\citeauthoryear{{Shen}, {Toki}, {Oyamatsu}  \&
  {Sumiyoshi}}{{Shen} et~al.}{1998b}]{Shen:1998NuPhA.637..435}
{Shen} H.,  {Toki} H.,  {Oyamatsu} K.,   {Sumiyoshi} K.,  1998b, \mn@doi
  [Nuclear Physics A] {10.1016/S0375-9474(98)00236-X}, \href
  {http://adsabs.harvard.edu/abs/1998NuPhA.637..435S} {637, 435}

\bibitem[\protect\citeauthoryear{{Shen}, {Horowitz}  \& {Teige}}{{Shen}
  et~al.}{2011a}]{Shen:2011PhRvC..83c5802}
{Shen} G.,  {Horowitz} C.~J.,   {Teige} S.,  2011a, \mn@doi [\prc]
  {10.1103/PhysRevC.83.035802}, \href
  {http://adsabs.harvard.edu/abs/2011PhRvC..83c5802S} {83, 035802}

\bibitem[\protect\citeauthoryear{{Shen}, {Horowitz}  \& {O'Connor}}{{Shen}
  et~al.}{2011b}]{Shen:2011PhRvC..83f5808}
{Shen} G.,  {Horowitz} C.~J.,   {O'Connor} E.,  2011b, \mn@doi [\prc]
  {10.1103/PhysRevC.83.065808}, \href
  {http://adsabs.harvard.edu/abs/2011PhRvC..83f5808S} {83, 065808}

\bibitem[\protect\citeauthoryear{{Shibata} \& {Nakamura}}{{Shibata} \&
  {Nakamura}}{1995}]{Shibata1995}
{Shibata} M.,  {Nakamura} T.,  1995, \mn@doi [\prd] {10.1103/PhysRevD.52.5428},
  52, 5428

\bibitem[\protect\citeauthoryear{{Steiner}, {Fischer}, {Gandolfi}  \&
  {Hempel}}{{Steiner} et~al.}{2012}]{Steiner:2012nuco.confE..38}
{Steiner} A.,  {Fischer} T.,  {Gandolfi} S.,   {Hempel} M.,  2012, in Nuclei in
  the Cosmos (NIC XII). p.~38

\bibitem[\protect\citeauthoryear{{Stergioulas} \& {Friedman}}{{Stergioulas} \&
  {Friedman}}{1995}]{Stergioulas1995}
{Stergioulas} N.,  {Friedman} J.~L.,  1995, \mn@doi [\apj] {10.1086/175605},
  \href {http://adsabs.harvard.edu/abs/1995ApJ...444..306S} {444, 306}

\bibitem[\protect\citeauthoryear{Stergioulas, Apostolatos  \& Font}{Stergioulas
  et~al.}{2004}]{stergioulas2004non}
Stergioulas N.,  Apostolatos T.~A.,   Font J.~A.,  2004, Monthly Notices of the
  Royal Astronomical Society, 352, 1089

\bibitem[\protect\citeauthoryear{{Sugahara} \& {Toki}}{{Sugahara} \&
  {Toki}}{1994}]{Sugara:1994NuPhA.579..557}
{Sugahara} Y.,  {Toki} H.,  1994, \mn@doi [Nuclear Physics A]
  {10.1016/0375-9474(94)90923-7}, \href
  {http://adsabs.harvard.edu/abs/1994NuPhA.579..557S} {579, 557}

\bibitem[\protect\citeauthoryear{{Sumiyoshi}, {Yamada}, {Suzuki}, {Shen},
  {Chiba}  \& {Toki}}{{Sumiyoshi} et~al.}{2005}]{Sumiyoshi_2005ApJ...629..922}
{Sumiyoshi} K.,  {Yamada} S.,  {Suzuki} H.,  {Shen} H.,  {Chiba} S.,   {Toki}
  H.,  2005, \mn@doi [\apj] {10.1086/431788}, \href
  {http://adsabs.harvard.edu/abs/2005ApJ...629..922S} {629, 922}

\bibitem[\protect\citeauthoryear{Suwa}{Suwa}{2014}]{Suwa_2014doi:10.1093/pasj/pst030}
Suwa Y.,  2014, \mn@doi [Publications of the Astronomical Society of Japan]
  {10.1093/pasj/pst030}, 66, L1

\bibitem[\protect\citeauthoryear{{Thompson}}{{Thompson}}{1971}]{Thompson1971}
{Thompson} P.~A.,  1971, \mn@doi [Physics of Fluids] {10.1063/1.1693693}, \href
  {http://adsabs.harvard.edu/abs/1971PhFl...14.1843T} {14, 1843}

\bibitem[\protect\citeauthoryear{Thompson, Kim  \& Carofano}{Thompson
  et~al.}{1986}]{Thompson_1986JFM...166...57}
Thompson P.~A.,  Kim Y.~G.,   Carofano G.~C.,  1986, Journal of Fluid Mechanics
  (ISSN 0022-1120), 166, 57

\bibitem[\protect\citeauthoryear{{Thorne}}{{Thorne}}{1980}]{Thorne_1980RvMP...52..299}
{Thorne} K.~S.,  1980, \mn@doi [Reviews of Modern Physics]
  {10.1103/RevModPhys.52.299}, \href
  {http://adsabs.harvard.edu/abs/1980RvMP...52..299T} {52, 299}

\bibitem[\protect\citeauthoryear{{Todd-Rutel} \& {Piekarewicz}}{{Todd-Rutel} \&
  {Piekarewicz}}{2005}]{Todd-Rutel:2005PhRvL..95l2501}
{Todd-Rutel} B.~G.,  {Piekarewicz} J.,  2005, \mn@doi [Physical Review Letters]
  {10.1103/PhysRevLett.95.122501}, \href
  {http://adsabs.harvard.edu/abs/2005PhRvL..95l2501T} {95, 122501}

\bibitem[\protect\citeauthoryear{{Toki}, {Hirata}, {Sugahara}, {Sumiyoshi}  \&
  {Tanihata}}{{Toki} et~al.}{1995}]{Toki:1995NuPhA.588..357}
{Toki} H.,  {Hirata} D.,  {Sugahara} Y.,  {Sumiyoshi} K.,   {Tanihata} I.,
  1995, \mn@doi [Nuclear Physics A] {10.1016/0375-9474(95)00161-S}, \href
  {http://adsabs.harvard.edu/abs/1995NuPhA.588..357T} {588, 357}

\bibitem[\protect\citeauthoryear{{Typel}, {R{\"o}pke}, {Kl{\"a}hn}, {Blaschke}
  \& {Wolter}}{{Typel} et~al.}{2010}]{Typel:2010PhRvC..81a5803}
{Typel} S.,  {R{\"o}pke} G.,  {Kl{\"a}hn} T.,  {Blaschke} D.,   {Wolter} H.~H.,
   2010, \mn@doi [\prc] {10.1103/PhysRevC.81.015803}, \href
  {http://adsabs.harvard.edu/abs/2010PhRvC..81a5803T} {81, 015803}

\bibitem[\protect\citeauthoryear{{Vaidya}, {Mignone}, {Bodo}  \&
  {Massaglia}}{{Vaidya} et~al.}{2015}]{Vaidya:2015A&A...580A.110V}
{Vaidya} B.,  {Mignone} A.,  {Bodo} G.,   {Massaglia} S.,  2015, \mn@doi [\aap]
  {10.1051/0004-6361/201526247}, \href
  {http://adsabs.harvard.edu/abs/2015A%26A...580A.110V} {580, A110}

\bibitem[\protect\citeauthoryear{Voss}{Voss}{2005}]{Voss2005}
Voss A.,  2005, PhD thesis, University of Wuppertal

\bibitem[\protect\citeauthoryear{{Weber}}{{Weber}}{2005}]{Weber2005}
{Weber} F.,  2005, \mn@doi [Progress in Particle and Nuclear Physics]
  {10.1016/j.ppnp.2004.07.001}, \href
  {http://adsabs.harvard.edu/abs/2005PrPNP..54..193W} {54, 193}

\bibitem[\protect\citeauthoryear{{Woosley} \& {Heger}}{{Woosley} \&
  {Heger}}{2007}]{Woosley2007}
{Woosley} S.~E.,  {Heger} A.,  2007, \mn@doi [\physrep]
  {10.1016/j.physrep.2007.02.009}, \href
  {http://adsabs.harvard.edu/abs/2007PhR...442..269W} {442, 269}

\bibitem[\protect\citeauthoryear{{Zdunik} \& {Haensel}}{{Zdunik} \&
  {Haensel}}{2013}]{Zdunik2013}
{Zdunik} J.~L.,  {Haensel} P.,  2013, \mn@doi [\aap]
  {10.1051/0004-6361/201220697}, \href
  {http://adsabs.harvard.edu/abs/2013A%26A...551A..61Z} {551, A61}

\bibitem[\protect\citeauthoryear{{Zel'dovich}}{{Zel'dovich}}{1946}]{Zeldovich1946}
{Zel'dovich} Y.,  1946, Zh. Eksp. Teor. Fiz., 4, 363

\bibitem[\protect\citeauthoryear{{Zhou}, {Zhou}  \& {Li}}{{Zhou}
  et~al.}{2018}]{Zhou_2018PhRvD..97h3015}
{Zhou} E.-P.,  {Zhou} X.,   {Li} A.,  2018, \mn@doi [\prd]
  {10.1103/PhysRevD.97.083015}, \href
  {http://adsabs.harvard.edu/abs/2018PhRvD..97h3015Z} {97, 083015}

\bibitem[\protect\citeauthoryear{{van Riper}}{{van Riper}}{1978}]{vanRiper1978}
{van Riper} K.~A.,  1978, \mn@doi [\apj] {10.1086/156029}, \href
  {http://adsabs.harvard.edu/abs/1978ApJ...221..304V} {221, 304}

\makeatother
\end{thebibliography}
\end{NoHyper}
\end{document}